\documentclass[letterpaper]{article}

\usepackage{amsmath,amssymb,amsthm}
\usepackage[letterpaper,margin=1in]{geometry}

\usepackage{xcolor}
\usepackage{graphicx}
\usepackage{needspace}
\usepackage{caption}
\usepackage{booktabs}
\usepackage{microtype}
\usepackage{lmodern}
\usepackage{enumitem}
\usepackage{mathtools}
\usepackage{float}
\usepackage{flafter}
\usepackage{aliascnt}
\floatstyle{ruled}
\newfloat{algorithm}{htbp}{loa}
\floatname{algorithm}{Algorithm}
\usepackage[
    colorlinks=true,
    linkcolor=blue!62!black,
    citecolor=green!48!black,
    urlcolor=blue!70!black,
    linktoc=page
]{hyperref}
\usepackage[nameinlink,noabbrev]{cleveref}
\newtheorem{theorem}{Theorem}
\newaliascnt{lemma}{theorem}
\newtheorem{lemma}[lemma]{Lemma}
\aliascntresetthe{lemma}
\newaliascnt{corollary}{theorem}
\newtheorem{corollary}[corollary]{Corollary}
\aliascntresetthe{corollary}
\newaliascnt{observation}{theorem}
\newtheorem{observation}[observation]{Observation}
\aliascntresetthe{observation}
\newaliascnt{proposition}{theorem}
\newtheorem{proposition}[proposition]{Proposition}
\aliascntresetthe{proposition}

\crefname{theorem}{Theorem}{Theorems}
\Crefname{theorem}{Theorem}{Theorems}
\crefname{lemma}{Lemma}{Lemmas}
\Crefname{lemma}{Lemma}{Lemmas}
\crefname{corollary}{Corollary}{Corollaries}
\Crefname{corollary}{Corollary}{Corollaries}
\crefname{observation}{Observation}{Observations}
\Crefname{observation}{Observation}{Observations}
\crefname{proposition}{Proposition}{Propositions}
\Crefname{proposition}{Proposition}{Propositions}
\crefname{algorithm}{Algorithm}{Algorithms}
\Crefname{algorithm}{Algorithm}{Algorithms}
\crefname{figure}{Figure}{Figures}
\Crefname{figure}{Figure}{Figures}
\crefname{section}{Section}{Sections}
\Crefname{section}{Section}{Sections}
\crefname{appendix}{Appendix}{Appendices}
\Crefname{appendix}{Appendix}{Appendices}

\allowdisplaybreaks

\newcommand{\one}{\mathbf 1}
\newcommand{\EE}{\mathbb E}
\newcommand{\tr}{\operatorname{tr}}
\newcommand{\diag}{\operatorname{diag}}
\newcommand{\res}{\operatorname{res}}
\newcommand{\cbar}{\bar c}
\newcommand{\eps}{\varepsilon}

\newcommand{\ipd}[2]{\langle #1,#2\rangle_d}
\newcommand{\proofstep}[1]{\par\smallskip\noindent\textit{#1}\enspace}
\newcommand{\opt}{\operatorname{opt}}
\newcommand{\congestion}{\operatorname{cong}}
\newcommand{\nd}[1]{\lVert #1\rVert_d}

\newcommand{\kcl}{K_{\rm cl}}
\title{Cut--Matching with Attenuation: Faster Expander Decompositions and Approximate Max Flow}
\author{Trevor Vaughn\thanks{Carnegie Mellon University. Email: tnvaughn@cmu.edu}}
\hypersetup{pdftitle={Cut–Matching with Attenuation: Faster Expander Decompositions and Approximate Max Flow},pdfauthor={Trevor Vaughn}}
\date{}
\begin{document}
\maketitle
\begingroup
\setlength{\leftmargini}{2.4em}
\begin{abstract}
For a weighted graph with $n$ vertices and $m$ positive-capacity edges or arcs,
we compute an $(\Omega(\phi),O(\phi\log^2 n))$-expander decomposition
for $0<\phi\le1/\log n$. Our algorithms are randomized and take $O(n+m\log^4 n+m\log n/\phi)$
time in undirected graphs and $O(n+m\log^4 n+m\log n\log\log n/\phi)$ in
directed graphs, where only backward arcs are charged. For polynomial capacity
ratios, a second undirected algorithm takes $O(n+m\log^{13/2} n\log\log n)$
time independent of $\phi$.

A set with little starting mass can trap walks after other walks mix. Our cut
player seeks a vertex score whose value at a walk's endpoint remains predictable
from its start, then requests a matching across the weighted median. Before each
lazy step we discard a small fraction of mass at vertices with unusually large
absolute scores. Half the remaining mass stays in place. The other half follows
the matching where possible and is lost where unmatched. For later scores we
fill vacancies from the average origin distribution, weakening the trapped set's
signal even when few walks escape. After a completed game, trimming flows use
the missing endpoint mass to extract a large induced expander.

For undirected graphs with polynomial capacity ratios, we also use the cut
player to build a congestion approximator. With forest contraction parameter
$\chi\ge2$, construction takes $O(n+m\log^2 n+\min\{n,m/\chi\}\log^7 n\log\log n)$
time and gives quality $Q=O(\chi\log^2 n\log\log n)$. The resulting tree gives
a $(1+\rho)$-fair cut in $O(n+mQ\log^2 n/\rho)$ time. For arbitrary positive
capacities, we route any componentwise balanced demand at congestion at most
$1+\varepsilon$ times optimal in $O(n+m\log^5 n\log\log n/\sqrt\varepsilon
+m\log^3 n\log\log n/\varepsilon)$ time.
\end{abstract}
\endgroup

\clearpage
\begingroup
\fontsize{10.3}{11.3}\selectfont
\tableofcontents
\endgroup
\clearpage

\section{Introduction}\label{sec:introduction}

An expander decomposition partitions a graph into induced subgraphs of
high conductance while cutting a small fraction of its total capacity.
In an undirected graph every edge between parts must be cut. In a
directed graph we can order the parts and keep every arc that points
forward. On unweighted graphs, Saranurak and Wang~\cite{SW19} gave a
near-linear-time undirected algorithm at inverse-polylogarithmic
conductance, and Fleischmann, G.~Z.~Li, and J.~Li~\cite{FLL25} matched
its running time and its bound on the fraction cut in the directed case.
We give faster algorithms in both settings, even for arbitrary positive
capacities. The fraction cut in either case matches the sharper undirected
bound of Agassy, Dorfman, and Kaplan~\cite{ADK23}, a logarithmic
improvement for unweighted directed graphs.

In the cut--matching game~\cite{KRV09}, the players build a sequence of
matchings routed in the graph. If a walk through the matchings mixes, the
routes give paths across every cut, and their congestion gives a lower
bound on conductance. If a matching request exposes a sparse cut with
substantial weight on both sides, we can recurse on its two sides. A set
of small total vertex weight poses a different problem.
Even after walks have mixed elsewhere, those that start in the set may stay
there. A matching can leave the set intact even if almost all its pairs
cross the cut requested by the player.

Our cut player looks for a vertex score that retains this dependence on
the origin. When the score is spread over substantial vertex weight,
matching across its weighted median brings different origin distributions
together. A trapped set too small to force such a match can still carry
much of the squared score. The player also discards a small fraction of
walk mass at vertices with unusually large absolute scores, a step we
call \emph{attenuation}. Filling the vacancies from the average origin
distribution weakens the trapped set's signal in subsequent rounds
even when no routed walk escapes.

Once the walk used for scoring
has mixed, consider a cut side carrying at most half the prescribed
starting mass. A constant fraction of that mass either reaches the
other side along a routed walk or is missing. A trimming flow uses the
missing starts as supply. It finds routes out of sets with too few
crossing walks or identifies a sparse piece to remove. In
a directed graph, counting missing ends and routing in reverse gives the
incoming bound as well.

Fleischmann, Li, and Li~\cite{FLL26weak} used weak expander decompositions
to build a congestion approximator for undirected maximum flow. We use our
cut player to build a tree that estimates the congestion needed to route
any demand. Within a prospective cluster the player tests how well the endpoints
of edges between groups can reach one another. A separate cut accounts
for edges leaving the cluster and determines which vertices can be
attached together in the tree. In the maximum-flow search, the tree
bounds how much extra congestion would complete a trial flow. This
lets us compare throughputs and route the unmet demand only for the
final choice.

We also compute \emph{fair cuts}, for which one feasible flow nearly
saturates every crossing edge toward the sink~\cite{LNPS23faircuts}.
In an undirected graph this edgewise guarantee lets us trim a completed
game in a single cut. The same tree serves all the matching requests
and the trim on a phase graph. For polynomial capacity spread this gives
an undirected decomposition algorithm whose running time is independent
of the target conductance.

\subsection{Results}\label{sec:results}

Let $G$ be a simple weighted graph with $n$ vertices and $m$ edges or arcs of
positive real capacities totaling $W$. The capacities have \emph{polynomial
spread} if the ratio of the largest to the smallest is at most
$(m+2)^{O(1)}$. The time and space bounds below hold on every execution.

\Needspace{6\baselineskip}
\begin{theorem}[Expander decompositions]\label{thm:main}
Fix $0<\phi\le1/\log n$. With high probability the algorithm returns a
partition in which every part $J$ induces a graph of conductance at least
$c_*\phi$, for a constant $c_*>0$. Conductance uses degrees within the part
and, for a directed cut, the smaller of its incoming and outgoing
capacities (\Cref{sec:prelim}). In an undirected graph the edges between
parts have total capacity $O(\phi W\log^2 n)$. In a directed graph the
algorithm orders the parts so that the arcs from later parts to earlier
ones have total capacity $O(\phi W\log^2 n)$.

If the input also specifies a nonnegative vertex measure $\mu$ with
$\mu(V)\le W$, then on the same event every part $J$ satisfies
\[
 c_{G[J]}(S,J\setminus S)
 \ge c_*\phi\min\{\mu(S),\mu(J\setminus S)\}
 \qquad(S\subseteq J).
\]
In a directed graph this inequality holds in both directions. The
algorithms use $O(n+m/\phi)$ space. Their respective running times are
\begin{equation}\label{eq:main-times}
 T_{\rm undir}=O\!\left(n+m\log^4 n+\frac{m\log n}{\phi}\right),\qquad
 T_{\rm dir}=O\!\left(n+m\log^4 n+\frac{m\log n\log\log n}{\phi}\right).
\end{equation}
\end{theorem}

For undirected graphs of polynomial capacity spread, a second algorithm gives
the same partition guarantees, including the guarantee for the optional
measure $\mu$, in $O(n+m\log^{13/2} n\log\log n)$ time independent of
$\phi$ (\Cref{thm:near-optimal-decomp}).

\paragraph{Congestion approximators.}
For a demand $b$ in an undirected graph, $b(v)$ is the prescribed net
outflow at $v$. Assume $b$ sums to zero on each connected component and
write $\opt_G(b)$ for the least congestion of a flow routing $b$.
A \emph{congestion approximator} of quality $Q$ is a matrix $R_G$ satisfying
\[
             \|R_Gb\|_\infty\le\opt_G(b)\le Q\|R_Gb\|_\infty
\]
for every such $b$. Our rows are given by a rooted tree whose terminals are
the graph vertices. An edge of capacity $u>0$ with terminal set $S$ below
it must carry $b(S)=\sum_{v\in S}b(v)$ units of net flow. Its row gives
$b(S)/u$. The sets $S$ are nested or disjoint.

\begin{theorem}[Congestion approximators]\label{ca:thm:tradeoff}
For an undirected graph of polynomial capacity spread and a contraction
factor $\chi\ge2$, the algorithm takes time $T_{\rm CA}$ and, with high
probability, returns a laminar congestion approximator with $O(n)$ rows and
quality $Q$, where
\begin{equation}\label{ca:eq:tradeoff}
 T_{\rm CA}=O\!\left(n+m\log^2 n+\min\{n,m/\chi\}\log^7 n\log\log n\right),
 \qquad Q=O(\chi\log^2 n\log\log n).
\end{equation}
\end{theorem}

Contracting a forest leaves $O(\min\{n,m/\chi\})$ nonisolated vertices.
We build the hierarchy on a sparsifier of this quotient. Routing inside
the contracted components when we restore the forest accounts for the
factor $O(\chi)$ in quality (\Cref{sec:compressed-queries}). The tree
also approximates cuts. Given the sides of any graph cut on its terminals,
the cheapest extension to the tree costs between the graph cut capacity
and $Q$ times that capacity (\Cref{sec:terminal-tree}).

\paragraph{Approximate maximum flow.}
Let $G$ be an undirected graph with arbitrary positive capacities.
Given a demand $b$ that sums to zero on each connected component and
$0<\varepsilon\le1/2$, our algorithm finds, with high probability, a
flow that routes $b$ exactly at congestion at most
$(1+\varepsilon)\opt_G(b)$. The running time is
\begin{equation}\label{eq:intro-flow}
 O\!\left(n+\frac{m\log^5 n\log\log n}{\sqrt\varepsilon}
                 +\frac{m\log^3 n\log\log n}{\varepsilon}\right).
\end{equation}

The contraction factor trades time spent building the tree against time
spent using it at the requested accuracy. \Cref{ca:cor:flow} gives the
bound for polynomial capacity spread, and \Cref{app:capacities} removes
the spread assumption.

\paragraph{Fair cuts.}
A $(1+\rho)$-fair cut has a source side $S$ and a single feasible flow
from $s$ to $t$ that sends at least $c_e/(1+\rho)$ out of $S$ along
each crossing edge $e$. Summing over those edges also bounds the cut
capacity by $1+\rho$ times the flow value.

\begin{theorem}[Fair cuts]\label{thm:fair-intro}
Let $G$ be undirected with polynomial capacity spread, let $s,t$ be
distinct vertices, and let $0<\rho\le1/2$. With high probability we can
compute a $(1+\rho)$-fair cut and its witnessing flow in time
\begin{equation}\label{eq:intro-fair}
 O\!\left(n+\frac{m\log^{11/2} n\log\log n}{\sqrt\rho}
           +\frac{m\log^4 n\log\log n}{\rho}\right).
\end{equation}
With a quality-$Q$ tree prepared as in \Cref{fc:thm:fair}, it suffices
that, for $m>0$, the total capacity incident to $s$ is at most $(m+2)^{O(1)}$
times the least positive edge capacity. The cut and its witnessing flow
then take $O(n+mQ\log^2 n/\rho)$ deterministic
time. The matrix products for the tree of \Cref{ca:thm:tradeoff} can
be prepared in $O(n+m)$ time (\Cref{ca:prop:query-bounds}).
\end{theorem}

\subsection{Prior work}\label{sec:prior-work}

\paragraph{Undirected expander decomposition.}
Khandekar, Rao, and Vazirani introduced the cut--matching game to embed an
expander in a graph~\cite{KRV09}. For unweighted undirected graphs,
Saranurak and Wang obtained induced expansion $\Omega(\phi)$ with $O(\phi
m\log^3 n)$ edges between parts in $O(m\log^4 n/\phi)$ time~\cite{SW19}.
Agassy, Dorfman, and Kaplan reduced the crossing bound to $O(\phi m\log^2
n)$ using a spectral cut player. Their algorithm takes $O(m\log^7 n+m\log^4
n/\phi)$ time~\cite{ADK23}. They also studied decompositions for nonuniform
measures~\cite{ADK25}. Using fair cuts, Li, Nanongkai, Panigrahi, and
Saranurak obtained expansion $\phi$ and crossing capacity
$O(\phi W\log^3 n)$ in $\widetilde O(m)$ time independently of
$\phi$~\cite{LNPS23faircuts}.

\paragraph{Directed expander decomposition.}
In unweighted directed graphs, Sulser and Probst Gutenberg initialize a
decremental decomposition in $O(m\log^{20}m/\psi)$ time. For $\psi\le
c/\log^{12}m$ and a fixed constant $c>0$, the parts have expansion
$\Omega(\psi/\log^{12}m)$~\cite[Theorem~2]{SG25}. Fleischmann, G.~Z.~Li,
and J.~Li showed how to continue the directed cut--matching game after removing
a sparse set. This gave induced expansion $\Omega(\phi)$ with $O(\phi m\log^3 n)$ backward
arcs in $O(m\log^4 n/\phi)$ time~\cite{FLL25}. For integer capacities in
$[1,U]$, they bounded the backward capacity by $O(\phi W\log^8(nU))$ in
$O(m/\phi+m\log^7(nU))$ time.

\paragraph{Congestion approximators and tree cut sparsifiers.}
R\"acke, Shah, and T\"aubig constructed hierarchical congestion
approximators of quality $O(\log^4 n)$ in $m^{1+o(1)}$ time~\cite{RST14}.
Peng constructed a congestion approximator in $\widetilde O(m)$
time~\cite{Peng16}. Henzinger, M\"unk, and R\"acke achieved quality
$O(\log^2 n\log\log n)$ for single-commodity demands in $\widetilde O(m)$
time~\cite{HMR25}. Agassy, Dorfman, and Kaplan obtained tree cut
sparsifiers of the same quality in near-linear time~\cite{ADK26}. Our
construction modifies the nested hierarchy of Henzinger, M\"unk, and
R\"acke and uses a grouping strategy similar to that of Agassy, Dorfman,
and Kaplan. We keep the cluster graph fixed so that one recursively
constructed tree serves all grouping tests, then use a separate fair
cut to trim the exterior boundary.

\paragraph{Approximate maximum flow.}
Sherman used congestion approximators to compute approximate undirected
maximum flows~\cite{Sherman13}. Peng obtained $\widetilde O(m)$ time for
fixed accuracy~\cite{Peng16}. For polynomially bounded integer capacities,
Fleischmann, G.~Z.~Li, and J.~Li used weak decompositions to construct a
quality-$O(\log^5 n)$ congestion approximator in $O(m\log^{12} n\log\log
n)$ time. Their maximum-flow algorithm takes $O(m\log^{12} n\log\log
n+\varepsilon^{-1}m\log^6 n)$ time~\cite{FLL26weak}. Our optimization
routine uses the box--simplex algorithm of Jambulapati and
Tian~\cite{JT23}.

\paragraph{Fair cuts.}
Li, Nanongkai, Panigrahi, and Saranurak introduced fair cuts and gave an
algorithm that finds a $(1+\rho)$-fair cut in $\widetilde O(m/\rho^3)$
time~\cite[Theorem~1.3]{LNPS23faircuts}. For
polynomially bounded integer capacities, Li and Li obtained $\widetilde O(m/\rho)$ time by improving the flow or changing the cut in a residual
graph~\cite{LL25faircuts}. Our queries can replace the entire flow
within the original edge capacities. This lets us keep one tree throughout
and use its rows to track the net outflow still to be corrected
(\Cref{app:fc}).

\subsection{Technical overview of the expander decomposition}\label{sec:overview}

Run one game on a vertex set $C$, starting $d(v)$ units of walk mass
at each $v$. Write $d(S)=\sum_{v\in S}d(v)$ and $M=d(C)$. If the game
completes, let $Q(S,T)$ be the mass of retained walks from $S$ to $T$.
We want little loss overall, yet for every set $S$ of weight at most
$M/2$, we want at least $\gamma d(S)$ of its starting mass to be
missing or end outside $S$, for a constant $\gamma>0$. This is a
lower bound on the return shortfall $d(S)-Q(S,S)$. Counting the
shortfall at each end gives
\[
d(S)-Q(S,S)
 = \underbrace{d(S)-Q(S,C)}_{\text{missing starts in }S}
      + Q(S,C\setminus S)
 = \underbrace{d(S)-Q(C,S)}_{\text{missing ends in }S}
      + Q(C\setminus S,S).
\]
The two crossing terms count walks that leave and enter $S$, respectively.
Trimming will use the
missing starts and ends as flow supply for the corresponding direction.

A later deletion can destroy a walk even if both its endpoints survive.
For example, a walk from $u\in S$ to $v\notin S$ may lose an
intermediate vertex. Its weight then moves from the mass crossing out
of $S$ to the missing starts at $u$. At $v$ it counts as a missing end. More generally, if $J$ is the
remaining vertex set and $Q_J$ counts the walks lying entirely in $J$,
then for every $S\subseteq J$ of weight at most $M/2$,
\[
         d(S)-Q_J(S,S)\ge d(S)-Q(S,S)\ge\gamma d(S).
\]
The prescribed weight $d(S)$ is fixed, and deleting walks can only
reduce the mass returning to $S$. Thus the lower bound survives every
deletion. To apply it inside $J$, we update the endpoint deficits.

\paragraph{Finding a predictable score.}
First consider a walk whose origin $X$ and endpoint $Y$ both have
distribution $\pi=d/M$. Let $A$ be a small block with probability
$p=\pi(A)$, and suppose the walk mixes perfectly within $A$ and within
$\bar A=C\setminus A$ but never crosses between them. The endpoint reveals
exactly one fact about the origin, namely its block. The mutual information
$I(X;Y)$ is the entropy of the block indicator and tends to zero with $p$.
Yet the origin predicts $\one_A(Y)$ exactly, however small $p$ is. We
therefore look for an endpoint label for which knowing the origin
explains a large \emph{fraction} of its variance.

For a nonzero centered label $y$ on the endpoints, the best prediction from
the origin is $z(X)=\mathbb E[y(Y)\mid X]$. Conditioning on $X$ gives the
mean squared prediction error
\[
 \mathbb E[(y(Y)-z(X))^2]
 =\mathbb E\!\left[\mathbb E[y(Y)^2\mid X]-z(X)^2\right]
 =\mathbb E[y(Y)^2]-\mathbb E[z(X)^2].
\]
Let $P(v,u)=\Pr[X=u\mid Y=v]$. Applying $P$ averages an origin label over
walks ending at each vertex. In the inner product $\langle
f,g\rangle_d=\sum_v d(v)f(v)g(v)$, the adjoint averages in the reverse
direction, since $P^*(u,v)=\pi(v)P(v,u)/\pi(u)=\Pr[Y=v\mid X=u]$. Thus
$z=P^*y$. Write $G=P-\one\pi^\top$ for the centered averaging operator
and set $\mathcal S=GG^*$. Since $y$
is centered, the fraction of variance explained by the origin is
\[
 \frac{\mathbb E[z(X)^2]}{\mathbb E[y(Y)^2]}
 =\frac{\|G^*y\|_d^2}{\|y\|_d^2}
 =\frac{\langle y,\mathcal S y\rangle_d}{\|y\|_d^2}.
\]
In the two-block walk $y=\one_A-p\one$ attains one, and a fraction $1-p$ of
its squared norm lies in $A$.

One predictable label can miss other traps. If the walk stays within
several blocks, a score supported on two of them can be fully predictable
while every remaining block ties at the median. The player samples all
predictable directions together. Fix an even round limit $r=\Theta(\log
n)$. In each round draw independent uniform signs $\xi_v$, set
$g(v)=\xi_v/\sqrt{d(v)}$, and use $x=\mathcal S^{r/2}g$ as the score. Along
any eigenvector of unit $d$-norm the component of $g$ has mean square one.
If the eigenvalue is $\beta$, the component of $x$ has mean square
$\beta^r$. Thus
\[
                   \mathbb E\|x\|_d^2=\tr(\mathcal S^r).
\]
A perfectly predictable direction contributes one, however small its
support. A direction with eigenvalue bounded away from one contributes
exponentially less as $r$ grows.

\paragraph{Computing scores when walks are lost.}
For choosing a cut we can fill each missing arrival from the current
average distribution of origins. To analyze the resulting scores we
use a completion that also keeps the origin marginal fixed. The two
rules give the same ordering of scores in every round.

Write $A_{vu}$ for the physical mass sent from $u$ to $v$ divided
by $d(v)$, and let $\Pi$ subtract the $d$-weighted mean from a
vector. Under the arrival-only rule, a fill has zero deviation from
the current average. Transporting the deviations and subtracting
their new average therefore updates the centered operator by
$\bar G'=\Pi A\bar G$.

For the completion, fill a missing arrival with the origin label
of a walk at a preceding vertex drawn according to $\pi=d/M$.
Distribute each missing departure over destinations according to $\pi$. If $\varepsilon M$ units were
missing, these additions bring both marginals to $(1+\varepsilon)d$.
The missing departures send the same score mass per unit of $d(v)$
to every destination, so $\Pi$ removes their contribution. Dividing
by $1+\varepsilon$ restores the prescribed marginals and gives
\[
                    G'=(1+\varepsilon)^{-1}\Pi A G .
\]
Starting from the same population, $G$ and $\bar G$ thus differ
only by a positive scalar after any sequence of steps. Their powered
probes give the same median order and the same marked vertices.
Only the physical transport uses routes in the graph.

\paragraph{Progress from attenuation and matching.}
In each round the player requests a matching across the median of the
scores weighted by $d$. Up to $\tau M$ of matched weight may miss the
median, meaning that both endpoints lie strictly on the same side.
Unmatched weight draws on a
separate allowance of $6\tau M$ for the whole game. A small trapped block
can fit inside the first allowance at every round. If its scores dominate
the probe, they must be unusually large at some vertices. Fix a small
constant $\eta>0$ and mark the set $U$ on which $x(v)^2$ exceeds $1/\eta$
times its $d$-weighted average. We discard a fraction $\kappa=1/(256r)$ of
the mass there before the lazy matching step. Since $d(U)\le\eta M$, at
most $\tau M$ is lost to attenuation over the entire game, where
$\tau=\eta/256$.

Return to the two-block walk example. Suppose $p\le\tau$, the player
marks $A$, and all mass is matched within its own block. Pairs in $\bar A$ meet the median
because all their scores tie there, and the pairs in $A$ fit the exception
allowance. The centered indicator has values $1-p$ on $A$ and $-p$ outside.
Attenuation reduces the first to $(1-\kappa)(1-p)$ and leaves the second
unchanged. Subtracting a common mean preserves the difference between the
two values. The completion then divides it by $1+\kappa p$, so one step
multiplies the difference by
\[
        \frac{1-\kappa(1-p)}{1+\kappa p}
                       =1-\frac{\kappa}{1+\kappa p}.
\]
This is the only predictable direction of the two-block walk. Its
contribution to $Z=\tr(\mathcal S^r)$ is multiplied by the $2r$th power of
this factor. Thus even a loss of order $1/r$ in marked departures removes a
constant fraction of their contribution to $Z$ without sending any walk
across the cut.

In a general round, condition on the history before drawing the probe
and put $F=\|x\|_d^2$. Vertex $v$ contributes
$d(v)\mathbb E[x(v)^2]=(\mathcal S^r)_{vv}$ to the trace. Attenuation
removes at least a $1-(1-\kappa)^{2r}$ fraction of the trace on the
marked vertices (\Cref{sec:player}). If much of the observed energy is
spread over the unmarked vertices, the matching makes progress in a
different way. Suppose they carry at least $F/2>0$. Their absolute
scores are at most $\sqrt{F/(\eta M)}$, so carrying this much squared
energy requires a large sum of absolute scores. Change the signs if
necessary to make the median nonnegative. The negative scores account
for half the absolute score sum because $x$ is centered, and a matched
negative score paired across the median has a nonnegative partner.
Even with the exceptions, the weighted sum $\mathcal D$ of absolute
gaps between partners is $\Omega(\sqrt{\eta MF})$.

Stop the score computation one pass early to obtain an origin label
$q$ with $x=Gq$. Thus $x(u)$ is the mean of $q$ among walks at $u$.
At the next step, walks arriving along $u\to v$ and walks held at
$v$ join the same population. Their old means differ by $x(u)-x(v)$.
To measure what the merge forgets, consider revealing that a walk
now at $v$ came from $u$. Its origin distribution is then the old
distribution at $u$. The average relative entropy of these old
distributions with respect to the mixtures they join is exactly the
mutual information lost in the step. This information concerns the
full origin, so its decreases telescope across rounds even when
the next probe chooses a different label $q$.

We can pair a constant fraction of each matched weight with equal
mass held at its destination, since at most $d(v)$ is matched into
$v$. Factoring each difference of origin probabilities as a difference
of squares relates the gap between the two old means to the distance
between the square roots of the two distributions. The second
moments of $q$ set the scale of this bound, and the divergences
from the common mixture control the distance. Summed over the
matched pairs, the divergences for both the arrivals and the stays
are paid for by the information lost in the step. Meanwhile each
matching marginal uses at most $d(v)$ at vertex $v$, so the total
label energy at either end is at most $\|q\|_d^2$. Cauchy--Schwarz
over the pairs combines these two bounds. The proof in
\Cref{sec:player} gives
\[
          \mathcal D^2\le32M\|q\|_d^2(I_{\rm old}-I_{\rm new}).
\]

Thus spread-out scores cost $\Omega(\eta F/\|q\|_d^2)$ of mutual
information. Since $x(Y)$ is the conditional mean of the centered
label $q(X)$, the ratio $F/\|q\|_d^2$ is the fraction of its variance
explained by the old vertex.

Combine the two gains in $Z\exp(cI)$, with $c$ proportional to
$1/\eta$. Attenuation can remove a constant fraction of $Z$ even
when little mass moves. An information loss of order $\eta$ removes
a constant fraction of the exponential factor. Both factors are
nonincreasing.

The origin label $q$ omits the last forward pass, so its mean
squared norm is $\tr(\mathcal S^{r-1})$. For logarithmic $r$ this
is within a constant factor of $\mathbb E F=Z$ while a nearly
predictable direction remains. To use these mean energies we must
also account for how the probe chooses where to attenuate.
The same probe chooses the marked vertices and measures their score
energy. Their contribution to $Z$, however, is the energy an
independent probe of the same state would put there in expectation.
The fourth-moment bound for a sum of independent signs limits how
much of the expected energy can come from unusually large draws.
Applied both to $x$ and to its origin label $q$, it gives a constant
expected fractional decrease in the combined potential until a cut is
returned or $\|G\|$ is bounded away from one. Initially $Z\le n$
and $I\le\log n$, whereas any direction of predictable variance
fraction $\beta$ contributes $\beta^r$ to $Z$. Taking the target
$\beta$ sufficiently close to one, $r=\Theta(\log n)$ rounds suffice
with high probability (\Cref{sec:player}).

\paragraph{From mixing to a bound on returning walks.}
Take $S$ of weight $0<d(S)\le M/2$ and write $p=d(S)/M$. In the
completed walk the mass that starts and ends in $S$ is
$M\Pr[X\in S,Y\in S]$. Independent endpoints would contribute
$Mp^2=p\,d(S)$. Their difference is $M$ times the covariance of the
two membership indicators. For the centered indicator $y=\one_S-p\one$,
this difference is $\langle y,Gy\rangle_d$ and
$\|y\|_d^2=(1-p)d(S)$. Thus
\[
     M\Pr[X\in S,Y\in S]=p\,d(S)+\langle y,Gy\rangle_d
                 \le [p+(1-p)\|G\|]d(S).
\]
Since $p\le1/2$, bounding $\|G\|$ away from one keeps a constant
fraction of the completed starting mass from returning to $S$,
simultaneously for all such sets.

A surviving physical path occurs in the completed walk with its
mass multiplied by at least $\prod_j(1+\varepsilon_j)^{-1}$,
where $\varepsilon_j$ is the fraction of transport missing in
round $j$. The total missing fraction from attenuation and unmatched
departures is at most $4\tau$, so this product is at least
$e^{-4\tau}\ge1-4\tau$. Retain a $1-4\tau$ fraction of each
surviving path. The return estimate now applies to the retained
physical mass $Q(S,S)$, and at either end at most $8\tau M$ is
missing in total (\Cref{sec:overlay}).

\paragraph{Routing the matchings.}
Split $d$ into equal halves at a median of the scores weighted by $d$,
dividing ties as needed. In an undirected graph we route from the lower
half to the upper and include each route in both directions. Each vertex then has
equal unused sending and receiving weights, which can be matched there.
A shortfall of $e$ produces self-pairs of total weight $2e$. If
$2e\le\tau M$, they fit the per-round exception allowance and no
weight remains unmatched. A larger shortfall gives a sparse cut with
at least $\tau M/2$ weight on each side.

In a directed graph we must route the two directions separately.
The unused sending and receiving weights can occur at different
vertices. Allowing a shortfall of $\tau M/r$ in each direction leaves
at most $2\tau M$ unmatched over $r$ accepted rounds. A failed request
exposes a set $S$ whose incoming or outgoing boundary in the live graph
has normalized capacity $O(\phi\,d(S)\log n)$. Its weight may be barely
more than $\tau M/r$. Instead of recursing immediately on
so unbalanced a cut, we retire the set from further requests and retry
the round. We discard the routes from the failed attempt. Routes from
earlier accepted rounds survive even if they pass through retired
vertices.

While the total retired weight is at most $\tau M$, we keep the
scores and recompute the median on the live vertices. The matching
argument only requires that the pairs straddle a common threshold, so
this new median suffices. We match each retired vertex to itself and
charge these pairs to the per-round exception allowance. Each failure
retires more than $\tau M/r$ fresh weight, so $O(r)$ requests suffice
in all. End the phase once the total retired weight exceeds $\tau M$.
The last piece weighed at most half the weight that was still live, so
a constant fraction of the phase weight also remains live. All pieces
charged for incoming capacity form one block before the live remainder.
Those charged for outgoing capacity form one block after it. A backward
arc between blocks is charged when the first of its endpoints is retired
(\Cref{sec:matching}).

\paragraph{Trimming to an induced expander.}
Trimming begins on all of $C$, including vertices retired during
matching. Let $\Lambda$ bound the congestion of the physical walks
in the normalized graph. On the current vertex set $J$, give each
vertex a supply $\Delta_J(v)$ that covers the larger of its missing
start and end masses, including the rounding in
\Cref{sec:endpoint-updates}. This supply is at most $d(v)$. Give $v$
room to absorb $\sigma d(v)$, where $\sigma=\gamma/2$, and multiply
the edge capacities by $4\Lambda/\gamma$ for the trimming flow.

Suppose the flow routes the supply. Take a side $S$ of a cut of $J$
with $d(S)\le d(J)/2$, and let $c_J^+(S)$ denote its outgoing capacity
before the multiplication. At least $\gamma d(S)$ starting mass fails to return.
The walks can account for at most $\Lambda c_J^+(S)$ of this shortfall
by leaving $S$. The rest is missing at its starts and covered by the
supply in $S$. Since only $\sigma d(S)$ of the supply can be absorbed
there, the trimming flow carries its surplus across the same boundary.
Together the walk and flow inequalities give
\[
 \gamma d(S)\le\Delta_J(S)+\Lambda c_J^+(S)
              \le\sigma d(S)+(1+4/\gamma)\Lambda c_J^+(S).
\]
Thus $S$ has outgoing capacity $\Omega(d(S)/\Lambda)$ even when few
walks leave it. For the incoming bound, count the missing ends and
route in the reverse graph. To return to input units, multiply both
sides of the expansion bound by $W/m$. Each input edge has capacity
at least $W/m$ times its rounded capacity, and $(W/m)d$ dominates the induced degrees and
the optional measure $\mu$ (\Cref{sec:normalization}). This gives
both expansion guarantees of \Cref{thm:main}.

If excess remains, the search exposes a piece whose supply pays for
its boundary in the direction searched and for a fixed fraction of
its weight. Remove each such piece, putting those with small incoming
boundary in $I$ and those with small outgoing boundary in $O$.
As long as the total removed weight is at most $M/2$, apply the return
bound to $I$ at its incoming boundary and to $O$ at its outgoing
boundary, using the walks \emph{before} the deletions. The charged
edges can account for at most half the required shortfall, even
counting an arc from $O$ to $I$ twice. The rest comes out of the
initial missing mass, which is small. Thus any light removed union
is in fact much lighter than $M/2$.

Could one deletion carry the union past $M/2$? A lost walk either
starts in $I$, ends in $O$, or traverses a backward arc in the order
$I,J,O$. The total missing start mass in $J$ is
$d(J)-Q_J(J,J)$, which also equals the total missing end mass there.
Relative to the original phase, $Q_J(J,J)$ has fallen by the mass
of all lost walks, while the prescribed weight has fallen by
$d(I)+d(O)$. This offset covers the walks starting in $I$ or ending
in $O$. Any further increase in the total deficit is bounded by the
load on the charged edges. Hence the supply remains small as
long as the removed union is small. Since the next piece must pay
for a fraction of its own weight out of that supply, it cannot take
the union past $M/2$. Once the remaining supply can be routed, we
finalize $J$ and recurse on $I$ and $O$ (\Cref{sec:trimming}).

To maintain the supply, disable the matching routes visiting a
deleted vertex. Passing mass forward through the surviving layers
counts ends, and passing it backward counts starts. Rounding the
survival fractions bounds the number of updates at each vertex
while keeping the supply an upper bound on both deficits
(\Cref{sec:endpoint-updates}).

\paragraph{Recursion and running time.}
Normalize the capacities once and fix integer vertex weights of
total $O(m)$ (\Cref{sec:normalization}). A phase of weight $M$
charges $O(\phi M\log n)$ in these units for the crossing edges in
an undirected graph or the backward arcs in a directed graph. Count
each such edge or arc at the split that first separates its endpoints.
Every recursive child has at most a constant fraction of
its parent's weight, so summing over the $O(\log n)$ levels and
restoring input units gives total charged capacity $O(\phi W\log^2 n)$.

Games may run at every level. Each of the $r$ scores takes $r$
passes through at most $r$ matching layers, giving $O(m\log^4 n)$
score work overall. Directed requests need shortfall $O(M/r)$ at
each round, where undirected requests can tolerate a constant
fraction of $M$. The routing time depends logarithmically on the
reciprocal of this relative tolerance, giving the extra $\log\log n$.
Trimming and endpoint updates cost $O(M\log n/\phi)$ in a phase,
but each successful trim finalizes at least $15/16$ of its weight.
The finalized parts are disjoint across levels, so the total weight
charged for trimming, including a last failed size check, is $O(m)$.
\Cref{sec:algorithm} gives the full work and space bounds.

\section{Preliminaries}\label{sec:prelim}

\paragraph{Cuts and conductance.}
A directed input graph may contain both $uv$ and $vu$. Auxiliary
graphs may have parallel edges or zero capacities. For disjoint sets
$S,T\subseteq V(H)$, let $c_H(S,T)$ be the total capacity between
them in an undirected graph, or from $S$ to $T$ in a directed graph.
Write $\bar S=V(H)\setminus S$. The edge boundary $\partial_H S$ of
an undirected cut has capacity $\delta_H(S)=c_H(S,\bar S)$. In a
directed graph the outgoing and incoming capacities are
$c_H^+(S)=c_H(S,\bar S)$ and $c_H^-(S)=c_H(\bar S,S)$.

For vertex weights or demands, write $w(S)=\sum_{v\in S}w(v)$.
For $w\ge0$, expansion with respect to $w$ is the minimum, over
cuts with positive weight on both sides, of the cut capacity divided
by $\min\{w(S),w(\bar S)\}$. In a directed graph the numerator is
$\min\{c_H^+(S),c_H^-(S)\}$. Conductance, denoted by $\Phi(H)$ or
$\Phi_{\rm dir}(H)$, uses the weighted degree $w=\deg_H^c$. This
counts all incident capacity, including both incoming and outgoing arcs
in a directed graph. We set conductance to one if no cut has positive
weight on both sides. In a part $J$ we compute degrees in $G[J]$
and use the restriction of any prescribed vertex measure to $J$. The
unweighted degree $\deg_H^0$ counts incident edges or arcs with
multiplicity.

\paragraph{Undirected flows.}
Fix an orientation of the edges of an undirected graph $G$. The
incidence matrix $B_G$ has $+1$ at the tail and $-1$ at the head, so
$B_Gf$ is the vector of net outflows. The sets $\partial^+S$ and $\partial^-S$ contain
the edges oriented out of and into $S$, respectively. Write
$C_G=\diag(c)$ and $\congestion_G(f)=\max_{e:c_e>0}|f_e|/c_e$,
with infinite congestion if a zero-capacity edge carries nonzero flow.

A demand $b$ is \emph{balanced} if $b(V)=0$ and \emph{componentwise
balanced} if its sum on each component of the positive-capacity
subgraph is zero. The latter are precisely the demands that $G$ can
route with finite congestion. Every flow routing $b$ sends net flow
$b(S)$ across each cut. For a componentwise balanced demand,
max-flow/min-cut gives
\[
 \opt_G(b)=\min_{B_Gf=b}\congestion_G(f)
          =\max_{S:\,\delta_G(S)>0}\frac{|b(S)|}{\delta_G(S)}.
\]
An empty maximum is zero. We set $\opt_G(b)=+\infty$ if $b$ is
imbalanced on any component.

\paragraph{Conventions.}
Vector inequalities and the product $\odot$ are coordinatewise.
Operator norms and adjoints use the indicated inner product, Euclidean
by default. Logarithms are natural unless a base is shown, and
$0\log0=0$. In size bounds we take logarithmic factors to be at least
two. The notation $\widetilde O$ suppresses polylogarithmic factors
in the input size. Accuracy parameters lie in $(0,1/2]$ unless specified otherwise.
Scalar arithmetic, exponentials, logarithms, rounding, and random
signs cost constant time. \emph{With high probability} means that for
any fixed $a>0$ the failure probability is at most $(m+2)^{-a}$, with
constants allowed to depend on $a$. Recursive calls use the original
input size for logarithmic factors and failure probabilities.

\Needspace{9\baselineskip}

\section{Cut player}\label{sec:player}

Start $d(v)>0$ units of walk mass at each vertex $v$ of a nonempty
set $V$, and write $M=d(V)$. Suppose a set $S$ of weight at most
$M/2$ retains almost all the walks that started there. Then knowing
that a walk started in $S$ tells us a great deal about where it ends.
The player looks for this dependence by predicting labels at one end
of a walk from its other end. In each round it asks for a matching
across a threshold in the predictions, with ties allowed on either
side. Failure to route the matching gives a sparse cut. On a successful
round, the player also attenuates a small amount of mass at vertices
with unusually large scores.

We compute predictions on a completed population with distribution
$d/M$ at both ends. The physical layers record the routes in the graph. From their
surviving walks we will retain a known fraction that also occurs in
the completed population. This will let us bound the physical mass
that returns to any set of weight at most $M/2$.

\subsection{Origin distributions and endpoint scores}

Let $X$ and $Y$ be the origin and current vertex of a random unit of
the completed population, and put $\pi=d/M$. At each vertex $v$ the
origins have distribution $p_v(i)=\Pr[X=i\mid Y=v]$. The matrix
$P(v,i)=p_v(i)$ averages origin labels at their destinations. Since
both marginals are $\pi$, $P\one=\one$ and $\pi^\top P=\pi^\top$. Write
\[
 D=\diag(d),\qquad J=\one\pi^\top,\qquad \Pi=I-J,\qquad
 G=P-J=\Pi P=P\Pi .
\]
In the inner product $\ipd fg=f^\top Dg$, the adjoint averages in the
opposite direction, since $P^*(i,v)=d(v)P(v,i)/d(i)=\Pr[Y=v\mid X=i]$.
Thus for centered origin and endpoint labels $q$ and $y$,
\[
 (Gq)(v)=\EE[q(X)\mid Y=v],\qquad
 (G^*y)(i)=\EE[y(Y)\mid X=i].
\]
For $y\ne0$, the fraction of its variance that the origin predicts is
\[
 \frac{\operatorname{Var}(\EE[y(Y)\mid X])}{\operatorname{Var}(y(Y))}
 =\frac{\nd{G^*y}^2}{\nd y^2}
 =\frac{\ipd y{GG^*y}}{\nd y^2}.
\]
If the origin determines which side of a nontrivial cut contains the
endpoint, its centered indicator attains one in this quotient, even
when one side carries arbitrarily little weight.

\paragraph{Completing a physical layer.}
Let $B(u,v)$ be the walk mass sent from $u$ to $v$ and write
$B(S,T)=\sum_{u\in S,v\in T}B(u,v)$. Both the sending and receiving
weights can fall short of $d$. Pairing these shortfalls directly could
keep the replacement mass inside the very set that trapped the walks.
Instead, send each missing departure to a destination drawn from $\pi$,
and fill each missing arrival from a source drawn from $\pi$.

\begin{lemma}[Stationary completion]\label{lem:player-completion}
Let $B\ge0$ have both marginals at most $d$. Set $A=D^{-1}B^\top$ and
\[
 e^+=d-B\one,\qquad e^-=d-B^\top\one,\qquad
 \varepsilon=1-B(V,V)/M,\qquad t=(1+\varepsilon)^{-1}.
\]
The completed transport $\widetilde B$ has both marginals $d$, and its
conditional-expectation operator $K$ satisfies
\begin{equation}\label{eq:completion}
 \widetilde B=t\bigl[B+e^+\pi^\top+\pi(e^-)^\top\bigr],
 \qquad
 K=D^{-1}\widetilde B^\top=J+t\Pi A\Pi .
\end{equation}
For the current population $P=J+G$, the update $P'=KP$ has centered
part $G'=t\Pi AG$.
\end{lemma}
\begin{proof}
Each added transport fills one deficit and contributes $\varepsilon d$
to the opposite marginal. Rescaling by $t$ therefore gives both
marginals $d$, so $K\one=\one$, $\pi^\top K=\pi^\top$, and
$K-J=\Pi K\Pi$. In the transpose update
\[
 K=t\bigl[A+\one(e^+)^\top/M+D^{-1}e^-\pi^\top\bigr],
\]
the missing departures contribute a constant across destinations, and
the missing arrivals use the mean across sources. The left projection
kills the constant and the right projection kills the mean. Hence
$K-J=t\Pi A\Pi$. Applying $K$ to $P=J+G$ gives $G'=t\Pi AG$.
\end{proof}

Thus the player need only transport a label along the physical layer,
subtract its weighted mean, and rescale. The added transports require
no routes in the input graph.

\subsection{Matching and attenuation}

A set of small vertex weight can carry much of the score's squared
norm. Attenuating departures there uses little walk mass. When the
squared norm is spread over substantial vertex weight, matching
across the median brings together populations with different means
for the sampled origin label.

\paragraph{What the player requests.}
Given scores $x$, the matching routine returns either the sparse cut
required by its application (an ordered split in the directed
decomposition) or a nonnegative matrix $N$. The weight matched from $u$
to $v$ is $N(u,v)$, with $N\one\le d$ and $N^\top\one\le d$. Fix
\begin{equation}\label{eq:constants}
 \lambda=2^{-14},\qquad \eta=\lambda/64,\qquad \tau=\eta/256 .
\end{equation}
Our mixing target is $\|G\|^2\le1-2\lambda$ and the player marks at
most $\eta M$ vertex weight per round. All but $\tau M$ of the matched weight must cross or meet a common
threshold $c$. For a pair from $u$ to $v$, this means $(x(u)-c)(x(v)-c)\le0$.
The applications choose a median of the scores weighted by $d$,
restricted to the live vertices when some have been retired.

A pair that misses $c$ still carries walk mass. For instance,
unused sending and receiving weight at the same vertex can be paired
there without using any edge. This pair misses the threshold unless
its score is $c$. Unmatched weight draws on a separate allowance over
the accepted rounds of the game,
\begin{equation}\label{eq:matching-loss-budget}
 \delta_j=M-N_j(V,V),\qquad \sum_j\delta_j\le6\tau M .
\end{equation}

\paragraph{Running the game.}
Fix a size bound $n_{\rm p}\ge |V|$ and a failure allowance
$0<\epsilon_{\rm fail}<1$. Choose the even round limit $r=2k$ by
\eqref{eq:round-choice}, so $r=\Theta(\log(n_{\rm p}+1)+
\log(1/\epsilon_{\rm fail}))$, and set $\kappa=1/(256r)=\tau/(r\eta)$.
To find an endpoint label that the origin predicts well, draw
$g(v)=\xi_v/\sqrt{d(v)}$ with independent uniform signs $\xi_v$.
This gives unit variance in each $d$-orthonormal coordinate. A pass by $G^*$ takes an endpoint
label back to the origins. A pass by $G$ averages those predictions
at the endpoints. Repeating the pair $k$ times multiplies the
component in an eigendirection of $GG^*$ by $\beta^k$, where $\beta$
is its predictable fraction of variance. Perfectly predictable
directions are preserved, and those with larger $\beta$ decay
more slowly.

Consider the last pair of passes. The backward pass leaves a label
$q$ at the origins. The forward pass computes the local means of $q$, giving
the factorization
\begin{equation}\label{eq:origin-label}
 q=G^*(GG^*)^{k-1}g,\qquad x=Gq=Pq=(GG^*)^kg .
\end{equation}
The endpoints that the matching pairs across a large score gap
therefore have different predictions of the same origin label.

The player marks vertices whose squared scores exceed $1/\eta$ times
the weighted average. They have total weight at most $\eta M$.
Attenuating a $\kappa$ fraction of their departures in each round
therefore costs at most $r\kappa\eta M=\tau M$ over the game.
Half the retained mass stays at its vertex to mix with arrivals.
The other half follows the matching, losing any unmatched portion.
This gives the departure scaling and the lazy physical layer in
\Cref{alg:cut-player}.

\begin{algorithm}[htbp]
\caption{Cut player on vertex weights $d$}\label{alg:cut-player}
\begin{enumerate}[leftmargin=2.4em,itemsep=2pt,topsep=3pt]
\item Initialize the centered operator $G\gets\Pi$.
\item For $j=1,\ldots,r$, perform the following steps.
\begin{enumerate}[label=(\alph*),leftmargin=2em,itemsep=2pt,topsep=2pt]
\item Draw fresh independent uniform signs $\xi_v$. Compute the scores
and the marked set
\begin{equation}\label{eq:probe}
 g(v)=\xi_v/\sqrt{d(v)},\qquad x=(GG^*)^k g,\qquad
 U=\{v:\eta Mx(v)^2>\nd x^2\}.
\end{equation}
\item Request a matching for $x$. If the routine returns a sparse split,
return it immediately. Otherwise let $N_j$ be the returned matching.
\item Set $W_U=\diag(\one_U)$ and $T_U=I-\kappa W_U$.
Store the physical layer $B_j=T_U(D+N_j)/2$ and update the centered
operator by
\begin{equation}\label{eq:physical}
 \varepsilon_j=1-B_j(V,V)/M,\qquad t_j=(1+\varepsilon_j)^{-1},\qquad
 G\gets t_j\Pi D^{-1}B_j^\top G .
\end{equation}
\end{enumerate}
\item Return the physical layers $(\zeta B_1,B_2,\ldots,B_r)$,
where $\zeta=1-4\tau$.
\end{enumerate}
\end{algorithm}

Write $A_N=(I+D^{-1}N^\top)/2$. A round with matching $N$ updates
$G$ to $G'=t\Pi A_N T_U G$. Indeed, the diagonal matrices $D,T_U$
commute, so $D^{-1}B^\top=A_NT_U$. Each factor after $G$ is a
contraction. For the matching factor, applying Cauchy--Schwarz at each
destination and summing over departures gives
\[
 \nd{D^{-1}N^\top y}^2
 =\sum_v\frac{(\sum_u N(u,v)y(u))^2}{d(v)}
 \le\sum_v\frac{\sum_u N(u,v)}{d(v)}\sum_u N(u,v)y(u)^2
 \le\sum_u d(u)y(u)^2 .
\]
The arrival bound cancels the denominator, and the departure bound
controls the remaining sum. In particular $\|G'\|\le\|G\|\le1$.

\paragraph{Output of a completed game.}
The product $Q_{\rm raw}=B_1D^{-1}B_2\cdots D^{-1}B_r$ counts the
walks that survive all $r$ layers, with origins as rows and destinations
as columns. Scaling the first layer retains a fraction $\zeta$ of every
surviving walk, so the returned layers represent the endpoint coupling
\begin{equation}\label{eq:physical-output}
 Q=\zeta Q_{\rm raw}.
\end{equation}
Thinning once compensates for every layer's rescaling. Between any
origin and destination, the retained physical mass is then at most the
completed mass (\Cref{sec:overlay}). Applying the bound on $\|G_r\|$ to
the indicator of a set with weight at most $M/2$ then bounds the
physical mass that returns there.

When trimming by successive deletions, we store the route of each matched pair.
Deleting a vertex can destroy one route and leave another with the
same endpoints intact. The hierarchy and the undirected algorithm of
\Cref{sec:near-optimal-decomp} trim with a single fair cut on the
original phase graph. They keep only the weighted endpoint pairs and
use the missing start mass at $v$ as the capacity of a spoke from the
trimming source. Thinning keeps this capacity between $4\tau d(v)$
and $d(v)$ (\Cref{sec:game-implementation}).

\begin{lemma}[Cut player]\label{lem:player}
Suppose the matching routine satisfies the two allowances and the
marginal bounds above. Conditional on the history before the game,
the probability that the game completes all $r$ rounds and misses
the mixing target
\begin{equation}\label{eq:terminal}
                     \|G_r\|^2\le1-2\lambda
\end{equation}
is at most $\epsilon_{\rm fail}$. The matching routine may choose
its response after seeing the current scores.

On every completed game, the missing physical mass satisfies
\begin{equation}\label{eq:player-mass-loss}
 \delta_{\rm loss}:=M-Q(V,V)\le8\tau M,\qquad
 d-Q\one\ge4\tau d,\qquad d-Q^\top\one\ge4\tau d .
\end{equation}
With exact matching marginals, $\delta_{\rm loss}\le5\tau M$.
Each routed pair contributes at most half its weight to its
physical layer. If a matching factor and its adjoint each cost
$O(s)$ to apply, including vertex operations, the player takes
$O(sr^3)$ work beyond the matching calls. The sparse layer factors
and vertex vectors suffice for all its computations.
\end{lemma}

For applications with a common outer size $n_{\rm p}$ and an
inverse-polynomial failure allowance, $r=\Theta(\log n_{\rm p})$.

Start layer $j$ with its full vertex weights. Only the moving half can
be lost to unmatched weight $\delta_j$, and only marked departures
are attenuated. The missing mass $\varepsilon_j M$ is therefore at most
$\delta_j/2+\kappa d(U_j)$. Summing over the game gives
\begin{equation}\label{eq:loss-budget}
 \sum_j\varepsilon_j\le\frac{\sum_j\delta_j}{2M}+r\kappa\eta
                  \le3\tau+\tau=4\tau .
\end{equation}
With exact matching marginals the sum is at most $\tau$. If $s$ is
the physical mass vector before layer $j$ and $s\le d$, the next
vector is $B_j^\top D^{-1}s\le B_j^\top\one\le d$. The bound
therefore holds at every layer, starting from $s=d$. The loss in
layer $j$ is
\[
 \sum_v s(v)\left(1-\frac{(B_j\one)(v)}{d(v)}\right)
 \le \sum_v\bigl(d(v)-(B_j\one)(v)\bigr)=\varepsilon_j M .
\]
The inequality uses the departure bound $B_j\one\le d$. Thinning
costs at most another $4\tau M$ and leaves both marginals at most
$\zeta d$, proving \eqref{eq:player-mass-loss}.

\subsection{Progress from attenuation and matching}

We prove the mixing guarantee of \Cref{lem:player}. Fix the history
before a new probe. Put $\mathcal S=GG^*$ and $Z=\tr(\mathcal S^r)$.
An eigendirection with predictable fraction $\beta$ contributes
$\beta^r$ to this trace. Since $2k=r$, it also contributes the
same amount to the expected squared norm of the probe
$x=\mathcal S^k g$. The corresponding statement at each vertex will
let us see which part of the trace attenuation can remove.

For $e_v=\one_{\{v\}}/\sqrt{d(v)}$, the coefficient of $g$ along
$e_v$ is an independent uniform sign. In particular
$\EE[\ipd gy^2]=\nd y^2$ for every fixed $y$. Since
$\sqrt{d(v)}x(v)=\ipd g{\mathcal S^ke_v}$ and $2k=r$, we have
\begin{equation}\label{eq:probe-moments}
  d(v)\EE[x(v)^2]=\ipd{e_v}{\mathcal S^re_v}
      =(\mathcal S^r)_{vv}=: \nu(v),\qquad
  F:=\nd x^2,\qquad \EE F=Z .
\end{equation}
For the diagonal identity, the inner product is
$\sqrt{d(v)}(\mathcal S^re_v)(v)$, and the coordinate
$(\mathcal S^re_v)(v)$ is $(\mathcal S^r)_{vv}/\sqrt{d(v)}$.
The marked set $U$ depends on this probe. Its observed energy is
$H=\sum_{v\in U}d(v)x(v)^2$. Its share of the trace,
$\nu(U)=\sum_{v\in U}\nu(v)$, is the energy an independent probe
of the same history would put there in expectation. Attenuation
reduces $Z$ in proportion to $\nu(U)$. If the unmarked vertices
carry most of $F$, matching instead merges populations with different
means of $q$. The next probe will generally choose a different
label. We therefore charge the matching gaps to the information
the current vertex retains about the origin $X$, which can only
decrease as populations merge.

\paragraph{Trace loss on the marked vertices.}
In the update $G'=t\Pi A_N T_U G$, the factors after $T_U G$ are
contractions, so they cannot increase any singular value. In the
$d$-orthonormal vertex basis $(e_v)$, both $\mathcal S$ and $T_U$
are symmetric positive semidefinite. The Lieb--Thirring inequality
bounds the trace after attenuation by the old trace weighted at the marked vertices,
\[
 Z_{\rm new}
 \le\tr[(T_U\mathcal S T_U)^r]
 \le\tr(\mathcal S^r T_U^{2r})
 =Z-\bigl[1-(1-\kappa)^{2r}\bigr]\nu(U).
\]
The last equality uses diagonality of $T_U^{2r}$: it gives the
vertices in $U$ weight $(1-\kappa)^{2r}$ and all the others weight
one. The power $2r$ turns the attenuation $\kappa=1/(256r)$ into a
constant fractional charge against the old trace on $U$. In particular,
\begin{equation}\label{eq:trace}
       Z-Z_{\rm new}\ge (1-e^{-1/128})\nu(U)\ge\nu(U)/129 .
\end{equation}

\paragraph{Score separation across a matching.}
Suppose $H<F/2$. On the unmarked vertices each $|x(v)|$ is at most
$\sqrt{F/(\eta M)}$. Using this bound on one factor of $x(v)^2$ at each unmarked
vertex gives
\[
 \|x\|_{1,d}:=\sum_v d(v)|x(v)|
      \ge (F-H)\sqrt{\eta M/F}>\tfrac12\sqrt{\eta MF}.
\]
Let $\mathcal D=\sum_{u,v}N(u,v)|x(u)-x(v)|$. For this count we may negate the scores and the threshold $c$
together. This leaves all gaps unchanged and lets us assume $c\ge0$. Whenever a
negative score is paired across $c$, its partner is nonnegative,
and their gap pays for the magnitude of the negative score.
Since $x$ is centered, the negative scores at full marginal $d$
weigh $\|x\|_{1,d}/2$ at the departures and just as much at the
arrivals. Keep only the pairs that meet or cross $c$. At either end
the omitted marginal has total weight at most $e=\delta_j+\tau M
\le7\tau M$. For any omitted weights $0\le b(v)\le d(v)$ of total
at most $e$, Cauchy--Schwarz gives
$\sum_v b(v)|x(v)|\le\sqrt{e\sum_v b(v)x(v)^2}\le\sqrt{eF}$.
It follows that
\begin{equation}\label{eq:correlation-l1}
 \mathcal D\ge\|x\|_{1,d}-2\sqrt{eF}
      \ge(1/2-\sqrt7/8)\sqrt{\eta MF}\ge\sqrt{\eta MF}/6 .
\end{equation}

\paragraph{How much information a merge loses.}
Fix the history, the probe, and the returned matching. In the completed
population at an old vertex $u$, the origins have distribution $p_u$
and the label $q$ has mean $x(u)$. The next step depends only on the
current vertex, so arrivals from $u$ carry this same distribution.
The origin distribution at a new vertex $v$ is therefore the mixture
\[
              p'_v=\frac1{d(v)}\sum_u\widetilde B_j(u,v)p_u .
\]
Write $\mathcal I(P)=I(X;Y)$ and $\mathcal I(P')=I(X;Y')$
for the mutual information before and after the step. For a walk
with origin $i$ moving from $u$ to $v$, subtract the two log ratios
$\log(p_u(i)/\pi(i))-\log(p'_v(i)/\pi(i))=\log(p_u(i)/p'_v(i))$.
The probability of this triple is $\widetilde B_j(u,v)p_u(i)/M$.
For a step $u\to v$ of positive weight, the mean of the log ratio
under $p_u$ is $D_{\rm KL}(p_u\Vert p'_v)$. Weighting these means by
$\widetilde B_j(u,v)/M$ gives
\begin{equation}\label{eq:information-correlation}
 \Delta\mathcal I:=\mathcal I(P)-\mathcal I(P')
   =\frac1M\sum_{u,v}\widetilde B_j(u,v)D_{\rm KL}(p_u\Vert p'_v).
\end{equation}

For a matched pair $u\to v$, the score gap is the difference
between the mean of $q$ among arrivals from $u$ and its mean among
walks held at $v$. Put $e(u)=\sum_i p_u(i)q(i)^2$, the second
moment of the label at $u$. Factor $p_u(i)-p_v(i)$ as a difference
of squares and apply Cauchy--Schwarz. The first factor has squared
norm at most $2(e(u)+e(v))$ by $(a+b)^2\le2(a^2+b^2)$:
\[
 \begin{aligned}
 |x(u)-x(v)|
 &=\left|\sum_i q(i)
    \bigl(\sqrt{p_u(i)}+\sqrt{p_v(i)}\bigr)
    \bigl(\sqrt{p_u(i)}-\sqrt{p_v(i)}\bigr)\right|\\
 &\le\sqrt{2(e(u)+e(v))}\,\|\sqrt{p_u}-\sqrt{p_v}\|_2 .
 \end{aligned}
\]
To bound the distance between these square roots, compare both old
distributions with the mixture $p'_v$ they join. For distributions
$p,s$, use $-\log t\ge2(1-\sqrt t)$, which follows from
$\log z\le z-1$ at $z=\sqrt t$. Applying it at $t=s(i)/p(i)$
and summing with weights $p(i)$ over $p(i)>0$ gives
\[
 D_{\rm KL}(p\Vert s)\ge 2-2\sum_i\sqrt{p(i)s(i)}
                     =\|\sqrt p-\sqrt s\|_2^2 .
\]
The last equality uses $\sum_i p(i)=\sum_i s(i)=1$.
Apply the triangle inequality through $\sqrt{p'_v}$ and square using
$(a+b)^2\le2(a^2+b^2)$. For every pair with $N(u,v)>0$ this gives
\[
 \|\sqrt{p_u}-\sqrt{p_v}\|_2^2
 \le 2D_{\rm KL}(p_u\Vert p'_v)+2D_{\rm KL}(p_v\Vert p'_v).
\]
By \eqref{eq:loss-budget}, $t_j\ge1/(1+4\tau)$, and the
departure multipliers are at least $1-\kappa$. Since
$(1-\kappa)/(1+4\tau)\ge1/2$, the completed transport contains
$t_jT_U(D+N)/2\ge(D+N)/4$ entrywise. The matched arrivals supply $N(u,v)/4$
units for the first divergence. For the second divergence, sum the
requests for stay mass at $v$: they total $\sum_u N(u,v)/4\le d(v)/4$.
The two charges fit into the matching and stay contributions of
$\widetilde B_j$, including on the diagonal. Thus
\[
 \begin{aligned}
 \frac18\sum_{u,v}N(u,v)\|\sqrt{p_u}-\sqrt{p_v}\|_2^2
 &\le \frac14\sum_{u,v}N(u,v)D_{\rm KL}(p_u\Vert p'_v)
       +\frac14\sum_v d(v)D_{\rm KL}(p_v\Vert p'_v)\\
 &\le M\Delta\mathcal I .
 \end{aligned}
\]

The origin marginal is $d/M$, so $\sum_u d(u)p_u(i)=d(i)$.
Summing the second moments over $u$ therefore gives
$\sum_u d(u)e(u)=\sum_i q(i)^2d(i)=\nd q^2$. The matching's
departure and arrival bounds each contribute at most this sum:
$\sum_{u,v}N(u,v)(e(u)+e(v))\le2\sum_u d(u)e(u)=2\nd q^2$.
Applying Cauchy--Schwarz over the matched pairs now gives
\[
 \mathcal D^2
 \le 2\left(\sum_{u,v}N(u,v)(e(u)+e(v))\right)
       \left(\sum_{u,v}N(u,v)\|\sqrt{p_u}-\sqrt{p_v}\|_2^2\right)
 \le32M\nd q^2\Delta\mathcal I .
\]
When $H<F/2$, squaring \eqref{eq:correlation-l1} gives
$\mathcal D^2\ge\eta MF/36$. Combining the bounds and dividing
by $32M$ yields
\begin{equation}\label{eq:diffuse-progress}
 H<F/2\quad\Longrightarrow\quad
                   \nd q^2\Delta\mathcal I\ge\eta F/1152 .
\end{equation}

\subsection{Expected contraction and the round bound}

Attenuation has given a fractional loss of trace, while a matching
with spread-out scores has given an additive loss of information.
Exponentiating the latter converts both to fractional decreases.
In view of \eqref{eq:diffuse-progress}, take
\[
                        \mathcal W=Z\exp(1152\mathcal I/\eta).
\]
Both factors are nonincreasing. For the analysis set $\mathcal W=0$
when the game returns a cut or $\|G\|^2\le1-2\lambda$ first holds,
including at initialization. The player does not compute the norm.
All further updates preserve the target once attained. We will show
that while $\mathcal W>0$, the next probe removes a constant fraction
of it in expectation.

\paragraph{Progress on a single probe.}
Fix a history with $\mathcal W>0$, and let
$a=1-\mathcal W_{\rm new}/\mathcal W$ be the fractional decrease in
the next round. If the potential stays positive, then
\[
 1-a=\frac{Z_{\rm new}}{Z}\exp(-1152\Delta\mathcal I/\eta)
       \le\frac{Z_{\rm new}}{Z}.
\]
In particular, \eqref{eq:trace} gives $a\ge\nu(U)/(129Z)$.
If $H<F/2$, using the same identity with \eqref{eq:diffuse-progress} gives
\[
 a\ge1-e^{-1152\Delta\mathcal I/\eta}
   \ge1-e^{-F/\nd q^2}\ge F/(2\nd q^2),
\]
where $0<F=\nd{Gq}^2\le\nd q^2$. If $H\ge F/2$, the marked vertices
already account for at least half of $F$. We obtain two bounds on
the same draw,
\begin{equation}\label{eq:energy-accounting}
                  F\le2H+2a\nd q^2,\qquad
                  \nu(U)\le129Za .
\end{equation}
They hold on a stopping outcome as well, since then $a=1$,
$F\le\nd q^2$, and $\nu(U)\le Z$. Thus a cut returned before a matching
is supplied also satisfies \eqref{eq:energy-accounting}.

\paragraph{Averaging over the probe.}
Take expectations conditional on the history. If $a$ were a small
constant on every draw, the right side of
$F\le2H+2a\nd q^2$ could have mean $Z$ only if $H$ accounted
for much of the trace. There are two dependencies to handle.
The marked set selects unusually large coordinates of $x$, so its
observed energy can exceed its share $\nu(U)$ of the trace.
Also, the few draws on which $a$ is large might carry unusually
large $\nd q^2$. We bound the energy selected in either way with
the same fourth-moment estimate.

For a linear form $L$ in the fresh independent signs, expansion of
the fourth power gives $\EE[L^4]\le3(\EE[L^2])^2$, since every
surviving term pairs signs. If $0\le w\le1$ is chosen after the
probe, Cauchy--Schwarz and $w^2\le w$ give
\[
       \EE[wL^2]\le\sqrt{\EE[w^2]\,\EE[L^4]}
                      \le\sqrt{3\EE w}\,\EE[L^2].
\]
First apply this to $L=\sqrt{d(v)}x(v)$ and $w=\one_{\{v\in U\}}$.
Then $\EE[L^2]=\nu(v)$. Summing over $v$ and applying
Cauchy--Schwarz once to this sum gives
\[
 \EE H\le\sqrt3\sum_v\nu(v)\sqrt{\Pr[v\in U]}
       \le\sqrt{3Z\,\EE\nu(U)}
       \le Z\sqrt{387\EE a},
\]
using $\nu(U)\le129Za$. Now take $L=\sqrt{d(i)}q(i)$ with
$w=a$ at every origin. Summing the same estimate gives
$\EE[a\nd q^2]\le\sqrt{3\EE a}\,\EE\nd q^2$.

Omitting the last forward pass costs one trace power in the mean
energy. Substituting $q$ from \eqref{eq:origin-label} and using
$GG^*=\mathcal S$ gives
\[
 \nd q^2=\ipd{\mathcal S^{k-1}g}{\mathcal S\,\mathcal S^{k-1}g}
        =\ipd{g}{\mathcal S^{2k-1}g}.
\]
The last equality uses self-adjointness. Since $2k=r$, averaging
over the signs gives $\EE\nd q^2=\tr(\mathcal S^{r-1})$. Let $\beta_i$ be the
eigenvalues of $\mathcal S$, padded with zeros to $n_{\rm p}$
entries. Concavity of $t^{1-1/r}$ applied to the average of
$\beta_i^r$ gives
\begin{equation}\label{eq:origin-energy}
 \frac{\EE\nd q^2}{Z}
 =\frac{\sum_i\beta_i^{r-1}}{\sum_i\beta_i^r}
 \le\frac{n_{\rm p}^{1/r}}{Z^{1/r}}
 \le\frac{e^{1/4}}{1-2\lambda}<\frac43 .
\end{equation}
The last bound uses $r\ge4\log(n_{\rm p}+1)$ and
$Z^{1/r}\ge\|\mathcal S\|>1-2\lambda$ while the potential is
positive. Thus the observed energy in \eqref{eq:energy-accounting}
satisfies
\[
 Z/2\le\EE H+\EE[a\nd q^2]
       \le(\sqrt{129}+4/3)Z\sqrt{3\EE a}\le22Z\sqrt{\EE a}.
\]
Consequently $\EE a\ge c_0:=1/2048$. This proves, on every
history including those on which the potential is already zero,
\begin{equation}\label{eq:contraction}
 \EE[\mathcal W_{\rm new}\mid\text{history}]
                    \le(1-c_0)\mathcal W .
\end{equation}

\paragraph{Number of rounds.}
At the start $P_0=I$, so the origin is known from the current vertex.
Consequently $\mathcal I(P_0)$ is the entropy of $\pi$, at most
$\log n_{\rm p}$. Also $Z_0=\tr(\Pi)\le n_{\rm p}$, giving
$\mathcal W_0\le(n_{\rm p}+1)^{1+1152/\eta}$. By
\eqref{eq:contraction} the expected final potential is at most
$\mathcal W_0e^{-c_0r}$. On the event that all $r$ matchings arrive
and the norm still misses its target, one remaining eigendirection
alone gives
\[
 \mathcal W_r\ge Z_r\ge\|G_r\|^{2r}
                  >(1-2\lambda)^r\ge e^{-4\lambda r}.
\]
Markov's inequality therefore bounds the probability of this event by
$(n_{\rm p}+1)^{1+1152/\eta}e^{-(c_0-4\lambda)r}$.
Since $4\lambda\le c_0/2$, we can meet the prescribed failure
allowance by choosing the smallest even $r$ satisfying
\begin{equation}\label{eq:round-choice}
 r\ge\max\left\{4,\ 4\log(n_{\rm p}+1),\
 \frac{2}{c_0}\left[\left(1+\frac{1152}{\eta}\right)\log(n_{\rm p}+1)
                  +\log\frac1{\epsilon_{\rm fail}}\right]\right\}.
\end{equation}

\paragraph{Computing the scores.}
The player stores each matching and its departure multipliers. Applying
one update factor $t\Pi A_N T_U$ or its adjoint takes one sparse
matching product, diagonal scalings, and subtraction of a weighted mean.
After $j$ accepted layers, applying $G$ or $G^*$ through those factors
costs $O((j+1)s)$. The probe $(GG^*)^k g$ uses $2k=r$ such applications.
Summing over $j=0,\ldots,r-1$ gives $O(sr^3)$ work beyond the matching
calls, as claimed in \Cref{lem:player}.

\subsection{From a completed game to cut bounds}\label{sec:overlay}

The mixing bound concerns the completed walk. To apply it to a
physical walk, follow its route through the completed layers.
Layer $j$ retains a $t_j$ fraction of its physical transport and
adds nonnegative mass. The completed walk therefore contains a
copy of every physical path, scaled by the same product
$\prod_j t_j$. Writing $A_j=D^{-1}B_j^\top$ and multiplying
the layers from $P_0=I$ gives
\[
 P_r\ge\Bigl(\prod_j t_j\Bigr) A_r\cdots A_1
     =\Bigl(\prod_j t_j\Bigr)D^{-1}Q_{\rm raw}^\top,
 \qquad
 \prod_j t_j\ge\exp\!\Bigl(-\sum_j\varepsilon_j\Bigr)
             \ge e^{-4\tau}\ge\zeta .
\]
In particular, the retained walks from $u$ to $v$ weigh at most
$d(v)\Pr[X=u\mid Y=v]$ in the completed population. In matrix form,
\begin{equation}\label{eq:path-domination}
                         Q\le P_r^\top D .
\end{equation}

\begin{lemma}[Cut bounds from physical walks]\label{lem:physical-cut}
On completion of a game satisfying \eqref{eq:terminal}, every set
$S$ of weight $d(S)\le M/2$ satisfies
\begin{equation}\label{eq:historical-cut}
                      d(S)-Q(S,S)\ge(\lambda/2)d(S).
\end{equation}
On any completion, let $e^+=d-Q\one$ and $e^-=d-Q^\top\one$ be
the missing starts and ends. If the retained walks in a graph $H$
put load at most $\Gamma c_e$ on each edge, then for every set $S$
\begin{equation}\label{eq:physical-endpoints}
 \begin{aligned}
 d(S)-Q(S,S)&=e^+(S)+Q(S,\bar S)\le e^+(S)+\Gamma c_H^+(S),\\
 d(S)-Q(S,S)&=e^-(S)+Q(\bar S,S)\le e^-(S)+\Gamma c_H^-(S).
 \end{aligned}
\end{equation}
In an undirected graph both capacities equal $\delta_H(S)$.
Extend $d,Q$ by zero at any vertices used only for routing.
\end{lemma}
\begin{proof}
Put $p=d(S)/M$ and $y=\one_S-p\one$. Independent endpoints with
distribution $\pi$ would give return mass $p\,d(S)$. In the completed
walk, the difference from that value is $M$ times the covariance of the
two membership indicators. Since $G_r$ and its adjoint kill constants,
\[
 Q(S,S)\le M\Pr[X\in S,\ Y\in S]
       =\ipd{\one_S}{P_r\one_S}
       =p\,d(S)+\ipd y{G_r y}
       \le\bigl[p+(1-p)\|G_r\|\bigr]d(S),
\]
where $\nd y^2=(1-p)d(S)$. On the mixing event
$\|G_r\|\le\sqrt{1-2\lambda}\le1-\lambda$. Thus the return shortfall is
at least $(1-p)\lambda d(S)$, giving \eqref{eq:historical-cut} for
$p\le1/2$.

For \eqref{eq:physical-endpoints}, count the same shortfall from each
end. Of the mass starting in $S$ that fails to return, $e^+(S)$ is
missing and the rest crosses an outgoing edge. Counting at the ends in
$S$ gives the missing ends $e^-(S)$ and the mass arriving across an
incoming edge. In each direction, congestion bounds the crossing mass.
\end{proof}

\paragraph{Walk congestion from the stored routes.}
Let $\widehat B_1,\ldots,\widehat B_r$ denote the layers returned by
\Cref{alg:cut-player}. Thus $\widehat B_1=\zeta B_1$,
$\widehat B_j=B_j$ for $j>1$, and
\begin{equation}\label{eq:physical-product}
 Q=\widehat B_1D^{-1}\widehat B_2\cdots D^{-1}\widehat B_r .
\end{equation}
At $u$, attenuation multiplies both the moving mass and the mass held
in place by $1-\kappa\one_U(u)$. Before the final scaling, a route of
matching weight $w$ from $u$ to $v$ therefore has physical weight
$(1-\kappa\one_U(u))w/2$, and the stay at $u$ has weight
$(1-\kappa\one_U(u))d(u)/2$. The multiplier depends on the departure
vertex, so even a symmetric matching can give an asymmetric layer.

The crossing-mass bound in \eqref{eq:physical-endpoints} needs a
congestion bound for complete walks. We obtain it directly from the
matching layers. A route of stored weight $w_{\rm phys}$ takes a
fraction $w_{\rm phys}/d(u)$ of the mass that reaches its start $u$.
At most $d(u)$ reaches $u$, so the complete walks using the route
weigh at most $w_{\rm phys}$. Summing the stored route weights across
layers, once per traversal of an edge, bounds its load and hence
gives the congestion bound. During trimming, we discard routes
that visit deleted vertices and keep surviving weights fixed. The
same congestion bound then applies to the surviving walks.

\section{Flow routines for the decompositions}\label{sec:flow-routines}

A matching request tries to route flow across a median. If too much remains
unrouted, we need a sparse cut. In trimming, the missing endpoint mass
supplies a flow. If excess remains, we need a set whose supply covers its
boundary capacity and a fixed fraction of its weight. Both routines maintain
vertex levels and send flow along paths of
decreasing level. The level sets expose a cut with little outward residual capacity
when a matching request falls short by more than its tolerance, or
when trimming leaves any excess. The cut's outgoing capacity equals
the net flow already sent out plus that residual capacity. Each
median request starts from zero flow. Trimming keeps the flow and the
levels from one deletion to the next.

\subsection{Capacities and vertex weights}\label{sec:normalization}

We fix the vertex weights before rounding arc
capacities or taking induced subgraphs. For expansion, the weight at $v$ must
dominate its incident capacity and any prescribed measure, both expressed in
the same normalized units. For the running time it must also cover the cost of
scanning the incident edges.

If $m=0$, return singleton parts. Otherwise set aside the isolated vertices as
singletons and scale the capacities to $c^0_e=mc_e/W$, of total $m$. In these
units the prescribed measure is $\bar\mu=(m/W)\mu$, with $\bar\mu=0$ if no
measure is supplied. On the remaining vertices set
\begin{equation}\label{eq:d}
 d(v)=s_\star\left\lceil\deg_G^{c^0}(v)+\deg_G^0(v)+\bar\mu(v)\right\rceil,
 \qquad s_\star=1/\tau .
\end{equation}
Both degree sums are $2m$, the measure has total at most $m$, and rounding up
adds less than one per vertex on at most $2m$ vertices. Thus $M_0=d(V)\le
7s_\star m$. Since $s_\star$ is even, the total weight is even and we
can split it into two equal integer halves at the median. Also,
$d(v)\ge s_\star=1/\tau$, so rounding an endpoint deficit up to the
next integer adds at most $\tau d(v)$.

Set the routing scale $\vartheta=\phi\log n\le1$. A request may send up to $c_e^0/\vartheta$
units along an arc. Round this bound down to an integer routing capacity $k_e$
and write $\cbar_e$ for its value on the normalized capacity scale. Each edge
loses less than $\vartheta W/m$ in the original units, giving
\begin{equation}\label{eq:lattice}
 k_e=\lfloor c^0_e/\vartheta\rfloor,\qquad
 \cbar_e=\vartheta k_e,\qquad
 \sum_e\left(c_e-\tfrac Wm\cbar_e\right)<\vartheta W .
\end{equation}
Discard edges with $k_e=0$. An undirected edge supplies two opposite arcs of
capacity $k_e$ each. Restoring the capacities can only increase cut capacities
inside a final part, and its vertex weights $d$ stay fixed. On edges between
parts, the total increase is less than $\vartheta W$.

Every phase inherits the same $d$, $\cbar$ and $\vartheta$. For a phase on $C$
of weight $M=d(C)$, we will use
\begin{equation}\label{eq:domination}
 \begin{gathered}
 d\ge s_\star\deg_G^{c^0},\qquad d\ge s_\star\deg_G^0,\qquad
 d\ge s_\star\bar\mu,\qquad d(v)\ge s_\star,\\
 |C|+|E(G[C])|=O(M),\qquad \cbar(E(G[C]))=O(M).
 \end{gathered}
\end{equation}
On a surviving vertex set $J\subseteq C$, write $\cbar_{G[J]}^+(S)$ and
$\cbar_{G[J]}^-(S)$ for the outgoing and incoming capacities of $S$ in $G[J]$.
\Cref{sec:near-optimal-decomp} works directly with the input capacities
and real vertex weights.

\subsection{Routing along decreasing levels}

Give each arc an integral capacity $U_e\ge0$ and each vertex an integral
source $\Delta(v)\ge0$. A flow $0\le f\le U$ leaves balance $\Gamma=\Delta-B_G
f$. Each arc $e=vw$ has its own residual pair, of capacity $U_e-f_e$ from $v$
to $w$ and $f_e$ from $w$ to $v$. Thus the outward residual capacity
$\res^+(S)$ of a set $S$ counts both unused capacity leaving $S$ and flow that
can be canceled on arcs entering $S$. With $\bar S$ the complement in the
current graph,
\begin{equation}\label{eq:residual-accounting}
 U^+(S)-\res^+(S)
 =f(S,\bar S)-f(\bar S,S)=\Delta(S)-\Gamma(S).
\end{equation}

\paragraph{Downhill search from a root.}
Let $z(v)$ be the integer balance at which $v$ stops absorbing flow.
For a routing request, $z(v)$ is the sink capacity at $v$. A vertex
with $\Gamma(v)<z(v)$ can absorb another $z(v)-\Gamma(v)$ units.
Give each vertex an integer level $j(v)$ between $0$ and $H$, where
$H\ge1$ is an integer. On every positive
residual arc $vv'$ maintain $j(v)\le j(v')+1$. We call such an arc
\emph{downhill} if $j(v)=j(v')+1$. A vertex $w$ is an \emph{eligible root}
if it has excess $\Gamma(w)>z(w)$ and room to rise, $j(w)<H$. The search
grows a path of downhill arcs from $w$. Its last vertex is the \emph{tip},
initially $w$. \Cref{alg:downhill} either delivers some excess to a vertex
below its threshold or raises the root to $H$.

\begin{algorithm}[htbp]
\caption{Downhill search from an eligible root $w$}\label{alg:downhill}
\begin{enumerate}[leftmargin=2.4em,itemsep=2pt,topsep=3pt]
\item Start a path of length zero at $w$.
\item Repeat the following steps with $v$ the current tip.
\begin{enumerate}[label=(\alph*),leftmargin=1.8em,itemsep=1pt,topsep=2pt]
\item If $\Gamma(v)<z(v)$, augment along the path by
\[
 \min\left\{\Gamma(w)-z(w),\ z(v)-\Gamma(v),\
                  \min_{e\text{ on the path}}\res(e)\right\}.
\]
Update the residual capacities and both endpoint balances. Return.
\item If a downhill arc leaves $v$, append it to the path and continue
from its head.
\item Otherwise raise $j(v)$ by one. If $v\ne w$, pop its entering arc
and continue from the preceding vertex. If $v=w$, return when $j(w)=H$ and
continue from $w$ otherwise.
\end{enumerate}
\end{enumerate}
\end{algorithm}

At a dead end every positive residual arc out of the tip leads to a vertex at
least as high. Raising the tip therefore preserves the level inequality. A
nonroot tip always has room to rise. Augmentation creates reverse capacity
only along uphill arcs and changes no interior balance. These observations
apply also to signed balances and thresholds, since the search compares only
their differences.

Every residual arc leaving $S_i=\{v:j(v)\ge i\}$ must go from
level $i$ to level $i-1$. To make the outward residual capacity small
relative to one side, it therefore suffices to find a level of small
weight relative to that side. We use the following
observation here and in trimming.

\begin{observation}[Geometric level selection]\label{lem:levels}
Let $R_0\supseteq\cdots\supseteq R_t$ satisfy $d(R_0)\le M$ and $d(R_t)\ge
a>0$. If $0<\delta<1$ and $\delta t>\log(M/a)$, then $d(R_i\setminus
R_{i+1})\le\delta d(R_i)$ for some $i<t$.
\end{observation}
\begin{proof}
Otherwise $d(R_t)<M(1-\delta)^t\le Me^{-\delta t}<a$.
\end{proof}

\begin{lemma}[Partial routing or a sparse cut]\label{lem:static-routing}
Let $M=d(C)$. For each vertex $v$ of $J\subseteq C$, let $a(v)$ be
the source supply and $z(v)$ the sink capacity. Both are nonnegative
integers, with $a+z\le d|_J$ and $a(J)=z(J)=b$. Fix
$0<\epsilon\le1/2$ and set
\[
 H=2\left(1+\left\lceil(4/\vartheta)\log(1/\epsilon)\right\rceil\right).
\]
In $O(MH)$ work the routine returns $b-e$ unit walks in $G[J]$, with $e$ units
unrouted. Their starting and ending weights are bounded by $a$ and $z$,
respectively, and their total length is $O(M/\vartheta)$. Each arc carries at
most its integer capacity. An undirected edge $uv$ therefore carries at most
$2\cbar_{uv}/\vartheta$ traversals in total.

If $e>\epsilon M$, the routine also returns a side $S$ with $e\le d(S)\le
d(J)/2$ whose normalized incoming or outgoing capacity in $G[J]$ is at most
$\tfrac32\vartheta d(S)$. The tolerance uses the original phase weight $M$,
even after some vertices have been removed.
\end{lemma}
\begin{proof}
Start with zero flow and levels, arc capacities $U=k$, and source
supply $\Delta=a$. Run \Cref{alg:downhill} with sink capacities $z$
until there is no eligible root. Each
augmentation reduces excess at its root and deficit at its tip by the same
amount. No other balance changes. Consequently $\alpha=(\Gamma-z)_+$ and
$\beta=(z-\Gamma)_+$ decrease coordinatewise from $(a-z)_+$ and $(z-a)_+$. A
vertex still in deficit has never been relabeled, since it absorbs flow
whenever the search reaches it. At termination all deficit is at level zero
and all excess at level $H$. A residual path from remaining excess to
remaining deficit would therefore need at least $H$ arcs. The balances
sum to $a(J)=z(J)$, so the remaining excess $\alpha$ and deficit $\beta$
have the same total $e$. Since $\Gamma-z=\alpha-\beta$, the final flow
has divergence
\[
 B_G f=a-\Gamma=(a-\alpha)-(z-\beta).
\]
The walk extraction below uses $a-\alpha$ as its starting weights and
$z-\beta$ as its ending weights. Each has total $b-e$.

\proofstep{A sparse cut when too much remains unrouted.}
Suppose $e>\epsilon M$. Write $L_i=\{j=i\}$ and cut between $S_i=\{j\ge i\}$
and $T_i=J\setminus S_i$. For $1\le i\le H$, all $e$ units of excess lie in
$S_i$ and all $e$ units of deficit in $T_i$. Since $\alpha\le a\le d$ and
$\beta\le z\le d$, each side weighs at least $e$. Substituting
$\Gamma(S_i)=z(S_i)+e$ into
\eqref{eq:residual-accounting} gives
\begin{equation}\label{eq:static-cut-identity}
 U^+(S_i)+e=a(S_i)-z(S_i)+\res^+(S_i).
\end{equation}
The net supply $a(S_i)-z(S_i)=z(T_i)-a(T_i)$ is at most
$\min\{d(S_i),d(T_i)\}$. For the residual term we can use either side
as well, since every residual arc leaving $S_i$ runs from $L_i$ to
$L_{i-1}$. The two residual capacities of each input
arc sum to $k_e$. Summing incident capacities on either level and applying
\eqref{eq:domination} gives
\[
 \res^+(S_i)\le (2/\vartheta)\min\{d(L_i),d(L_{i-1})\},
\]
where the factor two allows for the two copies of an undirected edge.

Start with the cut at level $K=H/2$ and move the threshold one level
at a time to shrink the side of smaller $d$-weight. If $d(S_K)\le d(T_K)$,
use $S_K\supseteq\cdots\supseteq S_H$, removing $L_i$ from $S_i$ at each step.
Otherwise use $T_K\supseteq\cdots\supseteq T_1$, removing $L_{i-1}$ from
$T_i$. The chain has at least $K-1$ steps. As the chosen side shrinks,
it continues to have at most half of $d(J)$, and even the last set weighs
at least $e$. Since
$(\vartheta/4)(K-1)\ge\log(1/\epsilon)>\log(M/e)$,
\Cref{lem:levels} gives a step whose removed level weighs at most
$\vartheta d(S)/4$, where $S$ is the set at that step of the chain.
The residual capacity leaving $S_i$ is at most $2/\vartheta$ times
the weight of that level, so $\res^+(S_i)\le d(S)/2$.
Then \eqref{eq:static-cut-identity} gives $U^+(S_i)\le\tfrac32 d(S)$.
Return $S$, charging its outgoing boundary if $S=S_i$ and its
incoming boundary if $S=T_i$. Multiplying the routing capacity by
$\vartheta$ gives the normalized capacity of that boundary. One sweep
of the level weights finds such a cut in $O(|J|+H)$ work.

\proofstep{Recovering walks from the final flow.}
Later searches may cancel flow on an earlier augmenting path, so use
the final flow $f$. If $b=e$, return no walks. Otherwise add a vertex
$h$ with $p(v)=a(v)-\alpha(v)$ units of flow from $h$ to $v$ and
$q(v)=z(v)-\beta(v)$ units from $v$ to $h$. Since $B_G f=p-q$, the
augmented flow is an integral circulation. All prescribed starts and
ends lie in the component of $h$ in its positive-flow support.

Take an Euler tour of this component from $h$, with one arc copy per
unit of flow. Splitting at visits to $h$ and dropping the auxiliary
arcs gives $b-e$ walks with the prescribed endpoints. An excursion
$h,v,h$ gives a walk of length zero. The walks use each arc at most
$f_e$ times, so their total length is at most
$\sum_e f_e\le\sum_e k_e=O(M/\vartheta)$ by \eqref{eq:domination},
counting both copies of each undirected edge. Constructing and
splitting the tour takes $O(|J|+|E(G[J])|+b+\sum_e f_e)=O(M/\vartheta)$
work.

\proofstep{Search work.}
There are at most $b$ integral augmentations, each on a path of length at most
$H$. A vertex rises at most $H$ times. Maintain lists of downhill arcs,
updating both endpoints when a level or a residual capacity changes. A relabel
scans its incident arcs, and an augmentation updates its path. Every path
extension is removed by a pop or cleared with an augmentation. Keep eligible
roots in a queue, retaining a root after an augmentation if it still has
excess. The search costs $O((|J|+|E(G[J])|+b)H)=O(MH)$, which covers both walk
extraction and the cut sweep.
\end{proof}

\subsection{Repairing a flow under deletions}\label{sec:du-ppr}

Trimming reuses its flow as sources increase and vertices are deleted,
following Fleischmann, Li, and Li~\cite[Section~7]{FLL25}. A repair
should leave us with either a flow whose balances fit within the sink
capacities $s(v)$ or a piece to delete. For a deletion we need the
piece's source supply to cover its outgoing capacity and a fixed
fraction of its weight. The second term lets us bound the size of the
piece by the supply still present in the graph.

A downhill search can find a superlevel set with little residual
capacity leaving it. By \eqref{eq:residual-accounting}, the balance
on that set is what lets its source supply pay for more than the
outgoing capacity. We therefore require a reserve at every positive
level. Each sink capacity $s(v)$ is a positive even integer, and we
use the reserve $u(v)=s(v)/2$. After each repair,
\begin{equation}\label{eq:flow-postconditions}
 j(v)>0\ \Longrightarrow\ \Gamma(v)\ge u(v),\qquad
 j(v)<H\ \Longrightarrow\ \Gamma(v)\le s(v).
\end{equation}
The upper bound confines any excess to level $H$. If there is no excess,
$\Gamma\le s$ gives $\Delta(S)\le s(S)+U^+(S)$ on every cut, even if
a level-zero balance is negative. If excess remains, search for a thin
boundary layer among the positive superlevel sets. For any such set $S$,
the reserve gives $\Gamma(S)\ge u(S)$, so
\eqref{eq:residual-accounting} yields
\[
 \Delta(S)\ge U^+(S)+u(S)-\res^+(S).
\]
Trimming takes $s=\sigma d$ and bounds the incident routing capacity at
each vertex by a common multiple of its reserve. At the height chosen
in \Cref{sec:trimming}, \Cref{lem:levels} gives a boundary layer thin
enough that $\res^+(S)\le u(S)/2$. Thus $\Delta(S)$ covers $U^+(S)$
and still has $u(S)/2=\sigma d(S)/4$ left to pay for the weight removed.

\paragraph{Pushing excess and restoring reserves.}
Deleting an arc that carries flow leaves its surviving tail with more
balance or its surviving head with less. A head can drop below its
reserve while keeping its positive level. A forward search pushes
excess away. A backward search repairs a depleted reserve by pulling
flow in or lowering the vertex to level zero, where the reserve is
optional. The gap between $u(v)$ and $s(v)$ also pays for relabeling in
the work analysis.

Initialize $f=j=0$. Between repairs, increase surviving sources by integral
amounts or remove a nonempty set and its incident arcs. Either update
preserves the level inequality on the surviving residual arcs. Write $\mathcal
R$ for the residual graph of the current flow. To restore
\eqref{eq:flow-postconditions}, repeatedly call
\Cref{alg:downhill} in either of the following two ways.
\[
\begin{array}{c|cccc}
 &\text{search graph}&\text{balance}&\text{balance cutoff }z&\text{level}\\\hline
 \text{push excess}       &\mathcal R&\Gamma&s&j\\
 \text{pull into reserve} &\mathcal R^{\rm rev}&-\Gamma&-u&H-j
\end{array}
\]
A push searches forward from a vertex with excess. A pull starts
at a depleted root with $j>0$ and $\Gamma<u$, and searches backward
along downhill residual arcs for a donor with $\Gamma>u$. Reversing
both the arcs and the levels lets us use the same search for a pull.
Whichever end we search from, the augmentation sends flow from donor
to recipient along downhill arcs of $\mathcal R$.

In the original levels, a push relabels upward where $\Gamma\ge s$, and a pull
relabels downward where $\Gamma\le u$. A push never takes its donor below $s$
or fills its recipient above $s$. A pull never drains its donor below $u$ or
fills its root above $u$. Thus fixing one violation creates none of the other
kind. Keep the eligible roots of the two searches in separate queues, updating
eligibility whenever a balance or level changes. Choose from either queue
after the current search has returned. Once both queues are empty,
\eqref{eq:flow-postconditions} holds. We call this procedure
\textsc{RepairByPaths}.

Directed trimming keeps a flow on the graph and a separate flow on its
reversal. A cut found in either state removes the same set $S$ from both. The
outgoing boundary paid for in the state that found the cut is the incoming
boundary in the other. The repair bound must therefore allow a deletion on the
current vertex set $J$ whenever
\begin{equation}\label{eq:du-source-paid-deletion}
 \min\{U(S,J\setminus S),U(J\setminus S,S)\}\le\Delta(S).
\end{equation}
Let $I$ be the initial supply plus all subsequent source increases.

\begin{lemma}[Bounded-height flow repair]\label{lem:du-kernel}
If $s(v)\ge\deg_G^0(v)$ for each vertex, all repairs and updates satisfying
\eqref{eq:du-source-paid-deletion} take $O(m+H[s(V)+I])$ work, excluding the
choice of deletion sets. Here $m$, the degrees and $s(V)$ refer to the initial
graph.
\end{lemma}
\begin{proof}
Let $b_v$ be the balance $\Gamma(v)$ clipped to $[u(v),s(v)]$.
The original balance has distance $p_v=|\Gamma(v)-b_v|$ from this interval.
An augmentation reduces the distance at its root. Relabels can change
direction at $v$ only after $b_v$ has moved from one end of the
interval to the other. We bound the total augmented volume $A$ and
the total distance $L$ traveled by the clipped balances, counting each
vertex only while it is live.

Write $P=\sum_{v\text{ live}}p_v$ and track $A+L+2P$.
When a balance changes by $a$, the clipped balance travels inside
the interval and its distance changes outside it, so
$|\Delta b_v|+|\Delta p_v|\le |a|$.
In an augmentation of $\delta$ units, the root moves $\delta$ toward
the interval without entering its interior. At the other endpoint
$p_v$ cannot increase and $b_v$ moves by at most $\delta$.
Thus $2P$ releases enough to pay for both $A$ and $L$.
Adding source can increase $L+2P$ by at most twice the amount added.

A deletion pays for flow in the cheaper direction. To cover the other
direction, use the imbalance on the vertices being removed. Write
$F^+,F^-$ for the outgoing and incoming flows erased on deleting $S$.
The smaller is at most $\Delta(S)$ by \eqref{eq:du-source-paid-deletion}, and
$F^+-F^-=\Delta(S)-\Gamma(S)$. Thus
\[
 F^++F^-\le3\Delta(S)+\sum_{v\in S}|\Gamma(v)|
          \le3\Delta(S)+s(S)+\sum_{v\in S}p_v .
\]
Erasing this flow increases $L+2P$ at the survivors by at most
$2(F^++F^-)$. Removing $S$ also erases $2\sum_{v\in S}p_v$ from the
account. The net increase is therefore at most
\[
 2(F^++F^-)-2\sum_{v\in S}p_v
     \le 6\Delta(S)+2s(S).
\]
Over all deletions, $s(S)$ sums to at most $s(V)$ and
$\Delta(S)$ to at most the total supply introduced, $I$. Initially
$P\le s(V)+I$, so $A+L=O(s(V)+I)$ on every finite prefix.

Each run of relabels in the same direction at $v$ has length at most $H$.
Between consecutive runs, $b_v$ travels the full distance
$s(v)-u(v)=s(v)/2$. As a relabel costs $O(s(v))$, all relabels together
cost $O(H[s(V)+L])$. The integral augmentations cost $O(HA)$.

\proofstep{Search work.}
Keep downhill arc lists for both searches and update the root queues
whenever a level or balance changes. A relabel scans the incident
arcs and an augmentation updates its path. Every extension of a search
path is popped at a relabel or cleared at an augmentation, apart from
a current path of length at most $H$. Deleting vertices and arcs costs
$O(|V|+m)$ in total. These bounds pay for every operation on any
finite prefix, which also proves termination.

\end{proof}

\section{Expander decompositions}\label{sec:decompositions}\label{sec:undirected}\label{sec:directed}

Run a cut--matching game in each phase $C$. When matching returns a balanced
sparse split, we recurse on its blocks. If all rounds complete, we trim $C$
using the routed walks. On the mixing event, trimming finalizes a large
induced expander. In a directed graph we also order the resulting parts so
that we can charge only backward arcs.

If $\phi\log^2 n\ge1$, singletons in any order suffice. They satisfy the
expansion requirement, and even charging the entire capacity $W$ meets the
boundary bound. Henceforth $\phi<1/\log^2 n$. Write $M=d(C)$ for the original
weight of the phase, take $r=\Theta(\log n)$ as in \Cref{lem:player}, and
recall that $\vartheta=\phi\log n$.

\subsection{Undirected matching}\label{sec:undirected-matching}

Split $d$ at a median of the scores weighted by $d$, assigning ties so that
the lower and upper weights $a,z$ each have total $M/2$ and $a+z=d$. Since
$d$ is even at every vertex, we can split the ties integrally. In an
undirected graph, a route from the lower half to the upper half also supplies
a route in reverse. Each endpoint then sends exactly as much as it receives,
so any unused weight there can be matched to itself.

\begin{corollary}[Undirected matching]\label{lem:undirected-matching}
An undirected request returns either a cut with both sides of weight at least
$\tau M/2$ and normalized capacity $O(\vartheta M)$, or a matching with exact
marginals $d$ and at most $\tau M$ weight that misses the median. Its physical
layer has congestion at most $2/\vartheta$. The matching work and total route
length are $O(M/\vartheta)$ per request and $O(M/\phi)$ over the phase.
\end{corollary}
\begin{proof}
Route $a$ to $z$ by \Cref{lem:static-routing} with tolerance $\tau/2$. A
shortfall $e>\tau M/2$ gives the required cut. Otherwise let $F(u,v)$ be the
mass routed from $u$ to $v$. Including the reversed routes and the self-pairs
gives
\begin{equation}\label{eq:undir-slack}
 N=F+F^\top+\diag(d-F\one-F^\top\one).
\end{equation}
Since $F\one\le a$ and $F^\top\one\le z$, the added diagonal is nonnegative
and both marginals equal $d$. The routed mass in each direction is $M/2-e$,
so the self-pairs weigh $2e\le\tau M$. Only those pairs can miss the median.

Each traversal of an edge in $F$ occurs twice in $N$, once in each direction.
Each copy has at most half its matching weight in the physical layer.
Together they thus contribute at most the original traversal's weight. The
load on an edge $uv$ is at most $2\cbar_{uv}/\vartheta$ by
\Cref{lem:static-routing}. The median can be selected in $O(M)$ work, even by
listing $d(v)$ copies of each score. The routing tolerance is constant and
$r/\vartheta=O(1/\phi)$, giving the claimed work and route length.
\end{proof}

\subsection{Ordering pieces by their charged boundary}

A directed search can expose a piece with little incoming capacity and large
outgoing capacity. Put that piece before the remaining vertices, so that only
incoming arcs from the remainder point backward. A piece with little outgoing
capacity belongs after the remainder for the same reason. Matching and
trimming both group all pieces of each kind into a single block.

\begin{observation}\label{lem:order}
Remove disjoint pieces from $C$ one at a time, charging each piece for either
its incoming or its outgoing normalized capacity just before removal. Let
$I,O$ be the unions of the pieces charged for incoming and outgoing capacity,
respectively, and let $J=C\setminus(I\cup O)$. If the charges sum to $C_{\rm
ch}$, the ordered partition $(I,J,O)$ has backward capacity at most $C_{\rm
ch}$. In particular,
\begin{equation}\label{eq:end-unions}
    \cbar_C^-(I)\le C_{\rm ch},\qquad \cbar_C^+(O)\le C_{\rm ch}.
\end{equation}
Omit empty blocks. In an undirected graph take $I$ to be the union of all
removed pieces and $O=\varnothing$. The crossing capacity of $(I,J)$ is then
at most $C_{\rm ch}$.
\end{observation}
\begin{proof}
A backward arc $J\to I$ is charged when its head is removed, and an arc $O\to
J$ when its tail is removed. An arc $O\to I$ is charged whichever endpoint is
removed first. These are all the backward arcs. The arcs entering $I$ and
those leaving $O$ are two subsets of them, overlapping on $O\to I$. In an
undirected graph a crossing edge is charged when its endpoint in $I$ is
removed.
\end{proof}

\subsection{Directed matching and retirement}\label{sec:matching}

In a directed graph the two directions across a median require separate
routes. Unused sending weight and unused receiving weight may occur at
different vertices. Pairing each vertex with itself can therefore
leave a shortfall. Allow $\tau M/r$
units to go unrouted in each direction. This permits at most $2\tau M$
unmatched weight over $r$ accepted rounds. A failed request, however, may
expose a piece of weight just above $\tau M/r$. Splitting it off immediately
could leave a child of weight $M-O(\tau M/r)$. Instead we retire the piece
from further requests and retry the current round. Once the retired weight
exceeds $\tau M$, we split into the live block and the two blocks of retired
vertices. Each block then weighs at most $(1-\tau/2)M$.

Keep $I$ and $O$, the sets retired for incoming and outgoing capacity,
throughout the game, starting with both empty. The player still computes each
round's scores $x$ and marked vertices on the full phase $C$.
\Cref{alg:directed-round} changes only where the current pair of routing
requests is made. An earlier accepted layer keeps all its routes, including
those that pass through vertices retired later.

\begin{algorithm}[htbp]
\caption{Directed matching for one round with scores $x$}\label{alg:directed-round}
\begin{enumerate}[leftmargin=2.4em,itemsep=2pt,topsep=3pt]
\item\label{step:dir-retry} Set $T=I\cup O$ and $J=C\setminus T$. At a
weighted median of $x$ on $J$, split $d|_J$ into integral lower and upper
weights $a_0,b_0$ of equal total.
\item In $G[J]$ make separate calls to \Cref{lem:static-routing}
with tolerance $\tau/r$, first from $a_0$ to $b_0$, then from $b_0$ to $a_0$.
If a shortfall exceeds $\tau M/r$, stop at that call and take its sparse cut.
\item If both requests succeed, extend their endpoint couplings $F_1,F_2$
by zero to $C$ and return
\begin{equation}\label{eq:N}
                       N=F_1+F_2+\diag(d|_T).
\end{equation}
Here $d|_T$ is zero outside $T$.
\item Otherwise discard the routes of this attempt. Add the returned side
$S$ to $I$ if its incoming capacity was bounded, or to $O$ if its outgoing
capacity was bounded, charging that boundary in $G[J]$.
\item If $d(I)+d(O)>\tau M$, return the ordered split
$(I,C\setminus(I\cup O),O)$ with empty blocks
omitted. Otherwise return to step~\ref{step:dir-retry} with the same scores
and marked vertices.
\end{enumerate}
\end{algorithm}

The routed pairs in an accepted attempt all cross or meet the same live
median. The player requires only this common threshold, so a change of median
on a retry costs nothing in its analysis. The self-pairs on $T$ may miss it,
but they weigh at most $\tau M$ and fill both marginals there. On $J$, the
two sending weights sum to $d|_J$, as do the two receiving weights. Thus
$N\one,N^\top\one\le d$. In an accepted round $j$ only the two live routing
shortfalls $e_1,e_2$ contribute to the missing mass at either end. Summing
over the accepted rounds gives
\begin{equation}\label{eq:missing-mass}
 \delta_j=M-N_j(C,C)=e_1+e_2\le 2\tau M/r,
 \qquad \sum_j\delta_j\le 2\tau M.
\end{equation}
This meets the game-wide allowance in \eqref{eq:matching-loss-budget}. Each
request sends at most $\cbar_e/\vartheta$ along a directed arc $e$, so their
combined matched load is at most $2\cbar_e/\vartheta$. The moving half of the
physical layer therefore has congestion at most $1/\vartheta$.

\begin{corollary}[Directed matching]\label{lem:matching}
A routing request costs $O(M\log\log n/\vartheta)$. The matching work over a
phase is $O(M\log\log n/\phi)$ and the total extracted route length is
$O(M/\phi)$. When matching returns an ordered split, each block has weight at
most $(1-\tau/2)M$ and the backward capacity is $O(\vartheta M)$ in normalized
units.
\end{corollary}
\begin{proof}
A failed attempt retires a fresh piece $S$ with $\tau M/r<d(S)\le d(J)/2$ and
charged capacity at most $\tfrac32\vartheta d(S)$ in $G[J]$. Fewer than $r$
such pieces fit below the retirement threshold $\tau M$, and at most one
further retirement ends the phase. There are also at most $r$ successful
attempts. Thus $O(r)$ attempts suffice, each with an $O(M)$ median
computation and at most two routing requests. A request costs
$O(M\log(r/\tau)/\vartheta)$ and its routes have total length
$O(M/\vartheta)$ by \Cref{lem:static-routing}. Thus the smaller tolerance
adds a factor $O(\log\log n)$ to search work alone. Multiplying by $O(r)$ and
using $r/\vartheta=O(1/\phi)$ gives the two phase bounds.

If retirement ends the phase, write $t\le\tau M$ for the weight retired
before the last piece. That piece weighs at most $(M-t)/2$, so the final
retired weight is greater than $\tau M$ and at most $(1+\tau)M/2$. The live
block therefore weighs less than $(1-\tau)M$, and neither retired block
exceeds $(1+\tau)M/2$. Since $\tau<1/2$, all three meet the claimed upper
bound. The retired pieces are disjoint, so their total charge is at most
$\tfrac32\vartheta d(T)$, where $T$ is their final union.
\Cref{lem:order} bounds the backward capacity by this charge.
\end{proof}

For each accepted pair store its endpoints, its physical weight and its
route. In an undirected graph a route and its reversal can share the same
vertex list and carry different physical weights. Integrality gives $O(M)$
endpoint records per request, so a layer and its adjoint each take $O(M)$
work to apply. If matching returns a split, discard the game. Otherwise its
route lists now describe the physical walks through all $r$ layers.

\subsection{Endpoint deficits under deletions}\label{sec:physical}

Trimming begins on all of $C$, including vertices retired during matching.
For the rest of the phase $J$ denotes the vertices that survive
\emph{trimming}. A walk survives precisely when its entire route lies in
$G[J]$. Deleting an interior vertex can therefore destroy a walk whose two
endpoints remain. Let $Q_J$ count the walks kept, with their original
weights. Initially $Q_C=Q$ includes the final thinning. We use
\begin{equation}\label{eq:decomp-constants}
              \gamma=\lambda/2,\qquad \Lambda=2r/\vartheta .
\end{equation}
Each physical layer has congestion at most $2/\vartheta$, so $\Lambda$ bounds
the congestion of complete walks by \Cref{sec:overlay}.

On the mixing event, \eqref{eq:historical-cut} gives $d(S)-Q_C(S,S)\ge\gamma
d(S)$ for every $S\subseteq C$ with $d(S)\le M/2$. For a surviving
$S\subseteq J$ the prescribed weight $d(S)$ is the same and $Q_J(S,S)\le
Q_C(S,S)$. Thus its return shortfall is still at least $\gamma d(S)$. To
apply this bound inside $G[J]$ we need to know how much of the shortfall is
carried by walks that still cross the cut.

On $J$ the missing starts are $e_J^+=d|_J-Q_J\one$ and the missing
ends are $e_J^-=d|_J-Q_J^\top\one$. The two counts in
\eqref{eq:physical-endpoints} apply to these surviving walks with
the outgoing and incoming capacities of $G[J]$. A single supply
suffices for both directions if it dominates
\[
                   \Delta_J^*(v)=\max\{e_J^+(v),e_J^-(v)\}.
\]
Rounding the surviving walk fractions gives an integer upper bound
$\Delta_J$ for the flow routines (\Cref{sec:endpoint-updates}).

\begin{lemma}[Endpoint deficits of surviving walks]\label{lem:physical}\label{lem:ledger}
After any completed game one can maintain integer sources
$\Delta_J^*\le\Delta_J\le d|_J$, nondecreasing at each surviving vertex. On
the mixing event, for every deletion sequence and both orientations,
\begin{equation}\label{eq:defective-expansion}
 \gamma d(S)\le\Delta_J(S)+\Lambda\cbar_{G[J]}^\pm(S)
       \qquad(S\subseteq J,\ d(S)\le M/2).
\end{equation}
On every completion, let $C_{\rm ch}$ be the sum of the normalized boundary
capacities charged for deleted pieces, each measured in its chosen direction
just before deletion. Then
\begin{equation}\label{eq:common-budget}
                 \Delta_J(J)\le\eta M/2+2\Lambda C_{\rm ch}.
\end{equation}
Maintaining the sources over a phase costs $O(M\log^2 n+M/\phi)$ work and
$O(M/\phi)$ space.
\end{lemma}
\begin{proof}
The return bound and \eqref{eq:physical-endpoints} give
\eqref{eq:defective-expansion} with any common upper bound on the deficits. For the total supply, fix the current ordered
blocks $(I,J,O)$ of
\Cref{lem:order}. A walk is lost precisely when it visits $I$ or $O$.
If it traverses no backward arc between the blocks, its block sequence
is nondecreasing. It can then visit $I$ only by starting there, and $O$
only by ending there. The endpoint marginals bound these two classes by
$d(I)$ and $d(O)$. The remaining lost walks traverse a backward arc,
so their weight is at most $\Lambda C_{\rm ch}$ by congestion and
\Cref{lem:order}. In the undirected case order the deleted union before
$J$ and count edges between the two blocks.

At either end the total deficit is $e(J)=d(J)-Q_J(J,J)$. Passing from
$C$ to $J$ loses at most $d(I)+d(O)+\Lambda C_{\rm ch}$ walk mass,
but reduces the prescribed weight by exactly $d(I)+d(O)$. Thus
\[
 e(J)\le e(C)+\Lambda C_{\rm ch}\le8\tau M+\Lambda C_{\rm ch}
\]
by \eqref{eq:player-mass-loss}.

The larger deficit at a vertex is at most the sum of the two, so $\Delta_J^*(J)\le 2e(J)$. The
construction in \Cref{sec:endpoint-updates} adds at most $\tau d(v)+1$ at
each surviving vertex and keeps the cap $d(v)$. Hence
\[
 \Delta_J(J)\le 2e(J)+\tau d(J)+|J|
             \le 18\tau M+2\Lambda C_{\rm ch},
\]
using $1\le\tau d(v)$. This implies \eqref{eq:common-budget} since
$18\tau\le\eta/2$. The same construction gives monotonicity and the work and
space bounds.
\end{proof}

\subsection{Trimming}\label{sec:trimming}

For a cut side $S$ with $d(S)\le d(J)/2$, surviving walks can carry at most
$\Lambda\cbar_{G[J]}^+(S)$ of the return shortfall across the outgoing
boundary. The rest is missing at the starts in $S$. Give the vertices their
common supply $\Delta_J$ and allow $S$ to absorb only $\sigma d(S)$, where
$\sigma=\gamma/2$. Routing the surplus out of $S$ bounds its outgoing
capacity from below, even when few surviving walks cross there. A flow on the
reversed graph gives the corresponding incoming bound from missing ends. In
an undirected graph a single state with two arc copies per edge suffices.

Set $\kcl=4/\gamma$ and scale the normalized arc capacities by
$\kcl\Lambda$ in each flow state. If repair leaves excess, we will
find a set $S$ whose supply satisfies
\begin{equation}\label{eq:high-cut}
             \kcl\Lambda c_S+\frac{\sigma}{4}d(S)\le\Delta_J(S),
\end{equation}
where $c_S$ is the normalized boundary capacity in the direction searched.
Since $\Delta_J(S)\le d(S)$, the original walks could carry at most $\Lambda
c_S\le d(S)/\kcl$ across that boundary in the charged direction. We will test
the removed unions $I$ and $O$ against the original walks. An arc from $O$ to
$I$ can occur in both boundaries. With this choice of $\kcl$, even
counting twice, the charged boundaries explain at most half the required
shortfall $\gamma(d(I)+d(O))$. The term $\sigma d(S)/4$ also bounds
the weight of the next proposed deletion. Since
$(\sigma/4)d(S)\le\Delta_J(S)\le\Delta_J(J)$, its weight is at most
$4\Delta_J(J)/\sigma$.

Each state uses the following sink capacities, arc capacities and height
limit.
\begin{equation}\label{eq:repair-parameters}
 s=\sigma d,\qquad U_e=\kcl\Lambda\cbar_e=2\kcl r k_e,\qquad
 H=1+\left\lceil\frac{8\kcl\Lambda}{\sigma}(1+\log M)\right\rceil .
\end{equation}
The arc capacities are integral. Since $\sigma s_\star=2^{12}$, each sink
capacity is an even integer at least the incident arc count.

\begin{algorithm}[htbp]
\caption{$\textsc{Trim}(C)$}\label{alg:trim}
\begin{enumerate}[leftmargin=2.4em,itemsep=1pt,topsep=3pt]
\item Initialize $I=O=\varnothing$, $J=C$, and source $\Delta_C$
as in \Cref{lem:physical}. In each state start with zero flow and zero
levels.
\item\label{step:trim-repair} Run \textsc{RepairByPaths} in each
state. If no state has excess, report failure when $d(J)<15M/16$,
and otherwise return $(I,J,O)$ with $J$ final, omitting empty blocks.
\item Otherwise, in a state with excess, find a superlevel set $S$ satisfying
\eqref{eq:high-cut}.
\item Add $S$ to $I$ or $O$ according to its charged boundary, as
in \Cref{lem:order}. Remove $S$ and its incident arcs from all states. Set
$J\gets J\setminus S$, discard the routes meeting $S$, and raise the
surviving sources to $\Delta_J$. Return to step~\ref{step:trim-repair}.
\end{enumerate}
\end{algorithm}
Whenever the algorithm returns a part $J$, the check guarantees
$d(J)\ge15M/16$. On the mixing event we obtain the stronger conclusions
below.

\begin{lemma}[Trimming to an induced expander]\label{lem:cleanup}
Use the sources of \Cref{lem:physical}. If the walk bound
\eqref{eq:defective-expansion} holds throughout trimming, the final size
check passes, and the procedure returns a nonempty $J$ with
\begin{equation}\label{eq:small-removal}
 d(C\setminus J)\le\eta M/\gamma,\qquad
 \Delta_J(J)\le\eta M,\qquad
 C_{\rm ch}\le\eta M/(\kcl\gamma\Lambda).
\end{equation}
Every cut of $G[J]$ satisfies
\begin{equation}\label{eq:induced-expansion}
 \min\{\cbar_{G[J]}^+(S),\cbar_{G[J]}^-(S)\}
 \ge\frac{\gamma-\sigma}{(\kcl+1)\Lambda}
       \min\{d(S),d(J\setminus S)\}.
\end{equation}
In addition to maintaining $\Delta_J$, repair and cut selection cost
$O(M\Lambda\log n)=O(M\log n/\phi)$ on every execution.
\end{lemma}
\begin{proof}
\proofstep{Expansion on return.}
Take the side $S$ of a cut of $J$ with $d(S)\le d(J)/2$. In either flow state all
balances are at most $s$ on return, so
\[
 \Delta_J(S)=\Gamma(S)+f(S,J\setminus S)-f(J\setminus S,S)
      \le\sigma d(S)+\kcl\Lambda\cbar_{G[J]}^\pm(S).
\]
Here $+$ denotes the outgoing boundary of $S$ in $G[J]$. Running the same
calculation in the reversed state gives the incoming boundary $-$ in $G[J]$.
Since $d(S)\le d(J)/2\le M/2$, the walk bound applies to this same side and
direction. Combining the inequalities gives
\[
 \gamma d(S)\le\Delta_J(S)+\Lambda\cbar_{G[J]}^\pm(S)
       \le\sigma d(S)+(\kcl+1)\Lambda\cbar_{G[J]}^\pm(S).
\]
This proves \eqref{eq:induced-expansion}.

\proofstep{A piece to delete.}
If excess remains, the top level is nonempty. For $1\le i\le H$, the balance
on $S_i=\{v:j(v)\ge i\}$ is at least $\sigma d(S_i)/2$ by
\eqref{eq:flow-postconditions}. Its outgoing residual arcs leave
from $L_i=\{j=i\}$, with total capacity at most $2\kcl\Lambda d(L_i)$ by
\eqref{eq:domination}, including both arc copies of an undirected edge. Take
$\delta=\sigma/(8\kcl\Lambda)$. Since $d(S_H)\ge1$ and $\delta(H-1)>\log M$,
\Cref{lem:levels} gives a level with $d(L_i)\le\delta d(S_i)$. Its residual
boundary uses at most $\sigma d(S_i)/4$, half the guaranteed balance. For
$S=S_i$ the balance identity \eqref{eq:residual-accounting} now gives
\[
 \Delta_J(S)=\kcl\Lambda c_S+\Gamma(S)-\res^+(S)
       \ge \kcl\Lambda c_S+(\sigma/4)d(S),
\]
as required in \eqref{eq:high-cut}. The same deletion meets
\eqref{eq:du-source-paid-deletion} in both states, since the boundary
paid for is outgoing in one and incoming in the other.

\proofstep{Why the deleted pieces have small total weight.}
Let $p=d(I)+d(O)$ after any number of deletions and let
$b=\Lambda C_{\rm ch}$. Each deleted piece paid for its boundary out
of at most its own weight by \eqref{eq:high-cut}, so
$\kcl b\le p$. Put $D_0=\eta M/2$. Whenever $p\le M/2$, both $d(I)$
and $d(O)$ are at most half the original phase weight. Apply the walk
bound to the incoming boundary of $I$ and the outgoing boundary of $O$,
using the original supply. As the sets are
disjoint, together they receive at most $D_0$. By \Cref{lem:order},
their boundaries together cost at most twice the deletion charges,
since an arc from $O$ to $I$ can occur in both.
Therefore
\[
 \gamma p\le D_0+2b\le D_0+(\gamma/2)p,
 \qquad\text{so}\qquad p\le2D_0/\gamma .
\]
It remains to rule out a jump past $M/2$. As long as
$p\le2D_0/\gamma$, \eqref{eq:common-budget} gives $\Delta_J(J)\le D_0+2b\le2D_0$.
By \eqref{eq:high-cut} the next piece weighs at most $8D_0/\sigma$,
and $2D_0/\gamma+8D_0/\sigma<M/2$. Starting from zero, $p$ thus stays
at most $2D_0/\gamma=\eta M/\gamma<M/16$. The size test therefore
passes. Substituting in $\kcl b\le p$ and $\Delta_J(J)\le2D_0$ gives
the other two bounds in \eqref{eq:small-removal}.

\proofstep{Work over all deletions.}
Each surviving source only increases, and each is capped by its original
$d(v)$. The total source ever introduced in a state is therefore at most $M$.
Since the sinks sum to $\sigma M$ and dominate the incident arc counts,
\Cref{lem:du-kernel} bounds total repair work by $O(MH)$. Keep the vertex
list and weight at each level up to date on relabels and deletions. A sweep
takes $O(H)$ to find a thin level and its superlevel weight, and listing a
deleted set costs another $O(H+|S|)$. These sets are disjoint, so cut
selection also costs $O(MH)$ on every
execution. Finally $\Lambda=2r/(\phi\log n)=\Theta(1/\phi)$ and
$H=O(\Lambda(1+\log M))=O(\log n/\phi)$.
\end{proof}

\subsection{Maintaining the endpoint deficits}\label{sec:endpoint-updates}

Deleting a vertex invalidates every walk that visits it, including walks
whose endpoints both survive. We store the routed pieces of the walks,
one layer at a time. Deactivating a route removes its contribution to every
complete walk that used it. Two passes through the remaining layers count the
surviving mass at each end without listing those walks.

For layer $\widehat B_j$ of \eqref{eq:physical-product}, let $B_{j,J}(u,v)$
be the total physical weight of its routes from $u$ to $v$ lying entirely in
$J$. Include stays and self-pairs at surviving vertices and extend the matrix
by zero off $J$. All retained route weights and the denominators in
$D=\diag(d)$ stay fixed. In particular two routes with the same endpoints can
be removed at different times.

\paragraph{A forward and a backward pass.}
Let $f_j(v)$ be the fraction of $d(v)$ that reaches $v$ through the first $j$
surviving layers. Let $g_j(v)$ be the fraction of the prescribed mass at $v$
that would survive the remaining layers. The two passes satisfy
\[
 f_0=g_r=\one_J,\qquad
 f_j=D^{-1}B_{j,J}^\top f_{j-1},\qquad
 g_{j-1}=D^{-1}B_{j,J}g_j .
\]
Thus $Df_r=Q_J^\top\one$ counts surviving ends and $Dg_0=Q_J\one$ counts
surviving starts. The common supply $\Delta_J$ uses both counts even in an undirected
graph. Attenuation acts at departure, so the sending and receiving
weights of a layer can differ.

A fraction can fall by an arbitrarily small amount when a route disappears.
To limit how often a change must be passed to the next layer, take $K=r/\tau$
and round each output down to a multiple of $1/K$ before using it as an
input. Write $\tilde f,\tilde g$ for the rounded evaluations on $J$.
Both $D^{-1}B_{j,J}^\top$ and $D^{-1}B_{j,J}$ have nonnegative entries
and row sums at most one, since the mass arriving at or leaving $v$
is at most $d(v)$. For the forward pass, write $A_j=D^{-1}B_{j,J}^\top$.
Then
\[
 f_j-\tilde f_j
 =A_j(f_{j-1}-\tilde f_{j-1})+(A_j\tilde f_{j-1}-\tilde f_j).
\]
The row-sum bound limits the first term to the largest input error.
Rounding down adds at most $1/K$ in the second term. Applying this
calculation in both passes gives
\[
 0\le f_j-\tilde f_j\le (j/K)\one,\qquad
 0\le g_j-\tilde g_j\le ((r-j)/K)\one .
\]
The larger endpoint deficit is the one with the smaller surviving fraction.
At $v\in J$ use the integer source
\begin{equation}\label{eq:endpoint-counter}
 \Delta_J(v)=d(v)-
      \left\lfloor d(v)\min\{\tilde f_r(v),\tilde g_0(v)\}\right\rfloor .
\end{equation}
Both fractions are nonincreasing under deletions, so the source can only rise
at a surviving vertex and never exceeds $d(v)$. Its excess above
$\Delta_J^*(v)$ is at most $(r/K)d(v)+1=\tau d(v)+1$.

\paragraph{Updating the two passes.}
Each entry stores its weighted sum over active records and its rounded
value. Every active route has both endpoints in $J$, so fixing both
boundary inputs to one on $C$ gives the same sums as $\one_J$.

On deleting vertices, use their route lists to deactivate every record
that visits them, including stays and both directions of a shared
undirected route. An active flag prevents double removal. Before
changing any rounded values, subtract each removed record's contribution
from its sum in each pass and mark the affected entries.

Process marked entries in layer order, forwards for $\tilde f$ and
backwards for $\tilde g$. By the time an entry is reached, its sum
contains all changes to its inputs. Round it down to the grid. If its
value drops by $\delta$, subtract $\delta$ times the record coefficient
from each dependent sum along an active record and mark that entry.
The resulting values agree with a fresh rounded evaluation on $J$.
Rounding error therefore accumulates only across layers. Update
\eqref{eq:endpoint-counter} at the affected surviving endpoints.

The stored sums have simple integer representations. All matching weights and
$d$ are integral, $\kappa=1/(256r)$, and the first layer is thinned by
$\zeta=1-4/s_\star$. Thus $L_B=512r s_\star$ clears the denominators of every
physical weight. For the forward pass at $v$ store the numerator
\[
 \sum_u (L_B B_{j,J}(u,v))(K\tilde f_{j-1}(u)).
\]
Integer division by $L_Bd(v)$ gives $K\tilde f_j(v)$. The backward pass uses
outgoing records in the same way. Each numerator is at most $K L_Bd(v)$ by
the marginal bounds.

Each rounded coordinate drops at most $K$ times during a phase, so
the $O(Mr)$ records transmit at most $O(MrK)=O(M\log^2 n)$ changes.
Every mark follows a removed record or a changed input. Visiting the
$r$ layers per deletion costs $O(r|C|)$ in total. Building and scanning
the route index costs $O(Mr+M/\phi)$ for the records and their total
length. Listing each marked entry once takes $O(Mr)$ space, so the
total space is $O(M/\phi)$ since $\phi<1/\log^2 n$.

\subsection{Recursion and proof of the decomposition theorem}\label{sec:algorithm}

At the start of a call, finalize as singletons any vertices isolated in its
rounded graph. They can appear in any order at zero rounded boundary cost. In
a directed graph this applies only to vertices with no incident arc in either
direction. Run a fresh game on the remaining vertices if there are any. For a
matching split, recurse on its nonempty blocks in the returned order. After a
trim, put the output on $I$ first, then the finalized $J$, then the output on
$O$. A reported failure stops the algorithm.

Build each child from the graph at the \emph{start} of the phase by scanning
the parent's input once. Pieces grouped into the same child may have been
removed at different times. Every edge between them still belongs to that
child's induced graph. Release the parent state before descending.

\begin{proof}[Proof of \Cref{thm:main}]
Both matching routines meet the two allowances of \Cref{lem:player}. For any
fixed $a>0$, choose the game's failure allowance $(m+2)^{-a-8}$ and its size
bound $n_{\rm p}=n$. Conditional on the history when a game starts, the
probability that it completes and misses the mixing target is at most this
allowance. Every split has at least two nonempty children when finalized
parts are included, and a game that produces just one part finalizes it.
There are at most $2m$ initially nonisolated vertices, so at most $4m+1$
games can start, including a possible final failure. A union bound gives
probability at least $1-(m+2)^{-a}$ that every completed game mixes. On this
event \eqref{eq:defective-expansion} holds for every surviving set tested in
every trim. By \Cref{lem:cleanup} the algorithm finishes and each finalized
core satisfies \eqref{eq:induced-expansion}.

\proofstep{Charging games and trims.}
Every recursive child has weight at most $(1-\tau/2)M$. This follows from the
balance of a matching split in \Cref{lem:undirected-matching,lem:matching}.
After a trim, only the end blocks recur, and they have combined weight at
most $M/16$ by the size check. The weights are positive integers and
$M_0=O(m)=O(n^2)$, so recursion has $O(\log n)$ levels. Phases at each level
are disjoint. Trimming admits a stronger charge. A successful trim finalizes
a core $J_C$ of weight at least $15d(C)/16$. The finalized cores
are disjoint across \emph{all} levels, so their total weight is at
most $M_0$. Hence
\begin{equation}\label{eq:mass-sum}
 \sum_{\text{phases }C}d(C)=O(M_0\log n)=O(m\log n),
 \qquad \sum_{\text{successful trims }C}d(C)\le(16/15)M_0=O(m).
\end{equation}
The bounds on phase weights hold on every execution. An execution that stops
at a failed check pays for at most one additional trim of weight at most
$M_0$, so the total weight on which trimming is attempted is always $O(m)$.

\proofstep{Expansion and crossing capacity.}
On the mixing event, \eqref{eq:induced-expansion} gives each core
$\Omega(\phi)$ expansion with respect to $d$ in normalized capacities, since
$\Lambda=\Theta(1/\phi)$. In input units $c_e\ge(W/m)\cbar_e$, while $(W/m)d$
dominates both the induced input degrees and the prescribed measure by
\eqref{eq:domination}. The same cut bound therefore yields both expansion
guarantees. Singletons satisfy them directly.

A matching split charges $O(\vartheta M)$ in normalized units. A trim charges
$O(\eta M/\Lambda)=O(\vartheta M)$ by \eqref{eq:small-removal}. These charges
cover the crossing edges, or the backward arcs in the order of
\Cref{lem:order}. Each final crossing edge or backward arc is first separated
at a single split, whose block order is preserved in every descendant. Write
$E_{\rm ch}$ for the set of these edges or arcs. By \eqref{eq:mass-sum},
\[
 \cbar(E_{\rm ch})\le\sum_C C_{\rm ch}(C)=O(\vartheta m\log n).
\]
Each edge of $E_{\rm ch}$ contributes its rounding loss only once, including
any edge rounded to zero. Restoring the input units in \eqref{eq:lattice}
gives
\[
 c(E_{\rm ch})\le(W/m)\cbar(E_{\rm ch})+\vartheta W
                 =O(\phi W\log^2 n).
\]

\proofstep{Work and space.}
The player and its matching requests can run at every level of the recursion.
Endpoint maintenance and flow repair start only when a game completes.
Summing their costs against the corresponding phase weights gives
\begin{center}
\begin{tabular}{@{}lll@{}}
\toprule
Operation & Work on weight $M$ & Total charged weight\\
\midrule
Cut player & $O(M\log^3 n)$ & $O(m\log n)$\\
Undirected matching & $O(M/\phi)$ & $O(m\log n)$\\
Directed matching & $O(M\log\log n/\phi)$ & $O(m\log n)$\\
Endpoint updates and route indexing & $O(M\log^2 n+M/\phi)$ & $O(m)$\\
Trimming and level scans & $O(M\log n/\phi)$ & $O(m)$\\
\bottomrule
\end{tabular}
\end{center}
Together with the initial $O(n+m)$ scan these bounds give
\eqref{eq:main-times} on every execution. In a phase, the route
records and endpoint passes use $O(M/\phi)$ space. The matching height is
$O((1+\log r)/\vartheta)=O(1/\phi)$, and the trimming height is $O((1+\log
M)/\phi)=O(M/\phi)$, so their level lists fit in the same bound. Pending
child graphs are disjoint after a parent releases its state. Thus the whole
recursion uses $O(n+m/\phi)$ space.
\end{proof}

\section{The congestion-approximator hierarchy}\label{sec:ca}\label{sec:ca-overview}\label{ca:sec:nested}

In a tree, the demand determines the flow across every edge.
If deleting an edge separates the graph vertices in $D$ from the
others, the net demand $b(D)$ must cross it.
Give this edge capacity $\delta_G(D)$, the capacity leaving $D$ in
the graph. Every flow in $G$ that routes $b$ must also send net
$b(D)$ across this cut, so it has congestion at least
$|b(D)|/\delta_G(D)$ when $\delta_G(D)>0$.
We want a single tree for which the largest of these ratios estimates
the optimum to within a polylogarithmic factor for \emph{every} demand.

Consider a star with one leaf for each graph vertex $v$. Give its
leaf edge capacity $\deg_G^c(v)$. A graph cut colors
the leaves red and blue. Give the center the color with greater total
weighted degree. The tree cut then costs the smaller of the two degree
sums. On a graph of constant conductance this is within a
constant factor of the graph cut. A sparse cut can separate two sides
of large degree. If each side has its own subtree, the edge joining
them can represent this cut at its true capacity. We must also represent cuts that split the vertices within a
subtree. This leads to a different choice of weight at each
internal node.

\subsection{From congestion to the cost of extending a cut}\label{sec:terminal-tree}

In a terminal tree $T$ each graph vertex is a terminal. The remaining
nodes have zero demand. Root $T$, let $C_e$ be the terminals below an
edge $e$, and write $u_e$ for its capacity. The unique tree flow for
a balanced demand $b$ has congestion
\[
       a_T(b)=\|R_Tb\|_\infty,\qquad
       (R_Tb)_e=b(C_e)/u_e\quad(u_e>0).
\]
If a zero-capacity edge has $b(C_e)\ne0$, we set
$a_T(b)=+\infty$. In particular we may join different component trees
by zero-capacity edges. This definition allows internal terminals,
although the hierarchy we construct has all its terminals at leaves.

Fix a graph cut $X$. Its terminal colors are prescribed, but we may
choose the colors of the other tree nodes. Write $\lambda_T(X)$ for
the least capacity of a tree cut extending these colors. To prove
that $T$ has quality $q$, we will extend each coloring at cost at most
$q\delta_G(X)$. Indeed, a tree flow of congestion $a_T(b)$ must send
net $b(X)$ across every such extension. The bound on its cost gives
$|b(X)|\le q a_T(b)\delta_G(X)$ for every $X$ whenever $a_T(b)$ is
finite. These are precisely the graph cut conditions for routing $b$
at congestion $q a_T(b)$. Conversely, a maximum flow in $T$ from the
terminals of $X$ to the other terminals induces a demand of tree
congestion at most one and net outflow $\lambda_T(X)$ from $X$.
Routing it in $G$ at congestion $q$ requires
$\lambda_T(X)\le q\delta_G(X)$. Thus
\begin{equation}\label{ca:sc:eq:cut-characterization}
 \lambda_T(X)\le q\delta_G(X)\quad\hbox{for every }X
 \quad\Longleftrightarrow\quad
 \opt_G(b)\le q a_T(b)\quad\hbox{for every balanced }b .
\end{equation}

To bound the tree congestion from above, call a denominator
\emph{valid} if $u_e\ge\delta_G(C_e)$. Validity gives
$a_T(b)\le\opt_G(b)$ by the cut bound on each tree edge. It also
gives $\delta_G(X)\le\lambda_T(X)$. For each graph edge crossing
$X$, the path between its terminals in $T$ changes color at least
once, and its capacity is counted in $\delta_G(C_e)$ at that change.
With $u_e=\delta_G(C_e)$ each graph edge pays exactly its capacity
times the number of color changes on its path. Conversely, if
$\delta_G(X)\le\lambda_T(X)$ for every $X$, then each denominator
is valid. For $X=C_e$, cutting $e$ alone gives an extension of cost
$u_e$, so $\delta_G(C_e)\le\lambda_T(C_e)\le u_e$.

\begin{theorem}[Base construction]\label{ca:thm:base}
Let $G$ be an undirected graph with $n$ vertices, $m$ positive-capacity
edges and polynomial capacity spread. Fix $a>0$ and set $N=16(m+2)$.
The construction takes
\[
 O\!\left(n+m\log^7 N\log\log N\right)
\]
work on every execution. With probability at least $1-N^{-a}$ it
returns a terminal tree with $O(n)$ vertices, height
$O(\log N)$, cut denominators equal to the graph capacities of their
descendant cuts, and congestion quality
\[
 q=O\!\left(\log^2 N\log\log N\right).
\]
For $A=R_TB_GC_G$, multiplication by $A,A^\top,|A|,|A|^\top$ takes
$O(n+m)$ work per product after $O(n+m)$ preprocessing. Every recursive
call uses the original $N$. For a simple input, $\log N=O(\log n)$.
Constants may depend on the
chosen failure exponent and polynomial capacity-spread bound.
\end{theorem}

We prove the quality bound here, assuming the local flow and cut
queries succeed. \Cref{sec:ca-computation} implements the queries
and bounds their work and failure probability. A recursive input
with $k$ nonisolated vertices and $p$ edges may use any common
$N\ge\max\{16,k+p+2\}$ in which its capacity spread is polynomial.
Write $s_J=|V(J)|+|E(J)|$.

\paragraph{Boundary weights at an internal node.}
Use a hierarchy of nested vertex sets, with each terminal at its
singleton leaf, as in Henzinger, M\"unk, and R\"acke~\cite{HMR25}.
The edge above a child $D$ of $P$ has capacity $\delta_G(D)$.
We can distribute this capacity among the vertices of $D$ by
assigning each $v\in D$ the weight
\[
                    b_D(v)=c_G(\{v\},V(G)\setminus D).
\]
Thus the weight to use at $P$ is $w_P(v)=b_D(v)$ for $v$ in child $D$.
An edge between children contributes an endpoint in each. An edge
leaving $P$ contributes only its endpoint in $P$. Each endpoint keeps
the color of its own terminal, even when its child contains both colors.

Suppose $D$ and $P$ have different colors. Then each $v\in D$
disagrees with exactly one of these two nodes. Charge its share
$b_D(v)$ of the tree edge capacity to that node. This accounts for
the full cut edge and charges only terminals disagreeing with a
node. It remains to count how much one terminal can pay there.

\begin{observation}[Endpoint accounting]\label{ca:lem:nested-color}
Fix terminal colors, choose any color for each internal node $P$, and let
$M_P\subseteq P$ be the terminals disagreeing with that node. The resulting
tree cut costs at most $2\sum_P w_P(M_P)$.
\end{observation}
\begin{proof}
At $P$, a given $v\in M_P$ can be charged only for the edge above
$P$ and the edge to the child containing $v$. For the former, the
charge is at most $b_P(v)\le w_P(v)$. The latter can charge at
most $w_P(v)$. A leaf agrees with its terminal and receives no charge.
Summing over vertices and nodes pays for every cut tree edge.
\end{proof}

\subsection{Choosing groups inside a pending piece}\label{sec:ca-grouping}

We construct the tree one connected component at a time. Suppose
$S$ is a piece pending under a node $P$. Its vertices still have to
be assigned to children of $P$. Until we split $S$ or attach it, all
of $S$ counts as one piece at $P$, with boundary weight $b_S(v)$
at $v\in S$.

Suppose we were to attach all of $S$ as a child, with groups $D$ as
its first prospective children. For $v\in D$, let
$a(v)=c_G(\{v\},S\setminus D)$ be the capacity from $v$ to other
groups inside $S$. The weight of $v$ at the new node would be
\[
                   b_D(v)=a(v)+b_S(v).
\]
Coloring $S$ by majority under $b_D(v)=a(v)+b_S(v)$ leaves only the
color with smaller total weight to charge at $S$ in \Cref{ca:lem:nested-color}.
Grouping can change $a$, while $b_S$ stays fixed. At a whole
component $b_S=0$, so the entire initial charge is the smaller of the
two colors' intergroup weights. We therefore seek groups
such that
\begin{equation}\label{eq:internal}
 \min\{a(X),a(S\setminus X)\}\le q_{\rm in}\delta_{G[S]}(X)
 \qquad(X\subseteq S),\qquad q_{\rm in}=O(\log N).
\end{equation}
For a proper subset $S$, a cheap cut inside $S$ can have large
exterior weight on \emph{both} sides. Then neither choice of color
for $S$ need be affordable. Once we have grouped $S$ we may have to
leave some vertices under $P$ before attaching a child. Doing so
can itself increase the charge at $P$. If a split separates two
red endpoints of an edge at a blue parent, both shares of its
capacity now count toward disagreement at $P$, although the edge
does not cross the terminal cut. The exterior trim must pay for disagreements at the new child
and for these newly exposed endpoints at the old parent. During
grouping $S$ still counts as one piece at $P$. Each group must have at
most $|S|/2$ vertices when $|S|>1$, since the next recursive
inputs will each lie in a group.

Work throughout in the fixed graph $J=G[S]$. Start with singleton
groups, so each edge between distinct vertices contributes its capacity
to $a$ at each end. If a cheap cut separates two sides of large
intergroup weight, grouping one side can substantially reduce $a(S)$.
To \emph{group} a set $Z$, make it
one part and keep each nonempty remainder $D\setminus Z$ of an old
part. Edges inside $Z$ cease to contribute to $a$, but remain in
$J$ for later routing.

Let $a'$ be the intergroup weights after grouping $Z$. Only boundary
edges contribute inside the new group, so $a'(Z)=\delta_J(Z)$.
Outside $Z$, an endpoint can start to count only on an edge into $Z$.
Keeping the old weights outside and allowing for every such edge
gives
\[
 a'(S)=\delta_J(Z)+a'(S\setminus Z)
      \le\delta_J(Z)+a(S\setminus Z)+\delta_J(Z)
       =a(S)-a(Z)+2\delta_J(Z).
\]

A set carrying a constant fraction of $a(S)$ and satisfying
$\delta_J(Z)\le a(Z)/4$ therefore removes a constant fraction of
the weight. To preserve the halving of recursive inputs, it must
also have at most half the vertices. The decrease is measured in endpoint weight, and the recursion
depth in vertices.

\paragraph{A sparse cut or a core.}
Apply the following test to the current intergroup weight $\mu=a$
in $J$. Its answer selects a set to group. Every vertex and edge
of $J$ remains available for subsequent tests, whichever set is
grouped. After the final groups are known, a separate fair cut will
choose the child to attach to $P$.

\begin{lemma}[A balanced cut or an expanding core]\label{lem:measured-core}
Let $\mu\ge0$ be vertex weights in $J$ of total mass $M$, and fix a
routing scale $\theta>0$ and a failure exponent $b>0$. Use the
round limit $r=\Theta_b(\log N)$ of the player for the common size
bound $N$ and failure allowance $N^{-b}$. Put
$\alpha=\lambda/2$, $\Gamma=r/\theta$ and $\phi_0=\alpha/(4\Gamma)$.
Using the median routine of \Cref{ca:rc:lem:matching} and at most one
$3/2$-fair cut, the test returns one of the following.
\begin{itemize}[leftmargin=1.7em,itemsep=3pt,topsep=3pt]
\item A cut $X$ with weight at least $\tau M/4$ on each side under $\mu$ and
capacity at most $\theta\min\{\mu(X),\mu(V(J)\setminus X)\}$.
\item A core $A$ and its complement $R$ satisfying
\begin{equation}\label{eq:measured-test}
 \mu(R)\le M/8,\qquad
 \delta_J(R)\le \min\{10\tau M/\Gamma,\ \phi_0\mu(A)\}.
\end{equation}
If the game mixes, every $Y\subseteq A$ with $\mu(Y)\le\mu(A)/2$
also satisfies
\begin{equation}\label{eq:measured-light}
 \phi_0\mu(Y)\le\delta_{J[A]}(Y),\qquad
 c_J(Y,R)\le3\delta_{J[A]}(Y).
\end{equation}
\end{itemize}
If at most one vertex has positive weight, take $A=V(J)$ directly.
With the stated flow and cut routines, the conditional probability
of returning a core that fails \eqref{eq:measured-light} is at most
$N^{-b}$ given the history before the game.
\end{lemma}

We prove the lemma in \Cref{sec:ca-local}. For grouping choose a
constant $0<\theta\le1/4$, so $\phi_0=\Theta_b(1/\log N)\le1/4$.
Here both bounds in \eqref{eq:measured-light} refer to the weights
and edges \emph{before} we group anything. The first pays for old
intergroup endpoints of $Y$. The second will pay for endpoints newly
exposed if we put all of $R$ in one group.

\begin{lemma}[Internal grouping]\label{lem:internal}
On an execution with correct local guarantees, grouping takes $O(\log N)$
tests. If $|S|>1$, each final group has at most $|S|/2$ vertices, and
the final interpart measure satisfies \eqref{eq:internal}.
\end{lemma}
\begin{proof}
\proofstep{Grouping a set with few vertices.}

If $a(S)=0$, the desired inequality already holds. Otherwise test
the current weights. On a sparse cut, group the side $Z$ with fewer
vertices, breaking a tie arbitrarily. Both sides had substantial
weight, so $a(Z)\ge\tau a(S)/4$ and
$\delta_J(Z)\le\theta a(Z)\le a(Z)/4$. On a core response with
$|A|<|S|/2$, group $Z=A$. This time $a(Z)\ge7a(S)/8$ and
$\delta_J(Z)\le\phi_0a(Z)\le a(Z)/4$. In either case the endpoint
count above gives
\[
                  a'(S)\le a(S)-\tfrac12a(Z)
                             \le(1-\tau/8)a(S).
\]
Test again with $a'$. Initially $a(S)$ is at most twice the total capacity of $J$.
While $a(S)>0$, some edge still joins distinct groups, so $a(S)$
is at least the least positive edge capacity. The ratio of these bounds is polynomial in $N$, so at most
$O(\log N)$ tests can continue in this way.

\proofstep{When the core contains at least half the vertices.}
Here the core may be too large to use as a group. Its complement
$R$ has the required size. Group $R$ if it is nonempty, keep the
old groups restricted to $A$, and stop. This last step need not
decrease the weight. We show directly that the resulting weights
$a'$ satisfy \eqref{eq:internal}. Each set made into a group has at most $|S|/2$ vertices, and the
remainders of old groups only shrink.

Fix a cut of $S$. The $a'$-weight of either side bounds the smaller
of the two $a'$-weights. Choose the side $X$ for which $Y=X\cap A$
satisfies $a(Y)\le a(A)/2$ under the \emph{old} weights, and put $W=X\cap R$.
This choice lets us use both core inequalities on $Y$. An intergroup
endpoint in $Y$ either was already counted in $a(Y)$ or lies on an
edge into $R$, so $a'(Y)\le a(Y)+c_J(Y,R)$. All of $R$ is one group,
so $a'(W)=c_J(W,A)$. Split the latter into edges to $Y$ and edges
to $A\setminus Y$. The former can count at both endpoints and have
capacity at most $c_J(Y,R)$. Thus
\[
 a'(X)\le a(Y)+2c_J(Y,R)+c_J(W,A\setminus Y)
 \le (\phi_0^{-1}+6)\delta_{J[A]}(Y)+c_J(W,A\setminus Y)
       \le (\phi_0^{-1}+6)\delta_J(X).
\]
The last two capacities count disjoint sets of edges crossing $X$.
Thus $q_{\rm in}=\phi_0^{-1}+6=O(\log N)$ suffices. The same
calculation covers $R=\varnothing$.
\end{proof}

\subsection{Obtaining the core from a matching game}\label{sec:ca-local}\label{sec:ca-interfaces}

In a grouping test the matching game starts mass only at endpoints of
intergroup edges, but its routes may use every vertex and edge of $J$.
We prove \Cref{lem:measured-core} for any nonnegative starting measure
$\mu$ of total mass $M$ and routing scale $\theta>0$, so that the test
also applies in \Cref{sec:near-optimal-decomp}. If at most one vertex
has positive weight, take $A=V(J)$. Every cut then has a side of
weight zero under $\mu$, and $R$ is empty.

\paragraph{Matching the prescribed measure.}
Run the player on the positive-weight vertices with weights $\mu$.
In each round split these weights into two equal halves at the
$\mu$-weighted median of the scores, dividing ties as needed.
Route from the lower half to the upper in $J$ at
capacities $c_e/\theta$. The cut outcome of \Cref{ca:rc:lem:matching}, with shortfall
parameter $\tau$, gives the first alternative of \Cref{lem:measured-core}
by \eqref{short:eq:median-cut}. Otherwise we obtain $F\ge(1-\tau)M/2$ units of pairs between
the halves (\Cref{sec:game-implementation}). Use each routed pair
in both directions. Its two endpoints now send and receive the same
amount. Complete the matching marginal to $\mu$ at each vertex
with a self-pair. Only these self-pairs can miss the median, and
their total weight is $M-2F\le\tau M$. The lazy step sends half the
mass along the matching. This cancels the doubling of routed load,
so the physical walks have congestion at most $1/\theta$ in one
round.

For a fixed $b>0$, give the game failure allowance $N^{-b}$ and let $r=\Theta_b(\log N)$
be its round limit. In a completed game let $z(v)$ be the mass
starting at $v$ that is missing at the end. Every round matched its
full marginal, so the only losses are attenuation, at most $\tau M$
over the game, and the final thinning of a $4\tau$ fraction of the
survivors. Since the retained fraction from a start is at most
$1-4\tau$, we have
\begin{equation}\label{ca:rc:eq:z}
             4\tau\mu\le z\le\mu,\qquad z(V(J))\le5\tau M .
\end{equation}
Set $\alpha=\lambda/2$ and $\Gamma=r/\theta$. If the game mixed, then
for any $X\subseteq V(J)$ with $\mu(X)\le M/2$, at least $\alpha\mu(X)$
mass starts in $X$ and fails to end there (\Cref{lem:physical-cut}).
Its missing part has weight $z(X)$. Every surviving walk in this
count leaves $X$, and the
walks have total congestion at most $\Gamma$. Thus
\begin{equation}\label{ca:rc:eq:defective-cut}
 \alpha\mu(X)\le z(X)+\Gamma\delta_J(X)
             \qquad\bigl(\mu(X)\le M/2\bigr).
\end{equation}
The probability, conditional on the history before this game, that
it completes without this guarantee is at most $N^{-b}$.

\paragraph{Using the missing starts in a fair cut.}
Suppose a trim keeps $A$ and discards $R$. Take the lighter side $X$
of a cut in $A$, measured by $\mu$. Splitting its boundary into edges
to $A\setminus X$ and edges to $R$ gives
\[
 \alpha\mu(X)\le
       \Gamma\delta_{J[A]}(X)+z(X)+\Gamma c_J(X,R).
\]
The last two terms are exactly what an internal expansion bound
must account for. Place a source spoke of capacity $z(v)$ at each
vertex, scale $J$ by $\Gamma$, and let a fair cut choose $R$ on the
source side. Both the source spokes at $X$ and the edges from $R$
now bring flow into $X$ across the fair cut. If the sink spokes can
absorb only a small fraction of $\alpha\mu(X)$, the rest must leave
$X$ through $J[A]$. The same flow, counted on the other side of the
cut, bounds the measure discarded. We give the two conservation
bounds for arbitrary source weights $z$ so that we can use them
again to account for exterior edges.

\begin{lemma}[Fair-cut trimming]\label{lem:fair-core}
Let $\alpha,\Gamma>0$ and let $\mu,z$ be nonnegative measures on $V(J)$.
Form
\[
 K=\Gamma J+\{sv:z(v)\}+\{vt:(\alpha/8)\mu(v)\}.
\]
Let $R$ and $A$ be the vertices of $J$ on the source and sink sides
of a $3/2$-fair cut of $K$, respectively. Then
\begin{align}
 \Gamma c_J(Y,A)+(\alpha/8)\mu(Y)
 &\le2\Gamma\delta_{J[R]}(Y)+2z(Y)
                                      &&(Y\subseteq R),\label{st:sc:eq:source-fair}\\
 z(X)+\Gamma c_J(X,R)
 &\le2\Gamma\delta_{J[A]}(X)+(\alpha/4)\mu(X)
                                      &&(X\subseteq A).\label{ca:rc:eq:trim-local}
\end{align}
If $\alpha\mu(X)\le z(X)+\Gamma\delta_J(X)$ whenever
$\mu(X)\le\mu(V(J))/2$, then every $X\subseteq A$ with $\mu(X)\le\mu(A)/2$
satisfies
\begin{equation}\label{eq:trim-light-side}
 (\alpha/4)\mu(X)\le\Gamma\delta_{J[A]}(X),\qquad
 z(X)+\Gamma c_J(X,R)\le3\Gamma\delta_{J[A]}(X).
\end{equation}
\end{lemma}
\begin{proof}
On the sink side, take $X\subseteq A$. The source spokes at $X$ and
the edges from $R$ cross the fair cut into $X$ and carry at least
two thirds of their capacity in that direction. The sink spokes at
$X$ and the edges to $A\setminus X$ must carry this flow away.
On the source side, take $Y\subseteq R$. The edges to $A$ and the
sink spokes at $Y$ cross the cut out of $Y$. This time the source
spokes and the edges from $R\setminus Y$ supply the flow. The two
conservation bounds are
\[
 \tfrac23\bigl(z(X)+\Gamma c_J(X,R)\bigr)
    \le (\alpha/8)\mu(X)+\Gamma\delta_{J[A]}(X),\qquad
 \tfrac23\bigl(\Gamma c_J(Y,A)+(\alpha/8)\mu(Y)\bigr)
    \le z(Y)+\Gamma\delta_{J[R]}(Y).
\]
Multiplying by $3/2$ and rounding that factor up to two proves
\eqref{ca:rc:eq:trim-local} and \eqref{st:sc:eq:source-fair}.

For $X\subseteq A$ with $\mu(X)\le\mu(A)/2$, we also have
$\mu(X)\le\mu(V(J))/2$, so the assumed cut bound on $J$ applies.
Substituting the sink-side inequality gives
\[
 \alpha\mu(X)\le z(X)+\Gamma c_J(X,R)+\Gamma\delta_{J[A]}(X)
             \le(\alpha/4)\mu(X)+3\Gamma\delta_{J[A]}(X).
\]
Moving the weight term to the left gives
$(\alpha/4)\mu(X)\le\Gamma\delta_{J[A]}(X)$. Substituting this
back into \eqref{ca:rc:eq:trim-local} also bounds the combined
source capacity by $3\Gamma\delta_{J[A]}(X)$.
\end{proof}

\begin{proof}[Conclusion of the proof of \Cref{lem:measured-core}]
Apply \Cref{lem:fair-core} with the missing starts $z$. To bound the
\emph{whole} discarded part, take $Y=R$ in the source-side
conservation inequality. Its internal boundary vanishes, leaving
\[
       \Gamma\delta_J(R)+(\alpha/8)\mu(R)
                     \le2z(R)\le10\tau M .
\]
Thus $\mu(R)\le80\tau M/\alpha\le M/8$ by \eqref{eq:constants}
and $\delta_J(R)\le10\tau M/\Gamma$. Taking $X=A$ on the sink side
gives the other boundary estimate
$\Gamma\delta_J(R)\le(\alpha/4)\mu(A)$. This proves
\eqref{eq:measured-test} on every completed game. When it mixes,
\eqref{ca:rc:eq:defective-cut} supplies the cut bound required by
\Cref{lem:fair-core}, and \eqref{eq:trim-light-side} gives
\eqref{eq:measured-light}. The only remaining failure event is a
completed game that missed its mixing target, as in \Cref{lem:player}.
\end{proof}

\Needspace{12\baselineskip}
\subsection{Attaching a child and refining its parent}\label{sec:ca-exterior}

When a grouping game completes in \Cref{sec:ca-local}, its fair cut
uses the missing mass at each start as its source capacity. The sink
spokes are proportional to the intergroup weights of that test.
We group its core $A$ if $|A|<|S|/2$, and otherwise group its
complement if nonempty. After the last step the groups $D$ partition
all of $S$, and \Cref{lem:internal} gives \eqref{eq:internal} on
$G[S]$ for their final intergroup weights $a$.

If we attach all of $S$ as a child of $P$, the weight at its new node
is $a(v)+b_S(v)$ at each vertex. For an initial connected component,
$b_S=0$ and we can make this attachment directly. In a proper piece
even a cheap cut of $G[S]$ can have large exterior weight on both sides.
We seek a subset $B$ whose internal cuts can pay for exterior as well
as intergroup endpoints. Leaving $U=S\setminus B$ under the old
parent has a cost of its own. Edges inside $S$ that were invisible
at $P$ can now run between different pieces there.

\begin{figure}[htbp]
\centering
\setlength{\unitlength}{1pt}
\begin{picture}(402,118)
\linethickness{0.65pt}
\put(58,110){\makebox(0,0){\small\bfseries After grouping}}
\put(273,110){\makebox(0,0){\small\bfseries After choosing $B$}}
\put(125,67){\makebox(0,0){\color{black!65}$\longrightarrow$}}

\put(58,85){\makebox(0,0){$P$}}
\qbezier[11](58,77)(58,67)(58,57)
\put(58,49){\makebox(0,0){$S=D_1\cup D_2$}}

\put(273,85){\makebox(0,0){$P$}}
\qbezier(267,77)(229,67)(191,57)
\qbezier[11](273,77)(273,67)(273,57)
\qbezier[40](279,77)(317,67)(355,57)
\put(191,49){\makebox(0,0){$B$}}
\put(273,49){\makebox(0,0){$D_1\cap U$}}
\put(355,49){\makebox(0,0){$D_2\cap U$}}
\qbezier[19](188,41)(173,30)(158,19)
\qbezier[19](194,41)(209,30)(224,19)
\put(158,10){\makebox(0,0){$D_1\cap B$}}
\put(224,10){\makebox(0,0){$D_2\cap B$}}
\end{picture}
\caption{The attachment cut on a piece with two groups. When all four
intersections are nonempty, the new child $B$ has two pending pieces
and two pieces stay under $P$. The solid line is a tree edge. The
dotted lines indicate the parent under which each piece is pending.}
\label{fig:refine-parent}
\end{figure}
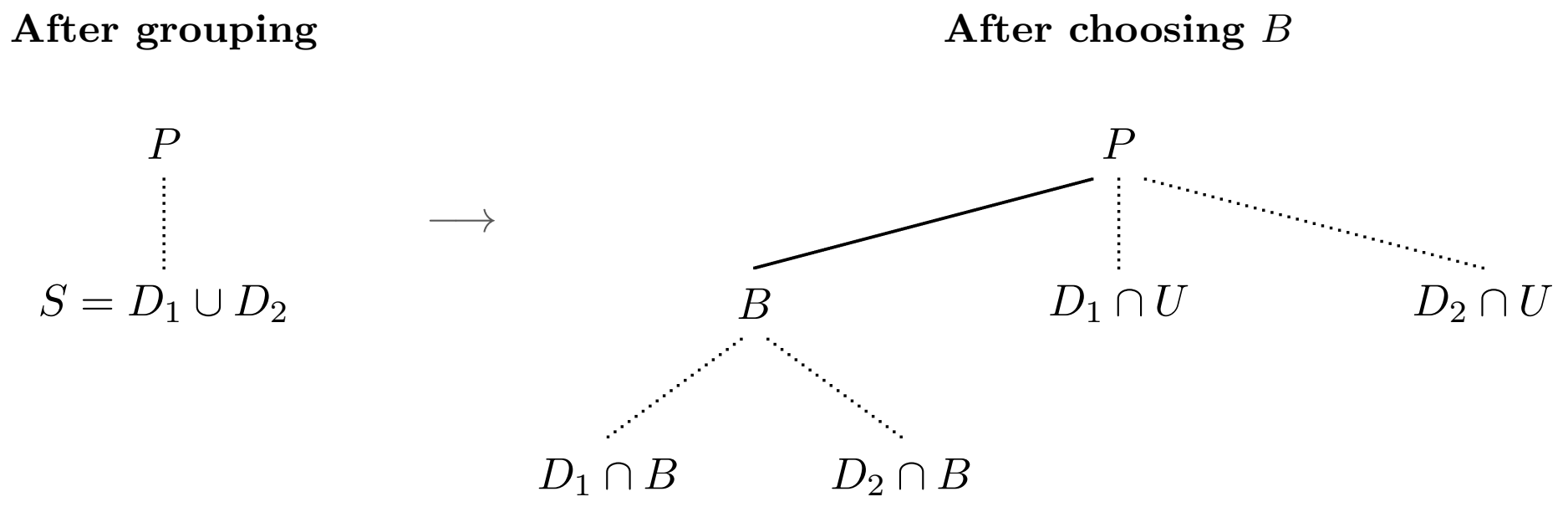

To choose $B$ we build a new network on the entire, unscaled graph $G[S]$.
Use the final intergroup weights on the sink spokes, at scale
$\omega=1/(8q_{\rm in})$ as prescribed by \Cref{lem:fair-core}.
Fix $0<\beta\le1$. When $v$ stays under $P$, we want only
$O(\beta)b_S(v)$ of the newly exposed weight to be charged to its
exterior endpoints. Recovering intergroup weight from a sink
spoke requires dividing by $\omega$, so give the source spoke at
$v$ capacity $\omega\beta b_S(v)$. With both sets of spokes in
place, compute a $3/2$-fair cut of
\begin{equation}\label{eq:boundary-network}
               G[S]+\{sv:\omega\beta b_S(v)\}+\{vt:\omega a(v)\}.
\end{equation}
Take $B$ to be the vertices of $S$ on the sink side and $U$ those on
the source side. The cut has capacity
\[
             \omega\beta b_S(B)+c_{G[S]}(U,B)+\omega a(U).
\]
Attaching a vertex cuts its exterior spoke. Leaving it at $P$ cuts
its intergroup spoke, and separating neighbors cuts their graph edge.
Attach $B$ to $P$ if it is nonempty. A singleton becomes a terminal
leaf. If $|B|>1$, the nonempty pieces $D\cap B$ are pending under $B$.
The nonempty pieces $D\cap U$ stay under $P$ (\Cref{fig:refine-parent}).

The flow witnessing this cut controls what happens to each subset
of $B$ and of $U$. On $B$ the flow enters through exterior spokes
and edges from $U$. Bounding the full exterior weight of an accepted
set from its spoke inflow costs a factor $O(1/(\omega\beta))$.
On $U$ the flow leaves through intergroup spokes and edges to $B$.
Here each unit of exterior weight can supply only $\omega\beta$
units of flow. These are the sink- and source-side bounds of
\Cref{lem:fair-core} with $\Gamma=1$, $\mu=a$, $\alpha=1/q_{\rm in}$
and $z=\omega\beta b_S$. Its hypothesis on $G[S]$ follows from
\eqref{eq:internal}. In particular \eqref{eq:trim-light-side} holds
inside $B$ for every $L\subseteq B$ with $a(L)\le a(B)/2$.

Fix a terminal cut $X$ and write $c_X(S)=\delta_{G[S]}(X\cap S)$.
By \Cref{ca:lem:nested-color} we must bound the weight of boundary
endpoints whose terminal color disagrees with their node. At a new
internal node $B$, choose the majority color under the initial
boundary weights $w_B^0$ of its pieces $D\cap B$. Keep that color
through later refinements, so the set of disagreeing terminals is
fixed. At $P$ the color is already fixed. Only newly exposed
endpoints can increase its charge.

\Needspace{6\baselineskip}
\begin{lemma}[Boundary charges of one refinement]\label{lem:retained}
At an accepted child $B$ the initial minority weighs at most
$O(q_{\rm in}/\beta)c_X(B)$ under $w_B^0$. Let $Y$ be the vertices of
$U$ whose color differs from that of $P$. The boundary weight
disagreeing with $P$ increases by at most
\begin{equation}\label{eq:parent-increment}
        (1+2/\omega)c_X(S)+2\beta b_S(Y).
\end{equation}
\end{lemma}
\begin{proof}
\proofstep{The charge at the new child $B$.}
Let $L$ be the terminal color class of smaller \emph{intergroup}
weight in $B$. The minority under $w_B^0$ weighs at most $w_B^0(L)$,
so it suffices to bound this quantity. Since $a(L)\le a(B)/2$ and
the edges from $L$ to $B\setminus L$ are exactly the edges of $X$
inside $B$, \eqref{eq:trim-light-side} gives
\[
       a(L)\le4q_{\rm in}c_X(B),\qquad
       \omega\beta b_S(L)+c_G(L,U)\le3c_X(B).
\]
Thus the intergroup weight on $L$ can still be paid for inside $B$.
The second bound also accounts for edges that leave $S$ or go to
vertices we left under $P$.

Indeed, for $v\in D\cap B$ an edge contributes to $w_B^0(v)$ precisely
when its other endpoint is outside $D\cap B$. It can leave $S$, go to
$U$, or stay in $B$ and go to another group. Hence
\[
 w_B^0(v)=b_{D\cap B}(v)=b_S(v)+c_G(\{v\},U)+c_G(\{v\},B\setminus D).
\]
The last term is at most $a(v)$. Restoring the exterior weight costs
a factor $1/(\omega\beta)\ge1$, giving
\[
 w_B^0(L)\le a(L)+
   \frac{\omega\beta b_S(L)+c_G(L,U)}{\omega\beta}
 \le \left(4q_{\rm in}+\frac{3}{\omega\beta}\right)c_X(B).
\]

\proofstep{The additional charge at the old parent $P$.}
Before the attachment $S$ counted as one piece at $P$. Every exterior
endpoint of $S$ was therefore already counted there. The increase comes
entirely from edges that the replacement of $S$ by $B$ and the
pieces $D\cap U$ has just separated.

Let $Y$ be the vertices of $U$ that disagree with $P$. Between two
pieces of $U$, each newly exposed disagreeing endpoint is counted
in $a(Y)$. For an edge between $U$ and $B$, inspect its endpoint
in $U$. If that endpoint belongs to $Y$, charge both endpoints to
the edge, at most $2c_G(Y,B)$ in all. Otherwise it agrees with
$P$, so a disagreeing endpoint in $B$ makes this an edge of $X$.
The total increase at $P$ is consequently at most
\[
                         a(Y)+2c_G(Y,B)+c_X(S).
\]
The first two charges can be large even on edges that do not cross
$X$. Apply source-side conservation \eqref{st:sc:eq:source-fair}
to $Y$. The fair flow leaving via its sink spokes and edges to $B$
must be supplied from $U\setminus Y$ or through its own source spokes,
which gives
\[
            \omega a(Y)+c_G(Y,B)
                \le 2\delta_{G[U]}(Y)+2\omega\beta b_S(Y).
\]
Upon division by $\omega$, the left side covers $a(Y)+2c_G(Y,B)$
because $1/\omega\ge2$. All edges from $Y$ to $U\setminus Y$
cross $X$, so $\delta_{G[U]}(Y)\le c_X(S)$. This proves
\eqref{eq:parent-increment}. Each endpoint counted in $b_S(Y)$
already disagreed with $P$ while $S$ was one pending piece. Thus
$2\beta b_S(Y)$ charges some of the increase to disagreement already
present at those vertices.
\end{proof}

Start with a common root and call $\mathsf{Refine}$ on each connected
component with $j=1$. The attachment edges at the root have capacity zero. In
\Cref{alg:hierarchy-refine} the index $j$ counts successive calls on
a branch under the same parent, including the current call. It starts at
one, increases when a piece stays under $P$, and restarts at one below
a new child $B$. \Cref{sec:hierarchy-allowance} uses these counts to
choose the allowances $\beta_j$.

\begin{algorithm}[htbp]
\caption{$\mathsf{Refine}(S,P,j)$ in the congestion-approximator hierarchy}
\label{alg:hierarchy-refine}
\begin{enumerate}[leftmargin=2.4em,itemsep=2pt,topsep=3pt]
\item If $S$ is a singleton, attach its terminal leaf to $P$ and return.
\item Compute the groups $D$ by the procedure of \Cref{sec:ca-grouping}
on the fixed graph $G[S]$. Let $a$ be their final intergroup weights.
\item On an initial component take $B=S$. Otherwise let $B$ be the
vertices of $S$ on the sink side of the fair cut in
\eqref{eq:boundary-network} with $\beta=\beta_j$. Put $U=S\setminus B$.
\item Attach a nonempty $B$ to $P$. A singleton is its terminal leaf.
For a larger $B$, call $\mathsf{Refine}(D\cap B,B,1)$ for each group
$D$ with $D\cap B\ne\varnothing$.
\item Call $\mathsf{Refine}(D\cap U,P,j+1)$ for every group $D$ with
$D\cap U\ne\varnothing$.
\end{enumerate}
\end{algorithm}

Even if the cut accepts all of $S$ or leaves all of it at $P$, every
next input lies in a single group $D$ of size at most $|S|/2$.
The next inputs and any directly attached singleton thus partition
$S$ into at least two smaller sets. The refinement has $O(n)$ calls
and depth $O(\log N)$ in each nontrivial component. The ancestors
of a terminal in the output tree are created only on acceptances
along its refinement branch, so the tree has height $O(\log N)$ as
well.

\subsection{Summing the boundary weights}\label{sec:hierarchy-allowance}

Fix a terminal cut $X$ and the majority color assigned to each node
at its creation. Let $W_X^0=\sum_P w_P^0(M_P)$ and
$W_X=\sum_P w_P(M_P)$ be the total weights disagreeing with their
nodes at creation and after all refinements. By
\Cref{ca:lem:nested-color} it suffices to bound $W_X$.
In \Cref{lem:retained}, a rejection charges a fraction $\beta_j$ of
an exterior endpoint, while an acceptance charges a price
proportional to $1/\beta_j$ for an edge of the terminal cut.
These two charges follow different objects along the same recursion.

\paragraph{A vertex that stays under its parent.}
Fix $v\in M_P$ and follow it until it belongs to a child $D$ of $P$.
Each pending piece $S$ on this part of its branch contains $D$.
An edge leaving $S$ at $v$ therefore leaves $D$ at $v$, so
$b_S(v)\le b_D(v)=w_P(v)$. As long as $v$ is rejected, the parent
stays $P$ and the successive calls have counters $1,2,\ldots$.
Thus the exterior term in \eqref{eq:parent-increment} charges this
vertex at most $2w_P(v)\sum_j\beta_j$. Summing first over the
vertices of each $M_P$, then over parents, gives
\[
 W_X-W_X^0\le (1+2/\omega)\sum_S c_X(S)
                      +2 W_X\sum_{j=1}^{d_{\max}}\beta_j .
\]
Each nontrivial component has at most $2m\le N$ vertices and the
pending pieces halve, so $d_{\max}=1+\lceil\log_2N\rceil$ bounds
the number of calls on a branch. We can absorb the second term if the allowances on
one such branch sum to at most $1/8$.

\paragraph{An edge that enters a new child.}
Fix an edge crossing $X$. Follow its endpoints together while
they belong to the same pending piece. If a child $B$ accepts both,
the edge contributes to $c_X(B)$ and incurs a charge
$O(q_{\rm in}/\beta_j)$ by \Cref{lem:retained}. The counter $j$
counts the calls since the parent of $B$ was created, up to and
including the call that creates $B$. If both endpoints go on to
the same piece under $B$, the next counter starts at one.
If they enter different pieces, this is the last accepted child
that can charge the edge. Consequently the counters at the charged
acceptances count disjoint stretches of one branch, and their sum
is at most $d_{\max}$. The initial component, which we accept directly,
costs at most $q_{\rm in}$ per unit of crossing capacity.

\paragraph{Choosing the allowance.}
A price proportional to $j$ can now be summed over the accepted
children of an edge. This suggests taking $\beta_j$ proportional to
$1/j$. The rejections at a fixed parent determine the normalization.
Put $H=\sum_{i=1}^{d_{\max}}1/i=O(\log\log N)$ and set
$\beta_j=1/(8Hj)$. With $j(S)$ the counter on the call for $S$,
the two bounds are
\begin{equation}\label{eq:allowance-sum}
 \sum_{S\text{ rejecting }v\text{ under }P}\beta_{j(S)}
       \le \tfrac18,\qquad
 \sum_{S\text{ charging }e}\frac{1}{\beta_{j(S)}}
       =8H\sum_{S\text{ charging }e}j(S)
                    \le 8H d_{\max}.
\end{equation}
The second sum ranges over the fair cuts whose accepted child
contains both ends of $e$. Applying the price bound to
each crossing edge and summing with its capacity gives
$W_X^0=O(q_{\rm in}H d_{\max})\delta_G(X)$. The common root has zero
boundary weight.

At any fixed recursion depth the pending pieces $S$ are disjoint.
Hence a crossing edge contributes to $c_X(S)$ at most once at each
depth. Using the rejection bound in the old-parent charges gives
\begin{equation}\label{eq:global-account}
 W_X\le W_X^0+(1+2/\omega)\sum_Sc_X(S)+\tfrac14W_X,
 \qquad \sum_S c_X(S)\le d_{\max}\delta_G(X).
\end{equation}
Since $1/\omega=8q_{\rm in}$ and $q_{\rm in}=O(\log N)$, the
extended tree cut costs at most
$2W_X=O(\log^2 N\log\log N)\delta_G(X)$. This proves the quality
bound in \Cref{ca:thm:base} by \eqref{ca:sc:eq:cut-characterization}.
\Cref{sec:tree-queries} computes the descendant-cut capacities and
prepares the tree products in linear work.

\section{Constructing the congestion approximator}\label{sec:ca-computation}

The hierarchy of \Cref{sec:ca} needs flows and fair cuts on each
induced graph $J=G[S]$ before it can decide which children to create.
The first such $J$ can be the very graph for which we are trying
to construct a tree. Throughout a call to $\mathsf{Refine}$ the edges
of $J$ stay fixed. Each request scales their capacities by a common
factor and specifies its own spokes at $s$ and $t$.

For a median request, the scores split $\mu$ into measures $\mu_{\rm L},\mu_{\rm R}$ of
mass $A$ each. In the query graph $K$, join $s$ to the left weights
and the right weights to $t$, and give the edges of $J$ capacities
$c_e/\theta$. A flow of value close to $A$ pairs almost all the
weight across the median. Suppose instead that a cut with original
vertices $X$ on its source side has capacity $A-\Delta$. Its
crossing spokes cost $A-\mu_{\rm L}(X)+\mu_{\rm R}(X)$, so
\begin{equation}\label{short:eq:median-cut}
       \Delta+\theta^{-1}\delta_J(X)
              =\mu_{\rm L}(X)-\mu_{\rm R}(X)
              =\mu_{\rm R}(S\setminus X)-\mu_{\rm L}(S\setminus X).
\end{equation}
The left side is therefore at most the weight of \emph{either}
side of the graph cut. Ask for throughput $(1-\xi/2)A$ when the
player can leave a fraction $\xi$ unmatched. Any cut violating the
request has $\Delta>\xi A/2$, so each side has weight greater than
$\xi A/2$ and the boundary costs at most $\theta$ times the lighter
weight. On a flow answer, the other half of the shortfall lets us
finish a tentative flow and scale it back within capacity.

We test flows on the full query graph $K$ against a tree of cut
rows. The test either exhibits a violated graph cut or leaves a
demand small on every row. On a median request this residual must
be routed on the edges of $K$ before we can give the player its
matched pairs.

To prepare $J$, contract a forest, build a tree recursively on a
sparsifier of the quotient $H$, and reattach the forest to obtain
a tree for $J$. Each request puts its own spokes on the scaled
tree. Absorbing the two hubs back into a tree gives rows for the
query, including either fair-cut trim. For a median residual,
undoing these absorptions recovers flows on the actual spokes.
They leave a demand balanced on every connected component of $J$.
Sum this demand on each forest component, route the sums on the
\emph{full} quotient $H$, and lift through the forest once to
balance the individual vertices. The sparsifier makes the
recursive construction small. The quality of the reattached tree
determines the accuracy required of the local solves. The quality
of the hierarchy built from their cuts and matchings follows from
\Cref{sec:ca}.

\subsection{Testing a demand with cut rows}\label{sec:tree-queries}

Orient the edges of $G$, put $B=B_G$ and $C=\diag(c)$, and write
$f=Cx$, $\|x\|_\infty\le1$, for a flow in the capacity box. For a
target demand $d$, the row $\one_U^\top/u_U$ measures how much net
outflow $U$ is still owed, in units of $u_U$. Its value on the
residual is positive if $U$ must send more and negative if it must
receive more. The largest absolute discrepancy is
\[
       \|R(d-BCx)\|_\infty
       =\max_{\|y\|_1\le1} y^\top R(d-BCx).
\]
Thus we can minimize the worst discrepancy by playing against a
distribution on the signed rows and the zero row. A fixed choice of
row weights $y$ gives vertex prices $\psi=R^\top y$. Sending one
unit from $u$ to $v$ pays $\psi(u)-\psi(v)$ toward the demand's
value $\psi^\top d$. Sending each edge's full capacity from higher to
lower price maximizes the total payment. The deficit still owed is
\begin{equation}\label{eq:price-deficit}
 D_d(\psi)=\min_{|f|\le c}\psi^\top(d-Bf)
          =\psi^\top d-\sum_{uv}c_{uv}|\psi(u)-\psi(v)|.
\end{equation}
At indicator prices $\psi=\one_U$ this is the cut deficit
$d(U)-\delta_G(U)$. At general prices the same saturating flow
maximizes the outflow of every upper level set simultaneously.
Consequently a positive price deficit reveals a violated cut even
if its side is absent from the rows of $R$.

\begin{observation}[Level-set cut identity]\label{obs:coarea}
Let $G$ be an undirected capacitated graph, $b$ a balanced demand, and
$\psi$ a vector of vertex prices. For $S_z=\{v:\psi(v)\ge z\}$,
\[
 b^\top\psi=\int_{\mathbb R}b(S_z)\,dz,\qquad
 \sum_{uv\in E(G)}c_{uv}|\psi(u)-\psi(v)|
       =\int_{\mathbb R}\delta_G(S_z)\,dz.
\]
If the first integral exceeds the second, we can find a threshold $z$
with $b(S_z)>\delta_G(S_z)$ in $O(|V(G)|+|E(G)|)$ work.
\end{observation}
\begin{proof}
For weights $w$ of sum zero it suffices to integrate between the
smallest and largest vertex prices. A vertex contributes $w(v)$
over a length $\psi(v)-\min\psi$, so $\int w(S_z)\,dz=w^\top\psi$.
Apply this first to $w=b$. For the second identity, let $f^\psi$
saturate each edge from high to low price, with zero flow on ties.
Every edge leaving $S_z$ is saturated outward, hence
$(Bf^\psi)(S_z)=\delta_G(S_z)$. Integrating with $w=Bf^\psi$ gives
the claimed sum over edges.

For the algorithm, compute $w=b-Bf^\psi$ once. Now $w(S_z)$ is the
deficit of the cut at every level, and $w^\top\psi>0$. We can find a
positive level without sorting the prices. At a median price $p$,
the integrals of $w(S_z)$ above and below $p$ are, respectively,
\[
       \sum_{\psi(v)>p}w(v)(\psi(v)-p),\qquad
       \sum_{\psi(v)<p}w(v)(\psi(v)-p).
\]
For the second sum, count the vertices outside $S_z$ and use
$w(V)=0$. Keep a half with positive integral. Clip the prices in the
discarded half to $p$ and combine all entries at $p$ into a single
entry of their total weight. This preserves every level sum in the
interior of the retained half and leaves at most
$\lfloor k/2\rfloor+1$ of the $k$ entries. With two distinct prices
left, any threshold strictly between them has positive deficit.
Median selection and the scans take linear work in total. A final
scan of the original prices returns the vertices of the cut.
\end{proof}

\begin{lemma}[Box feasibility or a cut]\label{ca:lem:box}
Let $G$ be an undirected capacitated graph with $m$ edges, $d$ a balanced
target, and $R$ cut rows of the form $\one_S^\top/u_S$ with $u_S>0$.
Put $A=RBC$. Suppose every absolute row sum of $A$ is at most one,
there are polynomially many rows in the outer size $N$, and products by
$A,A^\top,|A|,|A|^\top,R,R^\top$ take $O(m)$ work after setting aside
isolated vertices. Here $|A|$ is the entrywise absolute value. For
$0<e<1$, a deterministic algorithm takes $O(m\log N/e)$ work on the
active graph and returns either a flow $f$ with $|f_e|\le c_e$ and
$\|R(d-Bf)\|_\infty\le e$, or a cut $S$ with
\begin{equation}\label{ca:eq:box-cut}
                          d(S)>\delta_G(S).
\end{equation}
\end{lemma}
\begin{proof}
An initial scan deals with isolates. An isolate with positive demand
gives a violated singleton cut, and one with negative demand gives
the complementary cut. On the active graph, apply the box--simplex
solver of Jambulapati and Tian~\cite[Theorem~1]{JT23} to the payoff
above, using the distribution on signed rows. The absolute row-sum
bound is the solver's normalization. In $O(\log N/e)$ iterations of
the stated products it returns $x,y$ with gap at most $e$,
\[
 \max_{\|y'\|_1\le1}(y')^\top R(d-BCx)
      \ \le\ \min_{\|x'\|_\infty\le1}y^\top R(d-BCx')+e .
\]
The left side is the largest row error of $f=Cx$. The right side
before adding $e$ is $D_d(R^\top y)$, the deficit that remains
even when every edge pays as much as possible at these prices.
If it is positive, \Cref{obs:coarea} finds a cut satisfying
\eqref{ca:eq:box-cut} in $O(m)$ work. Otherwise
the displayed gap bounds the row error of $f$ by $e$.
\end{proof}

\paragraph{Applying the rows of a tree.}
Let $e$ be a tree edge with capacity $u_e>0$ and descendant terminals
$C_e$. The entry of $A=R_TB_GC_G$ for an oriented graph edge $uv$ is
\[
        A_{e,uv}=\frac{c_{uv}}{u_e}
                      \bigl(\one_{C_e}(u)-\one_{C_e}(v)\bigr).
\]
The column therefore follows the tree path from $u$ to $v$. Its sign
records which endpoint is below $e$. In particular the absolute
row sum is $\delta_G(C_e)/u_e$, at most one for every valid denominator.

Subtree sums aggregate these paths in linear work. For $Ax$, put
$t=c_{uv}x_{uv}$ at $u$ and $-t$ at $v$, accumulate over all graph
edges, and take subtree sums. Dividing by $u_e$ gives the signed sum
at $e$. For $|A|x$, put $t$ at both ends and $-2t$ at their lowest
common ancestor $z$. A subtree containing just one endpoint receives
$t$. A subtree containing both also contains $z$, so the three
contributions cancel. With $x_{uv}=1$ and the division omitted this
same computation gives all the capacities $\delta_G(C_e)$ needed for
the hierarchy in \Cref{sec:ca}.

For the transposes, let $p(v)$ be the sum of $y_e/u_e$ on the root
path to $v$, over edges of positive capacity. The root paths to $u$
and $v$ coincide through $z$, giving
\[
 (A^\top y)_{uv}=c_{uv}\bigl(p(u)-p(v)\bigr),\qquad
 (|A|^\top y)_{uv}=c_{uv}\bigl(p(u)+p(v)-2p(z)\bigr).
\]
The same subtree and root-path sums compute $R_Tb$ and $R_T^\top y$.
Linear-time preprocessing of the tree gives constant-time
lowest-common-ancestor queries~\cite{BFC00}. Recording one ancestor
per graph edge makes the preparation and each product
$O(|T|+|V(G)|+|E(G)|)$ work, also for internal terminals.
A zero-capacity tree edge requires $b(C_e)=0$. If the balance fails,
validity gives a zero-capacity graph cut, oriented by the sign of
$b(C_e)$.

\subsection{Routing a residual exactly with a supplied tree}

Fix a nonzero componentwise balanced demand $b$ and let
$a=\|Rb\|_\infty$. With a quality-$Q$ approximator, $b/(Qa)$ is routable within capacity, so the box test must return
a tentative flow. If each tentative flow halves the remaining row
norm, successive corrections cost at most $Qa,Qa/2,\ldots$ in
congestion. We finish the last residual on a maximum-capacity
spanning forest. This can lose a factor of at most the number of
graph edges, so logarithmically many halvings suffice at every
capacity spread.

We will use this exact solve on the full quotient $H$ at its own
tree quality $q_H$. The spanning forest that finishes its residual
lies in $H$. Once the quotient demand is completely routed, the
forest contracted from $J$ will carry the remaining imbalance
\emph{within} its components.

\begin{lemma}[Exact routing]\label{ca:lem:regression}
Let $R$ be a quality-$Q$ cut-row approximator for $G$, with valid
denominators and linear-time products for \Cref{ca:lem:box}.
Every componentwise balanced demand $b$ can be routed exactly at congestion
at most $3Q\|Rb\|_\infty$ in $O(n+mQ\log^2 N)$ work, where
$m=|E(G)|$ and $n=|V(G)|$.
\end{lemma}
\begin{proof}
Set aside isolates and return zero for $b=0$. Otherwise
$a=\|Rb\|_\infty>0$. At step $j$, the remaining demand $b_j$ is
componentwise balanced and has row norm at most $a_j=2^{-j}a$.
Tree quality makes $b_j/(Qa_j)$ routable within capacity, so the box
oracle must return an edge vector. Use tolerance $1/(2Q)$ and
multiply the result by $Qa_j$. The added congestion is at most
$Qa_j$, and the residual row norm is at most
$Qa_j/(2Q)=a_{j+1}$.

To determine how many halvings suffice, take a maximum-capacity
spanning forest of $G$ and root each component. For an edge $e$
of a rooted forest, write $S_e$ for the vertex set below $e$.
Every graph edge crossing the fundamental cut of $e$ has capacity
at most $c_e$ by the exchange property. Thus $\delta_G(S_e)\le m c_e$.
The unique forest flow for $b_j$ carries $|b_j(S_e)|$ on $e$,
and
\[
       |b_j(S_e)|\le\opt_G(b_j)\delta_G(S_e)
                   \le Q a_j m c_e .
\]
After $t=\lceil\log_2\max\{2,m\}\rceil$ halvings the forest
therefore finishes at congestion at most $Qa$. The total is at most
$Qa(\sum_{j<t}2^{-j}+m2^{-t})\le3Qa$.
Each halving takes $O(mQ\log N)$ work. Building the forest and
routing the last residual fit the claimed bound, independently of the
least positive capacity.
\end{proof}

\subsection{Preparing a router on the fixed graph}\label{sec:ca-preparation}

Fix the graph $J$ of one call to $\mathsf{Refine}$. To make the
recursive input smaller, contract a forest $F\subseteq J$. Let
$H=J/F$ be the quotient obtained by discarding loops and summing
the capacities of parallel edges. Write $P$ for the map that sums
a demand within each forest component.
The quotient has to carry $Pb$ between the components. Even when
$Pb=0$, there can still be substantial demand to route from one
vertex to another in the same component. The choice of $F$ has to
control this hidden demand as well as the size of $H$.

Take $F$ to be rooted. Lift a flow for $Pb$ from the quotient to $J$,
obtaining an edge vector $y$ of congestion $\gamma$.
The remaining demand $b-B_Jy$ sums to zero in each forest component,
so it has a unique routing on $F$. The amount this routing sends
across a forest edge $e$ is
\begin{equation}\label{eq:forest-lift-load}
  |b(S_e)-(B_Jy)(S_e)|
       \le |b(S_e)|+\gamma\delta_J(S_e).
\end{equation}
The quotient demand $Pb$ records only the total demand of each
forest component. That total can be zero while the subset $S_e$
still has net demand $b(S_e)$ to balance within the component.
The second term accounts for where the lifted flow enters and
leaves the component. Suppose we can choose $F$ so that
$\delta_J(S_e)\le\kappa c_e$. This bounds the second term by
$\kappa\gamma c_e$ and makes $\kappa c_e$ a valid capacity for a
tree edge above $S_e$. Its cut row then measures the first term in
the same units. We can achieve this while leaving only
$O(|E(J)|/\kappa)$ nonisolated quotient vertices.

\begin{lemma}[A forest for contraction]\label{simp:lem:forest}\label{ca:lem:forest}
For every $\kappa\ge2$, one can find in $O(n+m\log N)$ work a rooted
forest $F\subseteq G$ on all $n$ vertices such that $G/F$ has
$O(m/\kappa)$ nonisolated vertices and every edge $e$ of $F$ satisfies
\[
                           \delta_G(S_e)\le\kappa c_e.
\]
Every forest path below a root has at most $\kappa$ edges.
\end{lemma}
\begin{proof}
Fix a nontrivial connected component and find a maximum-capacity
spanning tree $T$ by Kruskal's algorithm, with a fixed order on
equal-capacity edges. In each fundamental cut of $T$, the tree edge has
capacity at least that of every crossing edge, by the exchange property.
We also need to bound the number of edges crossing the cuts we keep.

Root $T$ temporarily. In a bottom-up pass, add each vertex's
unweighted graph degree to the sum passed up by its children.
Whenever this sum exceeds $\kappa$, mark the vertex and pass zero
to its parent. Otherwise pass up the sum. In each component of $T$
left after the marks are removed, the top vertex's sum counts the
degrees of the whole component and is at most $\kappa$. If the
tree has no mark, mark its root.

We want each mark to be the root of a separate tree. Merely cutting
$T$ between marks gives small unmarked pieces, but can change the
descendant cut of a retained edge and lose its capacity bound.
Instead identify all marks in the graph component and run Kruskal
again with the same ordering. An edge rejected on the first pass
already had its endpoints connected by earlier edges. Identification
preserves that connection, so the second tree uses only edges of
$T$. It is a maximum-capacity tree in the identified graph.
Undoing the identification separates it into trees with one marked
root each. A path between two marks would have formed a cycle in
the identified tree.

Now fix a selected edge $e$. Its descendant side $S_e$ is connected
in $T$ and avoids the marks. Its total graph degree is at most
$\kappa$, so at most $\kappa$ edges leave it. The same set is a
fundamental-cut side in the identified graph, with exactly the same
crossing edges as before identification. The maximum-tree property
bounds each of their capacities by $c_e$. Thus
$\delta_G(S_e)\le\kappa c_e$. The nonroot vertices on a downward
path lie in one unmarked component and each has positive graph
degree, so the path has at most $\kappa$ edges as well.

With a single mark the whole component contracts to an isolate.
In any component with two or more marks, each mark accounts for
more than $\kappa$ units of degree that were then removed from the
bottom-up sum. These charges are disjoint and the total degree is
$2m$, giving $O(m/\kappa)$ nonisolated quotient vertices. Leave
input isolates as singleton trees. One sort, the two Kruskal passes,
and the tree traversal take $O(n+m\log N)$ work.
\end{proof}

\begin{lemma}[Forest attachment and flow lifting]\label{simp:lem:attach}\label{ca:lem:forest-lift}
Let $F$ be as in \Cref{simp:lem:forest}, $H=G/F$, and let $P$ sum
demands within forest components. For a componentwise balanced $b$, put
\[
       a_F(b)=\max_{e\in F}\frac{|b(S_e)|}{\kappa c_e},
\]
with value zero when $F$ has no edges. A quotient flow of divergence
$Pb$ and congestion $\gamma$ lifts in $O(n+m)$ additional work to a
flow of divergence $b$ and congestion at most $\kappa(a_F(b)+\gamma)$.

Given a quality-$q_H$ terminal tree $T_H$ on $H$ with valid
denominators, identify its terminals with the forest roots and give
each forest edge $e$ capacity $\kappa c_e$. The attached tree has
valid denominators and quality at most $\kappa(1+q_H)$.
\end{lemma}
\begin{proof}
Write $f_H$ for the quotient flow. Split the flow on each quotient edge
among its original edges in proportion to their capacities. The lift $y$ has congestion at
most $\gamma$. Summing net outflows over each forest component
cancels its internal edges and gives the net outflow of $f_H$ at
the corresponding quotient vertex. Hence
\[
             P(b-B_Gy)=Pb-B_Hf_H=0 .
\]
The residual is balanced on each forest component, so we route it
by subtree sums.
By \eqref{eq:forest-lift-load}, the load on $e$ is at most
$\kappa c_e(a_F(b)+\gamma)$. The lifted edges run between forest
components, so the two flows use disjoint edges. Their larger
congestion gives the first claim.

The rows of the attached tree measure exactly the two quantities we
can now route,
\[
           a_T(b)=\max\{a_{T_H}(Pb),\,a_F(b)\}.
\]
A quotient row is valid because its cut is a union of entire forest
components and has the same capacity before and after contraction.
Validity of a forest row is the cut bound in \Cref{simp:lem:forest}.
The quotient demand can be routed at congestion at most
$q_H a_{T_H}(Pb)$. Applying the first claim gives
$\opt_G(b)\le\kappa(1+q_H)a_T(b)$.
\end{proof}

\paragraph{Constructing the tree on the quotient.}
The quotient has $k=O(\min\{|V(J)|,|E(J)|/\kappa\})$ nonisolated
vertices but may still contain nearly every edge of $J$. We keep
its full edge set to route $Pb$. For the recursive construction,
however, the tree only needs to estimate cuts of $H$. A cut
sparsifier with $O(k)$ edges makes that construction small even
when $H$ is dense.

Merge parallel edges, let $p_H=|E(H)|$, and set the isolated
quotient vertices aside. Each represents a whole component of $J$.
We restore their terminals with zero-capacity component links,
including the original isolates of $J$, which may receive spokes
later. If $p_H=0$, there is no quotient flow to route and we use
$q_H=1$. If $0<k<C_b\log N$, a maximum-capacity spanning forest of
$H$ suffices for $T_H$. Give each forest edge $p_H$ times its old
capacity. Its fundamental cut in $H$ has at most $p_H$ edges,
each of capacity at most the original forest edge capacity, which
gives validity. For any cut, the crossing
forest edges belong to that same cut in $H$, giving the reverse
bound with factor $p_H$. Since $p_H\le k^2$, this tree has quality
$O_b(\log^2 N)$ and height $O_b(\log N)$. Here $b$ specifies the
desired failure allowance $N^{-b}$ in the sampling case.

For $k\ge C_b\log N$, use the spectral sparsification algorithm
of~\cite{VaughnSignDescent} at constant accuracy. With failure probability at
most $N^{-b}$ it returns a reweighted subgraph with $O(k)$ edges
and relative spectral error at most $1/2$. In particular every cut
weighs between one half and three halves of its capacity in $H$.
The work on every execution is
\[
             O_b\!\left(p_H\log^2 N+
                      k\log^3 N(\log\log N)^4\right).
\]
We justify the dependence on the common outer size $N$ in
\Cref{sec:sparsifier-confidence}.

We need the tree built on the sample to have valid denominators on
$H$. We also need the next base-construction input to have
polynomial capacity spread. For a retained edge of capacity $c'_e$
in the sample and $c_e$ in $H$, give it capacity $2c'_e+c_e$ in
the recursive input $\widehat H$. Doubling suffices for domination.
Adding back each retained edge's old capacity keeps the minimum
positive capacity from decreasing and adds at most one copy of
each original cut. Hence
\[
       \delta_H(X)\le\delta_{\widehat H}(X)\le4\delta_H(X)
                              \quad\text{for every }X\subseteq V(H).
\]
Summing the upper bound over singleton cuts also bounds the total
capacity of $\widehat H$ by four times that of $H$. A tree of
base quality $q$ on $\widehat H$ now satisfies
\[
  \delta_H(X)\le\lambda_{T_H}(X)
               \le q\delta_{\widehat H}(X)\le4q\delta_H(X).
\]
Consequently the tree has valid rows and quality $q_H\le4q$ on the
full quotient, where it will be used for every subsequent flow.

\paragraph{Routing with the prepared tree.}
Keep the forest, the full quotient $H$, its tree $T_H$, and the
attached tree $T_J$. For a demand $b$ on $J$, route $Pb$ on the
edges of $H$ by \Cref{ca:lem:regression} with the rows of $T_H$,
using zero flow if $Pb=0$. This exact solve costs congestion at
most $3q_Ha_{T_H}(Pb)$. Lift it to the original edges and send the
remaining, componentwise balanced demand along $F$. By
\Cref{ca:lem:forest-lift}, the resulting router has congestion at
most $\gamma_Ja_{T_J}(b)$ and work $D_J$, where
\begin{equation}\label{eq:quotient-router-work}
   \gamma_J=\kappa(1+3q_H),\qquad
   D_J=O\!\left(s_J+|E(H)|q_H\log^2 N\right).
\end{equation}
All residual halvings use the quotient rows of quality $q_H$.
The factor $\kappa$ enters when the finished quotient flow is
lifted, a single linear-time operation.

There is at most one recursive call, on $\widehat H$ with $O(k)$
edges and vertices. Apart from that call, choosing the forest,
preparing the quotient and the two trees' row products, and
sampling take
\begin{equation}\label{eq:preparation-local-work}
    O_b\!\left(|V(J)|+|E(J)|\log^2 N+
                     k\log^3 N(\log\log N)^4\right)
\end{equation}
work. On a successful return $T_J$ has quality at most
$\kappa(1+q_H)$ and height $O(\kappa+\log N)$, the sum of the
forest and quotient heights.

\subsection{Choosing spoke flows for the current request}\label{sec:median-router}

Let $T=T_J$ be the prepared tree. A request scales $J$ by $a>0$
and adds two sets of spokes, giving $K=aJ+E_s+E_t$. A demand $b$
balanced on $K$ may have nonzero sum on a component of $J$ that now exchanges flow with the hubs. Before we can use the
prepared router, we must decide which spokes supply this exchange.

Fix a spoke flow $g$ supplying the prescribed net outflows at $s$
and $t$. It leaves demand $d_g=(b-B_{\rm sp}g)|_{V(J)}$ on the
old vertices, where $B_{\rm sp}$ is the spoke incidence matrix.
We want both the spoke congestion and the tree norm of this demand
to be small. Put the very same spokes on $aT$, obtaining
$L=aT+E_s+E_t$, and extend $b$ by zero at the nonterminals.
Once $g$ is specified, the unique old-tree flow for $d_g$ has
congestion $a_T(d_g/a)$. Thus
\[
       \opt_L(b)=\min_{g\text{ supplying }s,t}
              \max\{\congestion_{\rm sp}(g),\ a_T(d_g/a)\}.
\]
A nonzero imbalance across a zero-capacity tree edge has infinite
cost. A cheap flow on $L$ therefore gives us actual spoke flows to
copy to $K$ and a demand the prepared router can carry on $aJ$.

When a hub has two spokes to the same tree, they form a cycle
with the path between their endpoints. Several allocations of spoke
flow can then give the same divergence. To choose one in linear
work, absorb the hubs, one at a time, to obtain a tree $T_K$.
Subtree sums now fix the flow on every edge.
Each absorption records enough information to send that flow back
through the original spokes at constant congestion loss. The cut
bounds for $T_K$ give the rows for testing $K$, and the reverse
passes give the spoke flows for correcting a median residual.

\begin{lemma}[A forest and one hub]\label{ca:lem:hub-tree}
Let $L$ consist of a forest $T$ and a hub with spoke capacity
$w_v\ge0$ at each forest vertex $v$. In linear work one can build a
tree $T^+$ on the same vertices with
\[
             \delta_L(X)\le\delta_{T^+}(X)\le9\delta_L(X)
                        \qquad(X\subseteq V(L)).
\]
Given any edge vector of congestion $\sigma$ on $T^+$, one can
compute in linear work an edge vector on $L$ with the same
divergence and congestion at most $9\sigma$.
\end{lemma}
\begin{proof}
Root each tree in the forest. At a vertex $v$ other than a root
consider the triangle formed by $v$, its parent, and the hub.
A missing spoke has capacity zero. Remove either the spoke at $v$
or its parent edge, and add its capacity to both remaining sides
of the triangle. Every cut that crossed the removed edge also
crosses its replacement path, so no cut loses capacity.

\proofstep{Combining the spokes.}
Work from leaves to root. At a nonroot $v$, the parent edge has its
original capacity $c_v$, while the current spoke capacity $s_v$
includes contributions already passed up by children. We may remove a
small spoke and move all of it up the parent edge. When the spoke is
large we retain it, passing up only $c_v$. Let $p_v$ be the capacity
added to the parent's spoke. The cutoff at $2c_v$ ensures that
$p_v\le2c_v$. When $v$ keeps its spoke, $p_v=c_v<s_v/2$, so more
than half of $s_v$ stays at $v$ to help redistribute its spoke
flow in the reverse pass. The resulting capacities are
\[
\begin{array}{c|ccc}
 & \text{retained parent edge} & \text{retained spoke}
 & \text{passed to parent }(p_v)\\ \hline
 s_v\le2c_v & c_v+s_v & 0 & s_v\\
 s_v>2c_v   & 0 & s_v+c_v & c_v .
\end{array}
\]
At a root retain the spoke and put $p_v=0$. Each vertex now has
one edge leading upward or directly to the hub. These edges form a
tree, including any zero-capacity links. Record how the original
spoke capacities were distributed. If $r_v$ is the amount retained
locally from $s_v$, then
\[
              w_v+\sum_{u\text{ child of }v}p_u=s_v=r_v+p_v .
\]

The spoke kept at $v$ has capacity at most $3r_v$, since
$s_v+c_v\le3(s_v-c_v)$ in the second case. Also $p_v\le2c_v$
and the kept parent edge has capacity at most $3c_v$ for every
nonroot $v$.

\proofstep{Recovering flows on the original spokes.}
Take an edge vector of congestion $\sigma$ on the retained tree.
Copy the flow on each retained parent edge directly, using at most
$3\sigma c_v$ of the original edge. Write $g_v$ for the signed
flow from $v$ to the hub on its retained spoke, or zero if there is
no such spoke. This may exceed what the original spoke at $v$ can
carry. Its capacity came partly from the children, and the stored
$p_u$ tell us how to send flow back down to those original spokes.

Make one pass from roots to leaves. Let $z_v$ be the signed flow
passed down from the parent, zero at a root. At $v$ we have to send
$z_v+g_v$ on to the hub. Send it to the original spoke at $v$
and down the child edges in proportion to the amounts they supplied,
\[
      t_v=\frac{z_v+g_v}{s_v},\qquad
      f(v,\text{hub})=w_vt_v,\qquad
      z_u=p_ut_v\quad(u\text{ a child of }v).
\]
Use zero for the ratio when $s_v=0$. The copied parent-edge flows
keep their original divergence. The remaining signed net outflow
at $v$ is
\[
     w_vt_v+\sum_{u\text{ child of }v}z_u-z_v=s_vt_v-z_v=g_v .
\]
This recovers the prescribed divergence at every forest vertex,
and hence at the hub.

The same proportions bound the load at every depth. Inductively
$|z_v|\le3\sigma p_v$, as holds at a root. The retained spoke
contributes $|g_v|\le3\sigma r_v$, so the total to distribute is
at most $3\sigma(p_v+r_v)=3\sigma s_v$. This justifies the zero
convention and gives $|t_v|\le3\sigma$. Each child $u$ receives
at most $3\sigma p_u$, proving the induction. The original spoke
carries at most $3\sigma w_v$. On the parent edge of $v$, the
copied tree flow and the downward transfer together use at most
\[
              3\sigma c_v+3\sigma p_v\le9\sigma c_v .
\]

To deduce the upper cut bound, choose a vector on $T^+$ saturating
every edge leaving $X$ outward. Its net outflow from $X$ is
$\delta_{T^+}(X)$, and the sweep realizes that outflow on $L$
at congestion at most nine. Both passes take linear work. Parallel
spokes at a vertex can be combined, and the recovered flow split
among them in proportion to capacity.
\end{proof}

\begin{lemma}[Adding a hub]\label{ca:lem:hub-extension}
Let $T$ approximate the terminal cuts of $J$ within a factor $q$.
After scaling $J,T$ together and adding a fresh hub, one can build a
terminal tree for the augmented graph in linear work, with quality
at most $9q$ and valid cut denominators. The same bound holds when a
specified set of terminals is identified as the hub. Two successive
hub additions give quality at most $81q$.
\end{lemma}
\begin{proof}
Place the fresh hub outside a terminal side $X$. Exactly the spokes
incident to $X$ cross, regardless of the chosen colors of the old
nonterminals. Before converting the augmented tree back to a tree
the two cut values are therefore
\[
  \lambda_{aT+\mathrm{spokes}}(X)=a\lambda_T(X)+w(X),\qquad
  \delta_{aJ+\mathrm{spokes}}(X)=a\delta_J(X)+w(X).
\]
Adding the same nonnegative quantity preserves both inequalities
$\delta_J(X)\le\lambda_T(X)\le q\delta_J(X)$ after scaling.
Apply \Cref{ca:lem:hub-tree} and minimize over the nonterminal
colors. The lower cut inequality gives valid denominators by
\Cref{sec:terminal-tree}, and the upper factor increases by at
most nine. For an identified hub, the allowed terminal cuts are
exactly those giving all the identified terminals the hub's color.
The old cut bounds therefore pass to the identified graph and
tree. Removing the hub from the latter leaves a forest with the
incident edges as spokes, to which the same construction applies.
Iterating gives the two-hub bound.
\end{proof}

\begin{corollary}[Extending a supplied router]\label{cor:hub-router}
Let $T$ approximate the terminal cuts of $J$. Suppose a router on $J$
routes each componentwise balanced demand $d$ in work $D_J$ at
congestion at most $\gamma_J\|R_Td\|_\infty$, where $\gamma_J\ge1$.
After scaling and adding two fresh hubs, one can route each demand
balanced on the new components in $D_J+O(s_J+|T|)$ work. Its
congestion factor relative to the augmented tree of
\Cref{ca:lem:hub-extension} is at most $81\gamma_J$.
\end{corollary}
\begin{proof}
Start with $aT$ and the actual spokes of $K$. Absorb the source
hub, then the sink hub, obtaining a tree $T_K$. Its terminal cuts
approximate those of $K$ by \Cref{ca:lem:hub-extension}.
Extend the demand $b$ by zero at the old nonterminals. Its unique
flow on $T_K$ has congestion $a_{T_K}(b)$ and is computed by subtree
sums. A zero-capacity tree edge has a zero-capacity descendant cut
in $K$ by validity, so componentwise balance forces zero flow there.

Undo the sink absorption with \Cref{ca:lem:hub-tree}. Set aside
the recovered sink spokes and undo the source absorption on the
remaining tree vector. The two sweeps preserve divergence and each
increases congestion by at most nine. We now have a flow on
$L=aT+E_s+E_t$ of congestion at most $\sigma=81a_{T_K}(b)$.
Copy its flows $g$ on the original spokes directly to $K$.

The old tree edges carry the residual $d_g$ at the old terminals.
Their capacities are $au_e$, so $a_T(d_g/a)\le\sigma$. In
particular, each component of $J$ has a zero-cost cut extension in
$T$. No flow crosses that extension, so $d_g$ sums to zero on
the component. Route $d_g/a$ on $J$
with the supplied router and multiply its flows by $a$. The result
has divergence $d_g$ and congestion at most $\gamma_J\sigma$ on
$aJ$. Its edges and the copied spokes are disjoint. Together
they supply exactly $b$ in $K$ with the asserted congestion.
\end{proof}

For our prepared tree, each new pair of spoke lists takes $O(s_J)$
work to absorb and prepare for row products. The resulting tree has
quality $O(\kappa q_H)$ on $K$ and the correction has parameters
\[
       \gamma_K=O(\kappa q_H),\qquad D_K=D_J+O(s_J).
\]
A core or exterior trim keeps its spokes fixed throughout its
fair-cut computation. Pass the augmented rows directly to
\Cref{fc:thm:fair}. A median changes its spokes at each round and
uses the exact correction once, as follows.

\subsection{Returning a cut or a matching to the player}\label{sec:game-implementation}

Split the positive weight at a median of the current scores, dividing
ties so that each half has mass $A=\mu(V)/2$. Use these halves as
the two lists of spokes in the network $K$ of \eqref{short:eq:median-cut}.

\begin{lemma}[Partial matching]\label{ca:rc:lem:matching}\label{cor:median-router}
In the network $K$ above, suppose the cut-row tree $T_K$ has valid
denominators and linear-time products. Let a supplied router route
every componentwise balanced demand $d$ exactly in work $D_K$, at
congestion at most $\gamma_K a_{T_K}(d)$, where $\gamma_K\ge1$.
For $0<\xi<1$, in $O(s_J\gamma_K\log N/\xi+D_K)$ work one can
either route at least $(1-\xi)A$ units from $s$ to $t$, or return
a separating cut of capacity below $(1-\xi/2)A$.
\end{lemma}
\begin{proof}
Return the zero flow for $A=0$, or the source component if $A>0$
and the hubs are disconnected. Otherwise apply \Cref{ca:lem:box}
to $d=(1-\xi/2)A(\one_s-\one_t)$ on the edges of $K$. A cut with
$d(X)>\delta_K(X)$ puts $s$ in $X$ and $t$ outside and has the
claimed capacity. For a tentative flow $f$, its residual
$b=d-B_Kf$ is balanced in each component. Use row tolerance
$\xi/(2\gamma_K)$ so that the supplied router corrects $b$ at
congestion at most $\xi/2$. The sum routes $d$ at congestion at
most $1+\xi/2$. Dividing by this factor gives a feasible flow of
value at least $A(1-\xi/2)/(1+\xi/2)\ge(1-\xi)A$.
\end{proof}

On a cut answer with $\xi=\tau$, \eqref{short:eq:median-cut} gives
both the balance and the sparse boundary required by
\Cref{lem:measured-core}. On a flow answer it remains to give the
player the endpoints that were paired.

\paragraph{Recovering just the matched endpoints.}
The oracle returns a flow of value $F$. The player needs a list of
pairs and the weight on each pair. We must know that these pairs
admit routes within the computed edge loads, but writing out the route
of every pair could cost more than the flow computation. We eliminate
the internal vertices while keeping only the two endpoint labels.

Orient every nonzero edge along its flow. On a source spoke
$s\to u$, record $u$ as the starting vertex in $J$. On a sink spoke
$v\to t$, record $v$ as the ending vertex in $J$. Maintain a
directed multigraph on the surviving vertices.
An arc represents a weighted walk whose internal vertices have already
been eliminated. To remove $v\notin\{s,t\}$, choose an incoming arc
$uv$ and an outgoing arc $vw$. Take the smaller of their weights
from both and put that amount on a new arc $uw$. Keep the tags at
the outer ends. The new arc represents the concatenation of the two
walks. We can split a walk's weight between its old arc and the new
one, so no original edge acquires additional load. Drop zero weights
and self-loops, and keep parallel arcs separate to preserve their
tags. Conservation at $v$ exhausts both lists together.

Each splice consumes at least one existing arc and creates at most
one. Thus the current multigraph always has at most its initial
number $p$ of arcs. Eliminating a vertex costs $O(1+\deg(v))$ at
its \emph{current} degree. To keep the sum of these costs small,
start a phase with $k$ internal vertices. Until half are gone, some
internal vertex has degree at most $4p/k$, since the sum of all
degrees is at most $2p$. Build a list of the eligible vertices in
one scan and update the affected degrees at each splice.
Eliminating half the vertices then costs $O(k+p)$, hence all phases
take $O(s_J\log N)$ work.

The remaining arcs join the two hubs. Their net weight from $s$ to
$t$ is still $F$. Discard the arcs from $t$ to $s$ and an equal
total weight of arcs from $s$ to $t$, splitting the last arc if needed.
Each surviving arc carries the ordered pair
$(u,v)$ from the source spoke $s\to u$ and sink spoke $v\to t$ of
its represented walk. That walk has no internal
hub, so erasing cycles in $J$ preserves its tags and can only save
capacity. This gives at most $p$ routed pairs without expanding the
walks.

Let $N_0(u,v)$ be the paired weight with tags $u,v$. The first
marginal is at most $\mu_{\rm L}$ and the second at most
$\mu_{\rm R}$ by the original spoke capacities. In particular their
sum is at most $\mu$, even at a vertex split between the two median
halves. The matching in round $j$ is
\[
 N_j=N_0+N_0^\top+
       \diag\bigl(\mu-N_0\one-N_0^\top\one\bigr).
\]
Both marginals equal $\mu$. Only the added diagonal weight can miss
the median, and its total is $\mu(V)-2F\le\xi\mu(V)$. The lazy
step takes half of each routed copy, so their combined edge load is
at most that of the original flow.

\paragraph{Recovering the missing starts for the trim.}
After a completed game we need $z(u)$, the mass \emph{from} $u$
that fails to reach the last layer. For a unit of mass placed at
$v$ just after round $j$, let $p_j(v)$ be its probability of
surviving the remaining rounds. At the final layer $p_r(v)=1$.
In round $j$ the fraction at $u$ that takes the physical step
to $v$ is $\widehat B_j(u,v)/\mu(u)$, where $\widehat B_j$ is the
stored pair-and-stay layer with attenuation and the thinning in
\eqref{eq:physical-product}. Missing fractions contribute zero to
survival. Conditioning on that one step gives, on the positive
support of $\mu$,
\[
       p_{j-1}(u)=\sum_v\frac{\widehat B_j(u,v)}{\mu(u)}p_j(v),
       \qquad z(u)=\mu(u)(1-p_0(u)).
\]
Set $z=0$ elsewhere. Working backward gives a separate survival
fraction for each starting vertex. Since the graph stayed fixed
throughout the game, the stored $O(s_J)$ endpoints and stays per
round give the entire computation in $O(rs_J)$ work.

\subsection{Construction time and success probability}

Consider a base input with $p>0$ edges after the isolate scan.
All calls use the common outer size $N$. Condition first on the
correctness of every sample and the mixing of every completed game.
There are two recursions to count. Within one hierarchy the
refinement calls pass disjoint vertex sets to the next depth.
Preparing a router at any such call can start a new \emph{base
construction} on a sparse quotient. The quotient trees need only
be good enough to answer the local queries. Once they do,
\Cref{sec:ca} gives the same output quality
$q=C_q\log^2 N\log\log N$ at every base construction.

\paragraph{Work within one refinement hierarchy.}
An original edge appears in at most one induced graph at any
depth. Thus $\sum_J |E(J)|=O(p\log N)$, and the sparse quotient
from $J$ has $O(|E(J)|/\kappa)$ edges. Taking
$\kappa=D\log N$ for a large enough constant $D$ bounds the total
size of all new base inputs by $p/2$. This choice also determines
the price of a local query. Each quotient tree has $q_H=O(q)$
on a successful recursive return. For a median with constant
shortfall $\xi=\tau$, the work bounds of
\Cref{ca:rc:lem:matching,cor:hub-router} and
\eqref{eq:quotient-router-work} give
\[
 \text{Tentative flow}\quad O(s_J\kappa q\log N),
 \qquad
 \text{Exact correction}\quad O\bigl(s_J+|E(H)|q\log^2 N\bigr).
\]
Both are $O(s_Jq\log^2 N)$, since $|E(H)|\le|E(J)|$. In the
correction the spoke sweeps take $O(s_J)$ work. Only the full
quotient requires an iterative solve, at its own quality $q_H$.

A game has $r=O(\log N)$ rounds. Endpoint extraction, the player's
$O(s_J r^3)$ work on the stored factors, and the backward survival
pass fit in the resulting $O(s_Jq\log^3 N)$ bound. A $3/2$-fair
trim costs the same order by \Cref{fc:thm:fair}, using the augmented
tree at quality $O(\kappa q)$. We verify the capacity bound for these trim networks below.

There are $O(\log N)$ grouping tests and at most one exterior trim
on $J$, for $O(s_Jq\log^4 N)$ work. At a fixed refinement depth
the pieces have at most $p$ edges and $2p$ vertices altogether,
since the base input had no isolates. Their vertex counts halve on
each branch, giving $\sum_Js_J=O(p\log N)$. Thus the total cost
of the tests and trims in this hierarchy is $O(pq\log^5 N)$.
To form their requests, subtract degrees in $J$ from the stored
base-input degrees to get the exterior weights. Scanning induced
edges suffices to update the intergroup weights and distribute
edges among the next pieces.

\paragraph{Summing the recursive preparations.}
The full quotients have $O(p\log N)$ edges in total and
$O(p\log N/\kappa)$ nonisolated vertices. By
\eqref{eq:preparation-local-work}, the preparation work outside
recursive calls is $O_b(p\log^3 N(\log\log N)^4)$, within the
cost of the tests. Write $p_i$ for the edge counts of the sparse
graphs sent to new base constructions. The choice of $\kappa$
gives
\[
   T(p)\le O(pq\log^5 N)+\sum_i T(p_i),\qquad
   \sum_i p_i\le p/2 .
\]
The edge counts at successive generations of these calls are at
most $p,p/2,p/4,\ldots$. Each generation uses the same outer $N$
in its costs and guarantees. Summing and adding the original
isolate scan gives $O(n+m\log^7 N\log\log N)$ work.

\paragraph{Capacity spread in the recursive calls and their requests.}
Restriction, contraction and merging parallel edges never decrease
the least positive capacity or increase total capacity. At a
successful sampling step the least positive capacity still does
not decrease and total capacity grows by at most four. A chain
starting with $p>0$ edges has at most $\lceil\log_2 p\rceil$
quotient calls, so the total-to-minimum ratio grows by at most
$4^{\lceil\log_2p\rceil}\le4p^2$. Thus the grouping-test count applies in every base input.

It remains to check the capacity range of the requests. Every
positive intergroup or exterior weight is a sum of capacities in
its base input, so it lies between the least positive capacity
and the total capacity. In a core trim, \eqref{ca:rc:eq:z} puts a
positive source spoke between $4\tau\mu(v)$ and $\mu(v)$, and a
sink spoke has capacity $(\alpha/8)\mu(v)$. In the exterior trim
\eqref{eq:boundary-network}, the scales are $\omega\beta_j$ on
exterior weights and $\omega$ on intergroup weights. Here the core
graph scale $\Gamma=r/\theta$ and $1/\omega$ are $O(\log N)$,
while $1/\beta_j=8Hj=O(\log N\log\log N)$. For both fair cuts
the ratio of source degree to least positive capacity is therefore
polynomial in $N$, as required by \Cref{fc:thm:fair}. Splitting a
median tie can create an arbitrarily small spoke. Those requests use
the box test and exact correction, whose work is independent of this
ratio.

\paragraph{Confidence of a quotient sample.}\label{sec:sparsifier-confidence}
A quotient with $k$ nonisolated vertices needs only $O(k)$ output
edges, but its failure probability must be small in the outer size
$N$. The algorithm of~\cite[Sections~3--5]{VaughnSignDescent}
first reduces the graph to $m_0=O_b(k\log N)$ edges.
Each subsequent phase reweights them so that a constant fraction
reach weight zero while the spectral error stays within its
allowance. Stop at a sufficiently large constant times $k$ edges.
The ratio between the initial count and this target is $O(\log N)$,
so at most $P=O(\log\log N)$ phases are needed.

Conditional on successful preprocessing and all preceding phases,
the partial-coloring bound of~\cite[Section~3]{VaughnSignDescent}
gives failure probability at most
\[
                        e^{-\ell}+e^{-p_t/64}
\]
in a phase with $p_t$ remaining edges. The first term bounds the
error from the random probes and we can reduce it by choosing the
even moment degree $\ell=\Theta_b(\log N)$. The second comes from the
random signs on the $p_t$ edges. It depends on how many edges the
phase starts with. Every executed phase has $p_t=\Omega(k)$, so
$k\ge C_b\log N$ makes \emph{both} terms at most $N^{-b-2}$.
Below this cutoff the direct spanning tree in
\Cref{sec:ca-preparation} already meets the required quality.

For preprocessing, use the outer size $N$ in the sampling
confidence parameters and projection dimensions of the graph
implementation in~\cite[Section~5]{VaughnSignDescent}. In constructing
the fixed approximate Laplacian inverse, stop recursive solves at
dimension $O_b(\log N)$ and factor those systems directly. The
factorizations cost $O_b(k\log^2 N)$ in total, and one application
uses $O_b(k\log N)$ work. The high-probability work bound for sampling
and constructing the inverse is $O_b(p_H\log N+k\log^2 N)$.
Cap both preprocessing calls at sufficiently large constant
multiples of their work and output-size bounds, and report failure
if a cap is exceeded. The
probability that preprocessing fails or violates its guarantees is
at most $N^{-b-2}$.

At constant accuracy the coloring work bound of
\cite[Section~5]{VaughnSignDescent} is
$O_b(\ell^2(m_0+k\log N\,P^4))$. Using $m_0=O_b(k\log N)$ gives
$O_b(k\log^3 N(\log\log N)^4)$ for all the phases. Together with
preprocessing this fits the stated sampling time. A union bound
over preprocessing and the phases gives failure probability at
most $N^{-b}$.

\paragraph{Adaptive calls and the work bound.}
If the isolate scan leaves no edges, return the isolated terminals
with zero-capacity links to a common root. Otherwise give the
remaining computation a budget $C p\log^7 N\log\log N$, with $C$
larger than the constant in the successful-execution bound. Abort
on a reported failure, an undefined operation, or exhaustion of
this budget. Give each sample and game fresh randomness. Conditional on the
history before a call with no earlier failure, a sample reports
failure or loses its cut guarantee with probability at most $N^{-b}$.
For a game the event with this probability bound is that it
\emph{completes} and fails to mix. A game returning a sparse cut
already provides the required local outcome.
The work cap permits only polynomially many calls, so the sum of
the conditional probabilities for the first failure is at most
$N^{-a}$ for sufficiently large $b$ relative to $a$. On the
complementary event, induction from the small quotient trees
justifies every router, the local outcomes give the quality
proved in \Cref{sec:ca}, and the work estimate precludes an abort.
The cap gives the claimed work on every execution. This proves
\Cref{ca:thm:base}.

\subsection{Trading construction time for tree quality}\label{sec:compressed-queries}

The choice $\kappa=\Theta(\log N)$ made all sparse quotients in
one refinement hierarchy fit in half the base input. We are now
free to apply a further contraction with any factor $\chi\ge2$ at
the outermost level. A larger $\chi$ reduces the size of the single
base call but allows more congestion on each forest edge when
lifting a flow. This time the attached forest is part of the output
tree, so this loss appears in its stated quality. The iterative
part of an exact query still runs on the full quotient at the base
quality.

\begin{proposition}[Construction and query bounds]\label{ca:prop:query-bounds}
For every $\chi\ge2$, the quality-$Q$ tree of \Cref{ca:thm:tradeoff}
has component height $O(\chi+\log n)$ on the same success event.
Write $R_G$ for its cut rows.

After $O(n+m)$ preprocessing, each of $A=R_GB_GC_G$,
$A^\top$, $|A|$ and $|A|^\top$ can be applied in $O(n+m)$ work.
On the same event, every componentwise balanced demand $b$ can be
routed exactly with congestion $O(Q)\|R_Gb\|_\infty$ in
$O(n+m\log^4 n\log\log n)$ work.
\end{proposition}
\begin{proof}[Proof of \Cref{ca:thm:tradeoff,ca:prop:query-bounds}]
To allow an invocation inside a larger graph, take an input with
$n'$ vertices, $m'$ edges and capacity spread polynomial in a common
outer size $N\ge\max\{16,3m'+2\}$. Fix a failure exponent $b>0$.
Use \Cref{sec:ca-preparation} with $\kappa=\chi$. Its one possible
base call has $O(k)$ vertices and edges for
$k=O(\min\{n',m'/\chi\})$. The base bound and the preparation cost
\eqref{eq:preparation-local-work} give total work
\begin{equation}\label{eq:outer-scale-construction}
 O\!\left(n'+m'\log^2 N+
                   \min\{n',m'/\chi\}\log^7 N\log\log N\right).
\end{equation}
Use failure allowance $N^{-(b+1)}$ for the sample and the base call
and cap the work at this bound. With probability at least $1-N^{-b}$
the quotient tree has valid denominators, height $O_b(\log N)$
and quality $q_H=O_b(\log^2 N\log\log N)$.

By \Cref{ca:lem:forest-lift}, the attached tree has quality at most
$\chi(1+q_H)$. Its paths gain at most $\chi$ forest edges, and its
$O(n')$ rows have the linear-work products of \Cref{sec:tree-queries}.
For a demand $d$, the router in \eqref{eq:quotient-router-work}
costs congestion at most $\chi(1+3q_H)a_{T_G}(d)$ and takes
$O(n'+m'+|E(H)|q_H\log^2 N)$ work. Substituting
$|E(H)|\le m'$ and taking $N=16(m+2)$ for the original simple
graph give the stated bounds.
\end{proof}

\Needspace{8\baselineskip}
\section{Approximate maximum flow and fair cuts}\label{sec:compression}

The tree from \Cref{ca:prop:query-bounds} gives a way to finish a
partially routed demand. Write $R$ for its cut rows and let $Q$
bound both its quality and the congestion factor of the prepared
router. A componentwise balanced residual $d$ can then be routed
with at most $Q\|Rd\|_\infty c_e$ additional load on each edge $e$.
In searching for maximum throughput we use this bound to decide
which trial flow is worth keeping. For a fair cut the eventual
correction must also preserve the outward flow on each crossing edge.

\subsection{Approximate maximum flow}

Fix a nonzero componentwise balanced demand $b$. We seek the
largest multiplier $a$ for which $ab$ can be routed within
capacity. By scaling, this throughput is $1/\opt_G(b)$. It can
be bounded from above by a cut and from below by a tentative flow,
both in the original capacities.

A cut $S$ with $b(S)>0$ gives the upper bound
$u=\delta_G(S)/b(S)$, since $a b(S)$ units must leave $S$.
For a lower bound $a$, keep an edge vector $y$ satisfying
\[
       \congestion_G(y)+Q\|R(ab-B_Gy)\|_\infty\le1 .
\]
The first term measures the largest fraction of any edge's capacity
already in use. The second bounds the additional congestion needed to
route the residual. Their sum certifies that $ab$ can be routed. We
need to compute the correction only for the final lower bound.

\begin{lemma}[Approximating minimum congestion]\label{ca:lem:opt}
Let $R$ be a quality-$Q$ cut-row approximator with valid denominators
and polynomially many rows in $n$. Suppose the products of
\Cref{ca:lem:box} for the full capacity box take $O(m)$ work after
setting aside isolates. For any nonzero componentwise balanced demand
$b$ and $0<\eps\le1/2$, one can find $a>0$ and a cut $S$ with $b(S)>0$
such that
\[
       a\le\frac{1}{\opt_G(b)}\le\frac{\delta_G(S)}{b(S)}
          \le(1+\eps)a .
\]
The search takes $O(mQ\log n/\eps)$ work on the active graph. Suppose
a router can route each componentwise balanced demand $d$ exactly at
congestion $O(Q)\|Rd\|_\infty$ in work $D$. One additional call then
produces a flow for $b$ at congestion at most $(1+\eps)b(S)/\delta_G(S)$.
Scaling this flow by $a$ gives a feasible flow of divergence $ab$.
\end{lemma}
\begin{proof}
Increase $Q$ by a constant if needed to cover the supplied router.
Tree quality alone suffices for the lower-bound certificate while
the search is running.

\proofstep{Starting the interval.}
Put $r=\|Rb\|_\infty>0$. With $y=0$ we can reserve the entire
capacity box for repair. This certifies $a=1/(Qr)$. For the upper
bound take a row attaining $r$. If the row is implicit,
$R^\top$ applied to its coordinate vector recovers its side.
Complement it if necessary so that $b(S)>0$. Validity of the row
denominator gives $u=\delta_G(S)/b(S)\le1/r$. Thus the initial
relative gap $g=u/a-1$ is at most $Q-1$.

\proofstep{Testing the midpoint.}
While $g>\eps$, test $tb$ at $t=(a+u)/2$ with
\Cref{ca:lem:box}. A violated cut $X$ has $b(X)>0$ and gives the
new upper bound $\delta_G(X)/b(X)<t$. Keeping the lower bound $a$
then leaves a relative gap below $g/2$.

On a flow answer, $f$ lies in the capacity box and
$\|R(tb-B_Gf)\|_\infty\le e$, where $e$ is our chosen row tolerance.
Its load and the congestion reserved for repair add to at most
$1+Qe$. Dividing both the trial demand and its tentative flow by
this factor gives a certified lower bound in the original box,
\[
          a'=\frac{t}{1+Qe},\qquad y'=\frac{f}{1+Qe},\qquad
          \congestion_G(y')+Q\|R(a'b-B_Gy')\|_\infty\le1 .
\]
The new gap splits at the tested value $t$,
\[
       \frac{u}{a'}-1
          =\frac{u-t}{a'}+\frac{t-a'}{a'}
          =\frac{u-t}{a'}+Qe .
\]
The interval above $t$ already has length $ag/2$. Give the loss
below $t$ another $g/4$ in relative gap by setting $e=g/(4Q)$.
Since $t/a=1+g/2>1+Qe$, the lower bound increases. The first term
is therefore at most $(u-t)/a=g/2$. A flow answer leaves gap at
most $3g/4$. Since $g$ starts at most $Q-1$ and decreases, this tolerance is
always below one.

A trial at gap $g$ costs $O(mQ\log n/g)$. Count backward from
the last tested gap, which exceeds $\eps$. Each tested gap is at most $3/4$ of its predecessor, so
\[
           \sum_{\rm trials}\frac1g
               \le\frac1\eps\sum_{j\ge0}(3/4)^j=\frac4\eps .
\]
At termination $u\le(1+\eps)a$. If an explicit flow is wanted,
route $ab-B_Gy$ once and add it to $y$. The certificate gives a
feasible flow for $ab$. Dividing by $a$ routes the original $b$ at
congestion at most
\[
            \frac1a\le\frac{1+\eps}{u}
            =(1+\eps)\frac{b(S)}{\delta_G(S)}
            \le(1+\eps)\opt_G(b).
\]
\end{proof}

\begin{corollary}[Approximate maximum flow]\label{ca:cor:flow}
Let $G$ be an undirected graph of polynomial capacity spread and let
$b$ be a componentwise balanced demand. For $0<\eps\le1/2$ and
$\chi\ge2$, with high probability the algorithm returns a flow $f$ with
$B_Gf=b$ and $\congestion_G(f)\le(1+\eps)\opt_G(b)$ in time
\begin{equation}\label{ca:eq:flow-runtime}
 O\!\left(n+\frac{m\log^7 n\log\log n}{\chi}
            +\frac{m\chi\log^3 n\log\log n}{\eps}\right).
\end{equation}
Optimizing $\chi$ gives
\[
 O\!\left(n+\frac{m\log^5 n\log\log n}{\sqrt{\eps}}
            +\frac{m\log^3 n\log\log n}{\eps}\right).
\]
For $\eps\gtrsim 1/\log^4 n$ this simplifies to
$O(n+m\log^5 n\log\log n/\sqrt{\eps})$.
\end{corollary}
\begin{proof}
Return zero for $b=0$. Otherwise set aside isolates, where $b$
vanishes, and use a common throughput multiplier on all components.
With contraction factor $\chi$, only the
$k=O(\min\{n,m/\chi\})$ nonisolated quotient vertices enter the
base construction. The search, however, tests the capacity box on
all $m$ original edges with tree quality
$Q=O(\chi\log^2 n\log\log n)$. By \Cref{ca:prop:query-bounds,ca:lem:opt},
the two costs depending on $\chi$ are consequently
\[
       O(k\log^7 n\log\log n)
       \quad\text{and}\quad
       O\bigl((m\chi/\eps)\log^3 n\log\log n\bigr).
\]
Preprocessing adds $O(n+m\log^2 n)$, and the single exact
correction costs $O(n+m\log^4 n\log\log n)$ independent of $\chi$.
With $k=O(m/\chi)$ the displayed terms give
\eqref{ca:eq:flow-runtime}. Their geometric mean is
$m\log^5 n\log\log n/\sqrt\eps$, which covers the other edge
costs. They balance at $\chi=\sqrt\eps\log^2 n$, so take
$\chi=\max\{2,\sqrt\eps\log^2 n\}$.

Alternatively, keeping $k\le n$ and choosing $\chi=2$ gives
\[
 O\!\left(n\log^7 n\log\log n+m\log^4 n\log\log n+
                 \frac{m\log^3 n\log\log n}{\eps}\right).
\]
After a successful preparation all steps are deterministic. Cap the
work at a sufficiently large constant times the applicable bound and
abort on an undefined operation. The algorithm then inherits the
success probability of the preparation and the work bound holds on
every execution.
\end{proof}

For arbitrary positive capacities, \Cref{app:capacities} obtains
the same search and routing bounds from two quotient trees. Their
stacked row norm bounds the flow in each tree. Each quotient cut
has a cheap extension in at least one of them, which suffices for
the required bound on every demand.

\subsection{Fair cuts}

The algorithm in \Cref{app:fc} reuses a single congestion approximator
on the original capacities. Each query asks to send half the current
cut's slack. It either finds a cut with less slack under the current
edge vector or supplies an approximate augmentation. A potential
tracks the slack and the congestion needed to route the accumulated
divergence error, which is repaired once at the end.

\begin{proof}[Proof of \Cref{thm:fair-intro}]
If $s,t$ are disconnected, take the
component of $s$ and the zero flow. Otherwise work on their component
and put all other vertices on the sink side. Set $N=16(m+2)$.
Polynomial spread ensures
$\delta_G(\{s\})/c_{\min}\le m c_{\max}/c_{\min}=N^{O(1)}$.

Prepare the tree of \Cref{ca:prop:query-bounds} with factor $\chi$.
By \Cref{fc:thm:fair}, the search and final correction take
$O(n+mQ\log^2 n/\rho)$ work.

Including construction and substituting the tree quality gives
\[
 O\!\left(n+m\log^2 n+
 \min\{n,m/\chi\}\log^7 n\log\log n+
                   \frac{m\chi\log^4 n\log\log n}{\rho}\right).
\]
Using $m/\chi$ for the quotient size, the construction and search terms balance at
$\chi=\sqrt\rho\,\log^{3/2} n$. Taking
$\chi=\max\{2,\sqrt\rho\,\log^{3/2} n\}$ gives \eqref{eq:intro-fair}.
As in \Cref{ca:cor:flow}, cap the work at the proved bound. The search
and final correction are deterministic after a successful preparation.
\end{proof}

\section{Undirected decomposition in time independent of conductance}
\label{sec:near-optimal-decomp}

Run the matching game on an undirected phase graph $J$ with
starting weights $d$. If all rounds finish, let $z(v)$ be the mass
starting at $v$ that fails to reach the final layer, and let
$\Gamma$ bound the congestion of the surviving walks. If the
game mixes, at least $\alpha d(X)$ of the mass started in a set
$X$ of at most half the phase weight fails to return to $X$.
This mass is either missing or reaches the last layer after
crossing the boundary of $X$. Thus \eqref{ca:rc:eq:defective-cut} gives
\[
        \alpha d(X)\le \Gamma\delta_J(X)+z(X)
                     \qquad(d(X)\le d(V(J))/2).
\]
We use a single fair cut to choose the part $A$ to finalize,
following~\cite{LNPS23faircuts}, and recurse on $R=V(J)\setminus A$.
Take the lighter side $X$ of a cut in $A$, measured by $d$.
Its original boundary includes edges to $R$, which disappear from
the induced graph on $A$. In the trim network, those edges and
the source spokes for the missing starts both cross the fair cut
into $X$. The fair flow sends at least two thirds of their combined
capacity $z(X)+\Gamma c_J(X,R)$ into $X$. It can leave $X$ only
through edges to $A\setminus X$ or through sink spokes of
total capacity $(\alpha/8)d(X)$. Conservation gives the bound
from \Cref{lem:fair-core},
\[
       z(X)+\Gamma c_J(X,R)
          \le 2\Gamma\delta_{J[A]}(X)+(\alpha/4)d(X).
\]
Writing $\delta_J(X)=\delta_{J[A]}(X)+c_J(X,R)$ in the return
bound and substituting the flow bound gives
$\Gamma\delta_{J[A]}(X)\ge(\alpha/4)d(X)$. This certifies
expansion inside the core of \Cref{lem:measured-core}. The
backward pass computes $z$ from the weighted matching endpoints
and the attenuation factors, and the routing capacities give
$\Gamma$. The trim can therefore use both quantities without
recovering the routes.

A smaller target conductance $\phi$ allows higher walk congestion.
The router of \Cref{sec:ca-computation} uses this allowance by
scaling the internal capacities of each request and scaling the
prepared tree by the same factor. This preserves the tree quality,
and a median request always allows the same relative shortfall.
For the trim, the initial slack is the total capacity $z(V(J))$
of the source spokes, which is bounded by the starting mass at
every graph scale.

\Needspace{8\baselineskip}
\begin{theorem}[Undirected expander decomposition]
\label{thm:near-optimal-decomp}
Fix $a,b>0$. Let $G$ be a simple undirected graph whose positive
capacities have spread at most $(m+2)^b$, and let $0<\phi\le1/\log n$.
A randomized algorithm takes $O_{a,b}(n+m\log^{13/2}n\log\log n)$
work on every execution. With probability at least $1-(m+2)^{-a}$
it partitions $G$ into induced $\Omega(\phi)$-expanders with total
crossing capacity $O(\phi W\log^2 n)$. If a nonnegative vertex
measure $\mu$ with $\mu(V)\le W$ is supplied, the parts also satisfy
the measure guarantee of \Cref{thm:main} on the same event.
\end{theorem}

\begin{proof}
Output each isolated vertex as a singleton and return if $m=0$.
On the remaining vertex set $V_+$, fix
\[
            d(v)=\deg_G^c(v)+\mu(v),\qquad M_0=d(V_+)\le3W ,
\]
taking $\mu=0$ if no measure was supplied. Keep these weights fixed
throughout the recursion. In a final part $A$, $d$ dominates both
the degrees in $G[A]$ and the prescribed weights $\mu|_A$. A lower
bound on cut capacity in terms of $\min\{d(X),d(A\setminus X)\}$
therefore implies both expansion guarantees. Put $N=16(m+2)$.
At each recursive call, separate the connected components at zero
boundary cost. Finalize singletons and run one phase on each larger
component.

\proofstep{One phase.}
On a connected input $C$, put $J=G[C]$ and $M=d(C)$, and use the
cut-or-core test of \Cref{lem:measured-core} with measure $d|_C$.
The test allows $r=\Theta_a(\log N)$ rounds and routes each matching
with capacities $c_e/\theta$, so the surviving physical walks have
congestion at most $\Gamma=r/\theta$. Take $\theta=\phi\log N$, so
$\Gamma=\Theta_a(1/\phi)$ and the induced expansion in
\eqref{eq:measured-light} is at least $\alpha/(4\Gamma)=\Omega(\phi)$.

If a median request returns a cut, both sides have weight at least
$\tau M/4$ and the boundary costs at most $\theta$ times the
smaller weight. Recurse on both sides. If all rounds finish, the
backward pass gives the missing mass $z(v)$ from every start. Run
the fair cut on
\[
              \Gamma J+\{sv:z(v)\}+\{vt:(\alpha/8)d(v)\}.
\]
Finalize the vertices $A$ of $C$ on the sink side and recurse on
$R=C\setminus A$. If the game mixes, every cut of $G[A]$ has
capacity at least $\alpha/(4\Gamma)$ times its smaller $d$-weight.
On every completion, \eqref{eq:measured-test} gives $d(R)\le M/8$
and $\delta_J(R)\le10\tau M/\Gamma$.

\proofstep{Crossing capacity and recursion depth.}
Assume the preparations succeed and the completed games mix. Count
each edge when its endpoints are first separated. On a median cut,
the boundary costs at most $\theta$ times the $d$-weight of either
side. Charge $\theta d(v)$ to each vertex on the side with fewer
vertices. Whenever a vertex is charged, its new part contains at
most half as many vertices as before. Thus the median cuts cost
at most $\theta M_0\log_2|V_+|$ in total. A trim costs at most
$10\tau M/\Gamma$ and finalizes a part of weight at least $7M/8$.
These finalized parts are disjoint, so the sum of their phase
weights $M$ is at most $8M_0/7$. The total crossing capacity is
\[
        O\bigl(\theta W\log n+W/\Gamma\bigr)
                            =O(\phi W\log^2 n),
\]
using $\log N=O(\log n)$ for a simple input.

The work bound needs a depth bound on both sides of every median
cut. Each median child has $d$-weight at most
$(1-\tau/4)M$, and the only child continuing after a trim has
weight at most $M/8$. Each active vertex has $d(v)\ge c_{\min}$
for the least positive input capacity $c_{\min}$, and
$M_0/c_{\min}\le3m(m+2)^b$. Thus every vertex appears in
$O_b(\log N)$ phase inputs. These inputs are disjoint at each
depth, so, writing $s_C=|C|+|E(G[C])|$, we have
$\sum_Cs_C=O_b(m\log N)$ after the isolate scan.

\proofstep{Work within one phase.}
Prepare the tree and router of \Cref{ca:prop:query-bounds} once for
$J$, with contraction factor $\chi\ge2$. The cost is
$O_{a,b}(s_C\log^2 N+(s_C/\chi)\log^7 N\log\log N)$ and the
quality is $Q=O(\chi\log^2 N\log\log N)$. The median queries and
the trim scale $J$ differently and attach different spokes.
In each case, scaling $J$ and the tree together preserves their
cut comparison. Adding the two hubs increases the quality factor
by at most a constant by \Cref{ca:lem:hub-extension}.

In each median request the allowed shortfall is a fixed fraction
$\tau$ of $M/2$, so the box test in \Cref{ca:rc:lem:matching}
costs $O(s_CQ\log N)$ at every graph scale. To correct the
tentative flow, two sweeps recover the spoke flows and leave a
demand on $J$ (\Cref{cor:hub-router}). We solve for its flow on
the full quotient at quality $O(\log^2 N\log\log N)$ and lift
the result through the forest once. By
\eqref{eq:quotient-router-work}, the correction costs
$O(s_C\log^4 N\log\log N)$ per request. Extracting endpoints costs
$O(s_C\log N)$ per round. The scores and the backward pass cost
$O(s_C r^3)$ and $O(s_C r)$ over the game. Thus all $r$ rounds
together cost
\begin{equation}\label{eq:independent-matching-work}
             O\!\left(s_C[Q\log^2 N+\log^5 N\log\log N]\right).
\end{equation}

For the trim, let $K$ be the augmented graph. The time bound in
\Cref{fc:thm:fair} depends on $\delta_K(\{s\})/c_{\min}(K)$,
which stays polynomial in $N$ even as the internal capacities
grow with $\Gamma$. The numerator is $z(C)\le5\tau M$.
By \eqref{ca:rc:eq:z}, each source spoke has
capacity at least $4\tau d(v)$, and each sink spoke has capacity
$(\alpha/8)d(v)$. Each internal edge has capacity at least
$\Gamma c_{\min}$ with $\Gamma=\Omega_a(1)$. Hence
\[
       c_{\min}(K)=\Omega_a(c_{\min}),\qquad
       \frac{\delta_K(\{s\})}{c_{\min}(K)}
                  = O_a(W/c_{\min}).
\]
Since $W/c_{\min}\le m(m+2)^b$, the bound is uniform over the
allowed values of $\phi$.
With $\rho=1/2$, the fair cut costs $O(s_CQ\log^2 N)$, already
covered by \eqref{eq:independent-matching-work}.

Combining preparation and queries gives phase work
\[
 O_{a,b}\!\left(s_C\log\log N
    \left[\frac{\log^7 N}{\chi}+\chi\log^4 N+\log^5 N\right]\right).
\]
The first term pays for building the quotient tree, and the second
for the queries using the reattached tree. They balance at
$\chi=\log^{3/2}N$, giving $O_{a,b}(s_C\log^{11/2}N\log\log N)$
work per phase. Summing over $C$ and including the isolate scan gives
$O_{a,b}(n+m\log^{13/2}N\log\log N)$.

\proofstep{Success probability.}
Give each preparation and game a failure allowance of $N^{-(a+2)}$
conditional on the preceding history. A game fails if it completes
without mixing. To count calls, stop the recursion at the first
failure and count finalized and pending parts as leaves. Each
phase that splits its input produces at least two nonempty parts,
including the finalized core when there is one. There are at most
$2m$ leaves, so at most $4m$ phases have started.
Each phase uses one preparation and at most one game, giving failure
probability at most $8mN^{-(a+2)}\le(m+2)^{-a}$ by a union bound.
After the isolate scan, cap the work at
$C_{a,b}m\log^{13/2}N\log\log N$ and abort on reported failure, an
undefined operation, or exhaustion of the budget. A sufficiently
large $C_{a,b}$ covers all work on the success event, and the cap
bounds the work on every execution.
\end{proof}

For polynomial capacity spread, choosing the faster of this algorithm
and the undirected algorithm of \Cref{thm:main} gives
\[
 O_{a,b}\!\left(n+m\log^4 n+
       m\min\{\log n/\phi,\ \log^{13/2}n\log\log n\}\right).
\]

\appendix
\section{Algorithm for fair cuts}\label[appendix]{app:fc}

Following Li and Li~\cite{LL25faircuts}, we ask to send more flow across
a cut until even its lowest-capacity edge is nearly saturated. An augmentation
fills some of the unused capacity. A cut that prevents the augmentation
already has less unused capacity under the old flow. We will recover
this alternative with approximate augmentations on the original graph,
so the same congestion approximator serves every request.

\begin{theorem}[Computing a fair cut with a supplied approximator]\label{fc:thm:fair}
Let $G$ be an undirected graph with $n$ vertices, $m$ positive-capacity
edges, and distinct terminals $s,t$. Fix $N\ge16$ with $m\le N^{O(1)}$.
If $m>0$, assume $\delta_G(\{s\})/c_{\min}\le N^{O(1)}$, where
$c_{\min}$ is the least positive capacity. After isolates are set aside,
let $R$ be a fixed cut-row approximator of quality $Q\ge1$ and size
linear in the active graph. Assume its denominators are valid and its
linear-time products from \Cref{ca:lem:box} are prepared. For
$0<\rho\le1/2$, a deterministic algorithm finds a $(1+\rho)$-fair cut
and its witnessing flow in $O(n+mQ\log^2 N/\rho)$ work.
\end{theorem}

\subsection{Augmenting across the current cut}\label{fc:sec:progress}

Suppose $s$ and $t$ are connected. Write $B=B_G$, $C=\diag(c)$, and
$b=\one_s-\one_t$. Fix a source side $S$ and an edge vector $f$ with
$|f_e|\le c_e$. An edge leaving a set $X$ can carry another
$c_e-f_e^{\rm out}$ units in the outward direction, where
$f_e^{\rm out}$ is its signed flow out of $X$. Thus
\[
 \Phi_f(X)=\sum_{e\in\partial X}(c_e-f_e^{\rm out})
          =\delta_G(X)-(Bf)(X)
\]
is the capacity leaving $X$ in the residual network of $f$. We call
it the \emph{slack} of $X$ and write $\Phi=\Phi_f(S)$.

\paragraph{An exact augmentation.}
Ask to send $a=\Phi/2$ more units from $s$ to $t$. If we can send
this much in the residual network, the slack of $S$ falls to $\Phi-a$.
Otherwise a source side of residual capacity less than $a$ is already
a better choice of cut. To use our fixed approximator, solve for the
replacement edge vector $g$ directly. Adding an augmentation of
divergence $ab$ to $f$ is exactly the request
\[
                      |g|\le c,\qquad Bg=d,\qquad d=Bf+ab .
\]
The capacities in this request are the original $c_e$ at every
iteration. For any $X$ the amount by which its requested outflow
exceeds its capacity is
\begin{equation}\label{fc:eq:obstruction}
       d(X)-\delta_G(X)=a\,b(X)-\Phi_f(X).
\end{equation}
Since $\Phi_f(X)\ge0$, a positive deficit forces $b(X)=1$ and
$\Phi_f(X)<a$. Thus any violated cut returned for $d$ is a source
side with less than half the slack of $S$, measured under the same
edge vector $f$.

\paragraph{What an approximate augmentation leaves behind.}
We start with $S=\{s\}$, $f=0$, and intended flow value $v=0$.
The difference $h=vb-Bf$ is the demand we would still have to route
to make $f$ a flow of value $v$. Once $h\ne0$, which divergence
should the next request use? If we asked for the full terminal demand
$(v+a)b$, any violated cut $X$ would still have $b(X)=1$. Its deficit
on that source side would be
\[
       (v+a)b(X)-\delta_G(X)=a+h(X)-\Phi_f(X).
\]
A violation would give only $\Phi_f(X)<a+h(X)$. Old unmet demand
in $X$ could account for the violation. The request
$d=Bf+ab=(v+a)b-h$ leaves that demand for the final repair, so
\eqref{fc:eq:obstruction} still gives $\Phi_f(X)<a$.

Suppose the replacement $g$ is within capacity and leaves unmet
demand $k=d-Bg$. Substitute $Bg=Bf+ab-k$ into the slack formula.
Keeping $S$ and raising the intended value to $v+a$ gives
\begin{equation}\label{fc:eq:update}
 \Phi_g(S)=\Phi-a+k(S),\qquad
 (v+a)b-Bg=(vb-Bf)+(d-Bg)=h+k .
\end{equation}
The scalar $k(S)$ is the shortfall in the transfer across this cut.
The vector $h+k$ records where flow still has to be sent. Switching
cuts keeps $f$ and $v$ fixed, so it adds nothing to this demand.

It is enough to make the final slack small in absolute units and the
final repair small in congestion. Take $\omega=\rho/4$ and
$\theta=\omega c_{\min}$. Once $\Phi\le\theta$, each crossing edge
has at most $\theta\le\omega c_e$ unused, since the edge slacks are
nonnegative and sum to $\Phi$. If a correction $f_h$ routes $h$ with
$|(f_h)_e|\le\omega c_e$, then $f+f_h$ sends at least
$(1-2\omega)c_e$ outward on each crossing edge and uses at most
$(1+\omega)c_e$ on every edge. Dividing by $1+\omega$ gives a feasible
flow of value $v/(1+\omega)$. Its outward flow meets the fairness
requirement because
\[
            \frac{1+\omega}{1-2\omega}
               =1+\frac{3\omega}{1-2\omega}
               \le 1+4\omega=1+\rho
            \qquad(\omega\le1/8).
\]
By \Cref{ca:lem:regression}, a correction of this size is available
whenever $3Q\|Rh\|_\infty\le\omega$. We can therefore defer routing
until $\Phi\le\theta$, provided we control this norm along the way.

\subsection{The cost of restoring conservation}\label{fc:sec:allowance}

A row $\one_U^\top/u_U$ of $R$ measures the net demand still to be
sent out of $U$, in units of its cut denominator $u_U$. Suppose a
transfer sends $a$ units from $s$ to a vertex $u\notin S$ and stops
there. All $a$ units have crossed $S$, but the error
$k=a(\one_u-\one_t)$ adds $a/u_U$ to each row with $u\in U$ and
$t\notin U$. If a later
step sends its newly requested $s$--$t$ transfer and also carries
these $a$ units from $u$ to $t$, its residual is
$-a(\one_u-\one_t)$. The two contributions cancel on every row. Charging
the largest absolute error after each individual transfer would
miss this cancellation and could charge a different row each time.

List both signs of the positive-denominator rows as $r_1,\ldots,r_p$,
so that $\max_j r_jh=\|Rh\|_\infty$. Give row $j$ weight
$e^{\lambda r_jh}$, with $\lambda>0$ to be chosen. An unmet demand
$k$ multiplies this weight by $e^{\lambda r_jk}$, so opposite errors
on the same row undo one another. Let $Z(h)$ be the total weight.
Comparing the sum to its largest term gives
\begin{equation}\label{fc:eq:smooth-max}
 Z(h)=\sum_j e^{\lambda r_jh},\qquad
 \lambda\|Rh\|_\infty\le\log Z(h)
                   \le\lambda\|Rh\|_\infty+\log p .
\end{equation}
Write $\pi_j=e^{\lambda r_jh}/Z(h)$ for the share held by row $j$.
The factor by which the total weight grows in one proposed update is
\begin{equation}\label{fc:eq:weight-growth}
        \frac{Z(h+k)}{Z(h)}=\sum_j\pi_j e^{\lambda r_jk}.
\end{equation}
The shares $\pi$ are known before the request. For a small transfer
they give the marginal cost of its error, since the derivative of
$\log Z(h+\tau k)$ at $\tau=0$ is $\lambda\sum_j\pi_j r_jk$.
The initial prices can understate the cost of a larger transfer
because its error can sharply increase a row's weight. At $h=0$,
for example, the two signs have equal weight, so the derivative
vanishes in every direction. Yet any
$k$ with $Rk\ne0$ increases $Z$ above $Z(0)$. The growth factor
in \eqref{fc:eq:weight-growth} accounts for a row that becomes
expensive during the proposed update.

With a maximum, a single row can set the charge for the entire
update. The standard solver guarantee therefore uses the same
accuracy against every row. The change in $\log Z$ measures how
much the update moves the total weight. A row with little current
weight can drive that change only once its error grows enough to
give it a significant share of the total. The solver can therefore
allow a larger payoff gap when a comparison gives such a row much
more weight than it currently has. That flexibility
saves a factor of $\log p$ in the iteration bound
(\Cref{fc:sec:prices}). At termination the total weight still
bounds every row's individual error.

\paragraph{Balancing slack against repair.}
When no error is left, an update reduces the slack by at least
$a=\Phi/2$. While $\Phi>\theta$ this removes more than a quarter
of the shifted slack $w=\Phi+\theta$. To allow an increase in
row weight when a flow saves slack, track their product. The
normalized logarithm of this product is
\begin{equation}\label{fc:eq:potential}
        \Psi(\Phi,h)=\log(1+\Phi/\theta)+\log\frac{Z(h)}p .
\end{equation}
Initially $h=0$ and $Z(0)=p$. For every later $h$, the weights for
opposite signs of a row sum to at least two, so $Z(h)\ge p$.
Also $w\ge\theta$ even when the current cut is completely saturated.
Thus $\Psi\ge0$ throughout, including a last update that reduces
$\Phi$ to zero while leaving demand to repair.

Let $\Phi_0=\delta_G(\{s\})$ and
$\Psi_0=\log(1+\Phi_0/\theta)$. Suppose each update decreases
$\Psi$. Since the slack factor $1+\Phi/\theta$ is always at
least one, the row weights can grow only to
\[
 e^{\lambda\|Rh\|_\infty}\le Z(h)
 \le (1+\Phi/\theta)Z(h)\le p(1+\Phi_0/\theta).
\]
Consequently $\lambda=(3Q/\omega)(\log p+\Psi_0)$ ensures
$3Q\|Rh\|_\infty\le\omega$. It remains to achieve a fixed
decrease in $\Psi$ whenever $\Phi>\theta$. An exact augmentation
would decrease its first term by $-\log(1-a/w)\ge a/w>1/4$.
We will keep at least $1/8$ of this gain or switch to a cut
with still less slack.

\subsection{Computing a flow or an obstructing cut}\label{fc:sec:prices}

For a candidate $g$, \eqref{fc:eq:update} gives the slack removed
as $a-k(S)$, and \eqref{fc:eq:weight-growth} gives the change in
row weight. Substituting into the potential and using $\log(1+x)\le x$
gives
\begin{align}
 \Psi(\Phi_g(S),h+k)-\Psi(\Phi,h)
     &=\log\!\left(1+\frac{k(S)-a}{w}\right)
                          +\log\frac{Z(h+k)}{Z(h)}
       \ \le\ -\frac aw+\mathcal E(k), \label{fc:eq:product}\\
 \mathcal E(k)&=\frac{k(S)}w+\log\sum_j\pi_j e^{\lambda r_jk}.
                                                  \label{fc:eq:weighted}
\end{align}
An exact answer has $\mathcal E(0)=0$. Any candidate with
$\mathcal E(k)\le1/8$ gives the required decrease, since $a/w>1/4$.
The first term charges a shortfall on $S$ and gives credit for
outflow beyond the request. The second records the effect of $k$ on
the final repair.

Suppose the query returns the old vector $g=f$. Then $k(S)=a$
and the slack has not decreased. An ordinary cut row for $S$
would measure only $a/\delta_G(S)$, which can be tiny when the cut
is nearly saturated. Dividing by $w$ measures the shortfall on
the scale of the slack we are trying to remove.
Normalizing a cut row by $w$ would give it absolute row sum
$\delta_G(S)/w$ on the capacity-scaled edge variables, which can
exceed the box solver's bound of one. In \eqref{fc:eq:game} we
charge the shortfall through a fixed vertex price on $S$. Only
the repair rows contribute to the interaction matrix, so its
row-sum bound still holds.

It suffices to compare the charge $\mathcal E$ of a candidate flow
with the price deficit $D_d(\psi)$ from \eqref{eq:price-deficit}.
A positive deficit identifies a cut obstructing the transfer.
At nonpositive deficit the following bound ensures that the
candidate's charge is small enough.

\begin{lemma}[A flow and prices for the same request]\label{fc:lem:weighted}
In $O(m(1+\lambda))$ work we can find $|g|\le c$ and vertex prices
$\psi$ such that
\begin{equation}\label{fc:eq:comparison}
                   \mathcal E(d-Bg)\le\lambda D_d(\psi)+1/8 .
\end{equation}
\end{lemma}

When $D_d(\psi)\le0$, the comparison gives
$\mathcal E(d-Bg)\le1/8$, so \eqref{fc:eq:product} bounds the flow
update's potential change by $-a/w+1/8<-1/8$. When $D_d(\psi)>0$,
switch to the obstructing cut. Its slack is less than
$a=\Phi/2=\Phi-a$, while its repair term is unchanged. Its
potential therefore falls by at least $a/w>1/4$. The prices that
certify these alternatives come from the same computation.

\begin{algorithm}[htbp]
\caption{One fair-cut update while $\Phi>\theta$}\label{alg:fair-step}
\begin{enumerate}[leftmargin=2.2em,itemsep=2pt,topsep=3pt]
\item Set $a=\Phi/2$, $w=\Phi+\theta$, $h=vb-Bf$, and $d=Bf+ab$.
Compute the shares $\pi_j=e^{\lambda r_jh}/Z(h)$.
\item Compute $g,\psi$ by \Cref{fc:lem:weighted} and evaluate
$D_d(\psi)$ by \eqref{eq:price-deficit}.
\item If $D_d(\psi)>0$, use \Cref{obs:coarea} to find a level set
$X$ of $\psi$ with $d(X)>\delta_G(X)$. Replace $S$ by $X$.
Otherwise keep $S$ and replace $(f,v)$ by $(g,v+a)$.
\end{enumerate}
\end{algorithm}

\begin{proof}[Proof of \Cref{fc:lem:weighted}]
A proposed flow with residual $k$ multiplies the total row weight
by $\sum_j\pi_j e^{\lambda r_j k}$. A row with a small current
share $\pi_j$ needs a larger signed error $r_j k$ for its term
$\pi_j e^{\lambda r_j k}$ to reach the same size. To express the
logarithm of the whole sum in terms of row prices, take any
distribution $y$ on the $p$ signed rows and write
$D_{\rm KL}(y\Vert\pi)=\sum_j y_j\log(y_j/\pi_j)$. Apply Jensen's
inequality to the logarithm with weights $y_j$ on the indices where
$y_j>0$ to obtain
\[
 \lambda\sum_j y_j r_j k-D_{\rm KL}(y\Vert\pi)
   =\sum_{j:y_j>0}y_j\log\frac{\pi_j e^{\lambda r_j k}}{y_j}
   \le \log\sum_{j:y_j>0}\pi_j e^{\lambda r_j k}
   \le \log\sum_j\pi_j e^{\lambda r_j k}.
\]
Equality holds when $y_j$ is proportional to $\pi_j e^{\lambda r_j k}$.
These maximizing weights are the shares the rows would have after
adding $k$ to $h$. Thus the increase in $\log Z$ is exactly the
largest weighted row error $\lambda\sum_j y_jr_j k$ after deducting
$D_{\rm KL}(y\Vert\pi)$.

Write $\widetilde R$ for the matrix of signed rows and
$k_x=d-BCx$ for the residual of the candidate $g=Cx$,
$x\in[-1,1]^m$. A distribution $y$ prices the residual at vertices
by $\widetilde R^\top y$. The other term of $\mathcal E(k_x)$
charges the shortfall $k_x(S)$ at the same rate for every $y$.
To include it, add $1/(\lambda w)$ to the price of every vertex
in $S$. This gives the box--simplex payoff
\begin{equation}\label{fc:eq:game}
 \psi_y=\widetilde R^\top y+\frac{\one_S}{\lambda w},
 \qquad F(x,y)=\psi_y^\top k_x,\qquad
 \lambda F(x,y)=\frac{k_x(S)}w+\lambda\sum_j y_jr_j k_x .
\end{equation}
At fixed $x$ the preceding maximization recovers the candidate's
charge. At fixed $y$, saturating each edge from higher to lower
price attains the least deficit at $\psi_y$, by
\eqref{eq:price-deficit}. Thus, with $\Delta_p$ the probability
simplex,
\begin{equation}\label{fc:eq:change}
 \mathcal E(k_x)=
    \max_{y\in\Delta_p}\{\lambda F(x,y)-D_{\rm KL}(y\Vert\pi)\},
 \qquad
 D_d(\psi_y)=\min_{z\in[-1,1]^m}F(z,y).
\end{equation}
It therefore suffices to compute $\widehat x,\widehat y$ satisfying
\begin{equation}\label{eq:fair-comparator}
   \lambda F(\widehat x,y)-D_{\rm KL}(y\Vert\pi)
        \le \lambda F(z,\widehat y)+\tfrac18
        \qquad(z\in[-1,1]^m,\ y\in\Delta_p).
\end{equation}
Indeed, maximize the left side over $y$ to assess the returned
flow at its updated row shares. Minimize the right side over $z$
to test the returned prices against every admissible flow. By
\eqref{fc:eq:change}, this gives \eqref{fc:eq:comparison}.

Against the current distribution $y=\pi$ we need additive error
$1/8$ in the scaled payoff. For a distribution concentrated on
row $j$, the relative entropy penalty is $\log(1/\pi_j)$.
A payoff gap of $1/8+\log(1/\pi_j)$ then leaves only $1/8$
after subtracting that penalty.
We will retain the corresponding allowance for every mixture of
rows by initializing the solver at $\pi$.

\proofstep{Computing the flow and the prices.}
The interaction matrix $A=-\widetilde R BC$ has absolute row sum
$\delta_G(U)/u_U\le1$ for either sign of the row for $U$. The fixed
price on $S$ adds an affine term in $x$ and leaves this matrix unchanged.
Fix each box coordinate in a zero column of $A$ at a minimizer of
its affine term. Run the box--simplex iteration of Jambulapati and
Tian~\cite[Algorithm~4 and Theorem~1]{JT23} at unit normalization
for $T$ steps, starting the remaining box coordinates at zero and
both simplex vectors at $\pi$. The prepared products by $A$, $|A|$,
and their transposes give $O(m)$ work per step.

For the returned averages, their telescoping argument~\cite[Lemma~2 and proof of Corollary~1]{JT23}
bounds the payoff gap by $6/T$ times the sum of the two initial
Bregman divergences. At our initial point
$(0,\pi)$, the quadratic term of the joint regularizer contributes
$y^\top|A|z^2$, with the square taken coordinatewise. Each simplex
initialization contributes $4D_{\rm KL}(y\Vert\pi)$. Thus for all
$z,y$ in the box and simplex,
\[
 F(\widehat x,y)-F(z,\widehat y)
   \le \frac6T\bigl(y^\top|A|z^2+8D_{\rm KL}(y\Vert\pi)\bigr)
   \le \frac6T\bigl(1+8D_{\rm KL}(y\Vert\pi)\bigr).
\]
The last inequality uses $z_e^2\le1$ and the row-sum bound. Take
$T=\lceil48\lambda\rceil$ and multiply by $\lambda$ to obtain
\eqref{eq:fair-comparator}. Return $g=C\widehat x$ and
$\psi=\psi_{\widehat y}$ in $O(m(1+\lambda))$ work.
\end{proof}

\begin{proof}[Proof of \Cref{fc:thm:fair}]
If $s,t$ are disconnected, return the component of $s$ and the
zero flow. Otherwise suppose first that $\rho\ge(m+2)^{-2}$. Start
from $S=\{s\}$, $f=0$, $v=0$. Each update of \Cref{alg:fair-step}
decreases the nonnegative potential by at least $1/8$, so the
algorithm reaches $\Phi\le\theta$ within $8\Psi_0$ updates.
At that point at most $\theta\le\omega c_e$ of any crossing edge's
outward capacity is unused. The same potential bound gives
$3Q\|Rh\|_\infty\le\omega$ by \Cref{fc:sec:allowance}.

Route the componentwise balanced demand $h=vb-Bf$ once by
\Cref{ca:lem:regression}. Adding this correction gives divergence
$vb$, uses at most $(1+\omega)c_e$ on every edge, and leaves at least
$(1-2\omega)c_e$ flowing outward on each crossing edge. Dividing by
$1+\omega$ gives a feasible witness, since
$(1-2\omega)/(1+\omega)\ge1/(1+\rho)$ for $\omega=\rho/4$.

In this range of $\rho$, the source-degree hypothesis gives
$\Phi_0/\theta=4\delta_G(\{s\})/(\rho c_{\min})=N^{O(1)}$.
Thus $\Psi_0=O(\log N)$, and $p=O(m)$ gives
$\lambda=(3Q/\omega)(\log p+\Psi_0)=O(Q\log N/\rho)$.
Each update costs $O(m(1+\lambda))$, including the level-set
search for a positive price deficit. Summing over $O(\log N)$
updates and adding $O(n+mQ\log^2 N)$ for the isolate scan and
the final correction gives $O(n+mQ\log^2 N/\rho)$.

For $\rho<(m+2)^{-2}$ the allowed work already covers an exact
maximum flow. The component containing $s,t$ has at most $m+1$
vertices, so Edmonds--Karp takes $O(n+m^3)=O(n+m/\rho)$ work.
Return its flow and a minimum cut. Equality of the net outflow
and the cut capacity forces every crossing edge to be saturated
outward.
\end{proof}

\Needspace{10\baselineskip}
\section{Approximate maximum flow with arbitrary capacities}\label[appendix]{app:capacities}

The capacity-spread assumption enters through the quotient tree
construction. We use the capacity windows of Bencz\'ur and
Karger~\cite[Section~8.6]{BK15} to build two quotient trees, so
each cut has a cheap extension in at least one of them.

\begin{corollary}[Arbitrary capacities]\label{ca:cor:real-capacities}
For every componentwise balanced demand on an undirected graph with
arbitrary positive real capacities, the algorithm achieves the congestion
and running-time guarantees of \Cref{ca:cor:flow}.
\end{corollary}
\begin{proof}
Contract the forest $F$ of \Cref{ca:lem:forest} with factor $\chi$.
The quotient $H=G/F$ has $p\le m$ edges and
$k=O(\min\{n,m/\chi\})$ nonisolated vertices. An isolated quotient
vertex represents an entire graph component and has net demand zero.
Set these vertices aside. If $p=0$, the forest rows and forest
correction alone suffice for the throughput search on $G$.
Otherwise set $N=16(m+2)$. Write $q=O(\log^2 N\log\log N)$ for
the quality in \Cref{ca:prop:query-bounds} at contraction factor two
and outer size $N$.

\proofstep{The two-tree guarantee.}
Give each tree edge the capacity of its descendant cut in the full
quotient $H$ and stack the resulting valid rows in $R_H$. For a
componentwise balanced demand $d$, the norm $a=\|R_Hd\|_\infty$
bounds the congestion of both tree flows. Since each flow carries
net demand $d(X)$ across every extension of $X$,
\[
 |d(X)|\le a\min\{\lambda_{T_1}(X),\lambda_{T_2}(X)\}.
\]
Bounding the minimum by $O(q)\delta_H(X)$ for every $X$ gives
$\opt_H(d)\le O(q)a$ by the graph cut formula.

\proofstep{Capacity windows.}
Put $K=N^3\ge kp$ and divide capacities into buckets
$[K^j,K^{j+1})$. Pair buckets $2i,2i+1$ in the first grouping
and $2i-1,2i$ in the second. For the largest crossing capacity $c>0$
of any cut, one grouping puts $c$ in the upper bucket of a pair
$[L,K^2L)$. The components formed by edges above the pair then
lie wholly on one side of the cut. Putting $c$ in the upper
bucket also gives $kpL\le KL\le c$, which will bound the total
cost of edges below the pair.

\proofstep{Building the trees.}
For each grouping, start with singleton trees and process occupied
pairs in decreasing order of capacity. At a pair $[L,K^2L)$,
contract the components formed by edges of capacity at least $K^2L$
and keep the edges of the pair, merging parallels and discarding
loops. Each contracted component is represented by an old tree.
Build a quality-$q$ tree for each nontrivial local component by
\Cref{ca:prop:query-bounds}. Identify the root of each old tree with
the local terminal representing it. The local root becomes the
root of the merged tree, and paths within an old tree stay intact.

A local graph on $t\ge2$ vertices reduces the component count by
$t-1$. Since $t\le2(t-1)$, the local graphs in one grouping have
$O(k)$ vertices altogether, and the assembled tree has $O(k)$ edges.
Join the final components by zero-capacity edges and assign the
full-quotient cut capacities to the tree edges. These capacities
can include contributions from several windows. To bound the cost of
an extension, count those contributions by graph edge. An edge
$uv$ pays $c_{uv}$ for each color change along its tree path,
by \Cref{sec:terminal-tree}.

\proofstep{Extending a cut.}
Fix a positive-capacity cut $X$ and choose the grouping that puts
its largest crossing capacity $c$ in the upper bucket of $[L,K^2L)$.
Color the terminals by membership in $X$. Every old component
has terminals of just one color, so give its entire tree that color.
Extend the colors of the contracted components through each local
tree of the chosen pair at cost at most $q$ times its local cut.
Keep the assigned colors through later attachments and color the
new nodes arbitrarily.

Edges of capacity at least $K^2L$ have paths entirely within
monochromatic old trees and pay zero. A pair edge can change color
only on the local tree path between its two old roots. If both
endpoints lie in one old component, it pays zero as well. At a
color change in a local tree, the pair edges using that tree edge
have total capacity equal to the capacity of its local descendant
cut. The original local tree capacity bounds that sum. Thus edges from the
chosen pair pay at most $q$ times the sum of its local cut
capacities, which count only crossing edges of $X$.

Each edge below $L$ pays less than $L$ at each of at most $O(k)$
color changes. There are at most $p$ such edges, so for the
selected tree $T$,
\[
              \lambda_T(X)\le q\delta_H(X)+O(kpL)
                          =O(q)\delta_H(X),
\]
since $kpL\le c\le\delta_H(X)$. A zero-capacity cut extends at
zero cost in either tree by giving each component its terminal color.

\proofstep{Work and routing.}
Sorting the occupied pairs and maintaining components by union--find
takes $O((k+p)\log N)$ work, independently of gaps between pairs.
Each grouping uses each edge once and $O(k)$ local vertices in total.
Even after merging parallel edges, each local input has capacity
spread at most $pK^2=N^{O(1)}$. Summing
\eqref{eq:outer-scale-construction} over the local calls gives
\[
 O\!\left(p\log^2 N+k\log^7 N\log\log N\right).
\]
Give each of the $O(k)$ calls failure allowance $N^{-(a+2)}$ so all
succeed with high probability. The tree products of
\Cref{sec:tree-queries} compute the full-quotient denominators and
prepare both trees in $O(k+p)$ work, also the cost of each product.

Reattach $F$ to both trees, giving its edges capacities $\chi c_e$.
With $P$ the map that sums demand on forest components, the new
rows are valid by \Cref{ca:lem:forest} and their combined norm is
$a(b)=\max\{a_F(b),\|R_HP b\|_\infty\}$. On $H$,
\Cref{ca:lem:regression} routes $Pb$ at congestion
$O(q)\|R_HP b\|_\infty$. Lifting through $F$ then routes $b$ at
congestion $O(\chi q)a(b)$ in $O(n+m+pq\log^2 N)$ total work by
\Cref{ca:lem:forest-lift}. The attached rows have linear-work
products. These are the construction and correction bounds used in
\Cref{ca:cor:flow}, including its bound using $k\le n$, so its
throughput search and work cap apply.
\end{proof}

\section*{AI methodology}

The author used OpenAI's GPT-6, in particular GPT-6.2 Astra Pro,
which contributed some of
the technical ideas in the congestion-approximator hierarchy and
its construction (\Cref{sec:ca,sec:ca-computation}) and in the flow
repair routine under deletions (\Cref{sec:du-ppr}). The model
developed the argument and wrote \Cref{app:capacities} in full.
It also assisted in simplifying the algorithms and analyses and
in drafting and revising the exposition throughout the paper.
The author assumes responsibility for all content.

\begingroup
\raggedright
\interlinepenalty=10000

\let\originalbibliography\thebibliography
\renewcommand{\thebibliography}[1]{%
  \originalbibliography{#1}%
  \setlength{\itemsep}{2.5pt plus 1pt minus 1pt}%
}
\bibliographystyle{alpha}
\bibliography{references}

@inproceedings{ADK23,
  author    = {Daniel Agassy and Dani Dorfman and Haim Kaplan},
  title     = {{\href{https://arxiv.org/abs/2205.10301}{Expander Decomposition with Fewer Inter-Cluster Edges Using a Spectral Cut Player}}},
  booktitle = {50th International Colloquium on Automata, Languages, and Programming (ICALP)},
  year      = {2023},
  url       = {https://arxiv.org/abs/2205.10301}
}

@misc{ADK25,
  author = {Daniel Agassy and Dani Dorfman and Haim Kaplan},
  title  = {{\href{https://arxiv.org/abs/2510.23913}{Expander Decomposition for Non-Uniform Vertex Measures}}},
  year   = {2025},
  note   = {arXiv:2510.23913}
}

@inproceedings{ADK26,
  author    = {Daniel Agassy and Dani Dorfman and Haim Kaplan},
  title     = {{\href{https://arxiv.org/abs/2511.06574}{Improved Tree Sparsifiers in Near-Linear Time}}},
  booktitle = {53rd International Colloquium on Automata, Languages, and Programming (ICALP)},
  year      = {2026},
  note      = {arXiv:2511.06574},
  pages = {7:1--7:23}
}

@article{BK15,
  author  = {Andr{\'a}s A. Bencz{\'u}r and David R. Karger},
  title   = {{\href{https://doi.org/10.1137/070705970}{Randomized Approximation Schemes for Cuts and Flows in Capacitated Graphs}}},
  journal = {SIAM Journal on Computing},
  volume  = {44},
  number  = {2},
  pages   = {290--319},
  year    = {2015},
  doi     = {10.1137/070705970}
}

@misc{FLL25,
  author = {Henry Fleischmann and George Z. Li and Jason Li},
  title  = {{\href{https://arxiv.org/abs/2507.09729}{Improved Directed Expander Decompositions}}},
  year   = {2025},
  note   = {arXiv:2507.09729}
}

@inproceedings{FLL26weak,
  author    = {Henry Fleischmann and George Z. Li and Jason Li},
  title = {{\href{https://doi.org/10.4230/LIPIcs.ICALP.2026.91}{Faster Weak Expander Decompositions and Approximate Max Flow}}},
  booktitle = {53rd International Colloquium on Automata, Languages, and Programming (ICALP)},
  pages     = {91:1--91:20},
  year      = {2026},
  doi       = {10.4230/LIPIcs.ICALP.2026.91}
}

@inproceedings{HMR25,
  author = {Monika Henzinger and Robin M{\"u}nk and Harald R{\"a}cke},
  title = {{\href{https://arxiv.org/abs/2511.03716}{An Improved Quality Hierarchical Congestion Approximator in Near-Linear Time}}},
  year = {2026},
  note = {Section numbers refer to the full version, arXiv:2511.03716},
  booktitle = {Proceedings of the 58th Annual ACM Symposium on Theory of Computing (STOC)},
  pages = {1417--1428}
}

@inproceedings{JT23,
  author = {Arun Jambulapati and Kevin Tian},
  title = {{\href{https://arxiv.org/abs/2303.15627v2}{Revisiting Area Convexity: Faster Box-Simplex Games and Spectrahedral Generalizations}}},
  year = {2023},
  note = {Algorithm and lemma numbers refer to the full version, arXiv:2303.15627v2},
  booktitle = {Advances in Neural Information Processing Systems},
  volume = {36},
  pages = {57583--57596}
}

@article{KRV09,
  author  = {Rohit Khandekar and Satish Rao and Umesh Vazirani},
  title   = {{\href{https://doi.org/10.1145/1538902.1538903}{Graph Partitioning Using Single Commodity Flows}}},
  journal = {Journal of the ACM},
  volume  = {56},
  number  = {4},
  year    = {2009}
}

@inproceedings{LL25faircuts,
  author = {Jason Li and Owen Li},
  title = {{\href{https://arxiv.org/abs/2411.19098}{A Simple and Fast Algorithm for Fair Cuts}}},
  year = {2025},

  booktitle = {Integer Programming and Combinatorial Optimization (IPCO)},
  pages = {400--411}
}

@inproceedings{LNPS23faircuts,
  author = {Jason Li and Danupon Nanongkai and Debmalya Panigrahi and Thatchaphol Saranurak},
  title = {{\href{https://arxiv.org/abs/2203.00751v2}{Near-Linear Time Approximations for Cut Problems via Fair Cuts}}},
  booktitle = {Proceedings of the ACM-SIAM Symposium on Discrete Algorithms (SODA)},
  year = {2023},
  note = {Full version: arXiv:2203.00751v2}
}

@inproceedings{Peng16,
  author = {Richard Peng},
  title = {{\href{https://arxiv.org/abs/1411.7631}{Approximate Undirected Maximum Flows in $O(m\,\mathrm{polylog}(n))$ Time}}},
  booktitle = {Proceedings of the ACM-SIAM Symposium on Discrete Algorithms (SODA)},
  pages = {1862--1867},
  year = {2016},
  doi = {10.1137/1.9781611974331.ch130}
}

@inproceedings{RST14,
  author = {Harald R{\"a}cke and Chintan Shah and Hanjo T{\"a}ubig},
  title = {{\href{https://doi.org/10.1137/1.9781611973402.17}{Computing Cut-Based Hierarchical Decompositions in Almost Linear Time}}},
  booktitle = {Proceedings of the ACM-SIAM Symposium on Discrete Algorithms (SODA)},
  pages = {227--238},
  year = {2014},
  doi = {10.1137/1.9781611973402.17}
}

@inproceedings{SG25,
  author    = {Aurelio L. Sulser and Maximilian Probst Gutenberg},
  title     = {{\href{https://arxiv.org/abs/2403.04542}{Near-Optimal Algorithm for Directed Expander Decompositions}}},
  booktitle = {ICALP},
  year      = {2025},
  note      = {arXiv:2403.04542}
}

@inproceedings{SW19,
  author    = {Thatchaphol Saranurak and Di Wang},
  title     = {{\href{https://arxiv.org/abs/1812.08958}{Expander Decomposition and Pruning: Faster, Stronger, and Simpler}}},
  booktitle = {Proceedings of the ACM-SIAM Symposium on Discrete Algorithms (SODA)},
  pages     = {2616--2635},
  year      = {2019},
  url       = {https://arxiv.org/abs/1812.08958}
}

@inproceedings{Sherman13,
  author = {Jonah Sherman},
  title = {{\href{https://arxiv.org/abs/1304.2077}{Nearly Maximum Flows in Nearly Linear Time}}},
  booktitle = {Proceedings of the 54th IEEE Symposium on Foundations of Computer Science (FOCS)},
  pages = {263--269},
  year = {2013},
  doi = {10.1109/FOCS.2013.36}
}

@misc{VaughnSignDescent,
  author = {Trevor Vaughn},
  title = {{\href{https://arxiv.org/pdf/2611.27907}{Linear-Size Spectral Sparsification by Sign Descent}}},
  note = {arXiv:2611.27907},
  year   = {2026}
}

@inproceedings{BFC00,
  author = {Michael A. Bender and Mart{\'i}n Farach-Colton},
  title = {{\href{https://doi.org/10.1007/10719839_9}{The {LCA} Problem Revisited}}},
  booktitle = {{LATIN 2000}: Theoretical Informatics},
  series = {Lecture Notes in Computer Science},
  volume = {1776},
  pages = {88--94},
  year = {2000},
  doi = {10.1007/10719839_9}
}
\endgroup
\end{document}